\RequirePackage[2020-02-02]{latexrelease}
\documentclass{article} 

\usepackage{preprint}

\usepackage{amsmath, amsthm, amssymb, amsfonts}

\usepackage{natbib}
\setcitestyle{square, comma, numbers, sort&compress}

\usepackage[utf8]{inputenc}	
\usepackage[T1]{fontenc}	
\usepackage{xcolor}		
\usepackage[colorlinks = true,
            linkcolor = purple,
            urlcolor  = blue,
            citecolor = cyan,
            anchorcolor = black]{hyperref}	
\usepackage{booktabs} 		
\usepackage{nicefrac}		
\usepackage{microtype}		
\usepackage{float}			
\usepackage{multicol}		

\usepackage{lipsum}		

\usepackage{newfloat}
\DeclareFloatingEnvironment[name={Supplementary Figure}]{suppfigure}
\usepackage{sidecap}
\sidecaptionvpos{figure}{c}

\usepackage{titlesec}
\titlespacing\section{0pt}{12pt plus 3pt minus 3pt}{1pt plus 1pt minus 1pt}
\titlespacing\subsection{0pt}{10pt plus 3pt minus 3pt}{1pt plus 1pt minus 1pt}
\titlespacing\subsubsection{0pt}{8pt plus 3pt minus 3pt}{1pt plus 1pt minus 1pt}

\usepackage{multirow}
\usepackage{graphicx}

\newcommand{\deHomTop}{\texttt{deHomTop808} }
\newcommand{\deHomTopns}{\texttt{deHomTop808}}

\newcommand\SecNum[1]{%
  \hyperref[#1]{\getrefnumber{#1}}%
}

\usepackage[font=small]{caption}
\usepackage[font=small]{subcaption}

\usepackage{placeins}

\usepackage{tikz,xcolor,hyperref}

\definecolor{lime}{HTML}{A6CE39}
\DeclareRobustCommand{\orcidicon}{
	\begin{tikzpicture}
	\draw[lime, fill=lime] (0,0) 
	circle [radius=0.16] 
	node[white] {{\fontfamily{qag}\selectfont \tiny ID}};
	\draw[white, fill=white] (-0.0625,0.095) 
	circle [radius=0.007];
	\end{tikzpicture}
	\hspace{-2mm}
}
\foreach \x in {A, ..., Z}{\expandafter\xdef\csname orcid\x\endcsname{\noexpand\href{https://orcid.org/\csname orcidauthor\x\endcsname}
			{\noexpand\orcidicon}}
}

\title{An 808 Line Phasor-Based Dehomogenisation Matlab Code For Multi-Scale Topology Optimisation}

\usepackage{xwatermark}
\newwatermark[firstpage,color=gray!90,angle=0,scale=0.28, xpos=0in,ypos=-5in]{*correspondence: \texttt{pdlj@dtu.dk}}

\usepackage{authblk}

\author[1]{Rebekka Varum Woldseth\orcidB{}}
\author[1]{Ole Sigmund\orcidE{}}
\author[1\thanks{\tt{pdlj@dtu.dk}}]{Peter D\o rffler Ladegaard Jensen\orcidA{}}

\affil[1]{Department of Mechanical Engineering, Technical University of Denmark. Koppels All\'{e}, B.404, 2800 Kgs. Lyngby, Denmark.}

\usepackage{graphicx}
\usepackage{array}
\usepackage{tabularx}
\usepackage{makecell}
\usepackage{cprotect}
\usepackage{textcomp}
\usepackage{appendix}
\usepackage{xurl}

\usepackage{placeins}

\usepackage{lmodern}
\usepackage{algorithm}
\usepackage{algpseudocode}

\usepackage{cancel}
\usepackage{bm}
\usepackage{amsmath}
\usepackage{amssymb}
\usepackage{relsize}
\usepackage{amssymb}

\usepackage{threeparttable}     

\usepackage{booktabs}

\usepackage{pgfplots}
\usepackage{tikz}
\usetikzlibrary{shapes.geometric, arrows,matrix}
\usepgfplotslibrary{fillbetween}

\pgfplotsset{compat=newest}
\pgfplotsset{plot coordinates/math parser=false}
\newlength\figureheight
\newlength\figurewidth
\usetikzlibrary{arrows.meta,
                chains,
                positioning,
                quotes,
                shapes.geometric,
                arrows,
                backgrounds,
                calc}
                
\definecolor{mycolor4}{rgb}{0.00000,0.44700,0.74100}%
\definecolor{mycolor5}{rgb}{0.85000,0.32500,0.09800}%
\definecolor{mycolor6}{rgb}{0.92900,0.69400,0.12500}%
\definecolor{mycolor3}{rgb}{0.46670,0.67450,0.18820}%
\definecolor{mycolor2}{rgb}{0.49400,0.18400,0.55600}%
\definecolor{mycolor1}{rgb}{0.63530,0.07840,0.18430}%

\definecolor{pycolor1}{HTML}{332288}%
\definecolor{pycolor2}{HTML}{88CCEE}%
\definecolor{pycolor3}{HTML}{44AA99}%
\definecolor{pycolor4}{HTML}{117733}%
\definecolor{pycolor5}{HTML}{999933}%
\definecolor{pycolor6}{HTML}{DDCC77}%
\definecolor{pycolor7}{HTML}{CC6677}%
\definecolor{pycolor8}{HTML}{882255}%
\definecolor{pycolor9}{HTML}{AA4499}%

\definecolor{bycolor1}{HTML}{4477AA}%
\definecolor{bycolor2}{HTML}{66CCEE}%
\definecolor{bycolor3}{HTML}{228833}%
\definecolor{bycolor4}{HTML}{CCBB44}%
\definecolor{bycolor5}{HTML}{EE6677}%
\definecolor{bycolor6}{HTML}{AA3377}%
\definecolor{bycolor7}{HTML}{BBBBBB}%

\definecolor{vycolor1}{HTML}{0077BB}%
\definecolor{vycolor2}{HTML}{33BBEE}%
\definecolor{vycolor3}{HTML}{009988}%
\definecolor{vycolor4}{HTML}{EE7733}%
\definecolor{vycolor5}{HTML}{CC3311}%
\definecolor{vycolor6}{HTML}{EE3377}%

\usepackage{enumitem}                

\usepackage{multirow}
\usepackage{graphicx}

\usepgfplotslibrary{groupplots, external}

\usepackage[normalem]{ulem}

\usepackage{lineno} 
\let\oldequation\equation
\let\oldendequation\endequation
\let\oldalign\align
\let\oldendalign\endalign

\definecolor{LightCyan}{rgb}{0.88,1,1}

\usepackage{listings}
\definecolor{LightCyan}{rgb}{0.88,1,1}
\definecolor{mygreen}{RGB}{28,172,0} 
\definecolor{mylilas}{RGB}{170,55,241}
\let\origthelstnumber\thelstnumber
\makeatletter
\newcommand*\Suppressnumber{%
  \lst@AddToHook{OnNewLine}{%
    \let\thelstnumber\relax%
     \advance\c@lstnumber-\@ne\relax%
    }%
}

\newcommand*\Reactivatenumber[1]{%
  \setcounter{lstnumber}{\numexpr#1-1\relax}
  \lst@AddToHook{OnNewLine}{%
   \let\thelstnumber\origthelstnumber%
   \refstepcounter{lstnumber}
  }%
}

\usepackage{cleveref}

\renewenvironment{equation}
  {\linenomathNonumbers\oldequation}
  {\oldendequation\endlinenomath}
\renewenvironment{align}
  {\linenomathNonumbers\oldalign}
  {\oldendalign\endlinenomath}

\newcolumntype{L}{>{\raggedright\arraybackslash}X}

\newenvironment{Figure}
  {\par\medskip\noindent\minipage{\linewidth}}
  {\endminipage\par\medskip}

\newenvironment{Table}
  {\par\medskip\noindent\minipage{\linewidth}}
  {\endminipage\par\medskip}

\begin{document}


\maketitle

\begin{abstract}


This work presents an 808-line Matlab educational code for combined multi-scale topology optimisation and phasor-based dehomogenisation titled \deHomTopns. The multi-scale formulation utilises homogenisation of optimal microstructures to facilitate efficient coarse-scale optimisation. Dehomogenisation allows for a high-resolution single-scale reconstruction of the optimised multi-scale structure, achieving minor losses in structural performance, at a fraction of the computational cost, compared to its large-scale topology optimisation counterpart. The presented code utilises stiffness optimal Rank-2 microstructures to minimise the compliance of a single-load case problem, subject to a volume fraction constraint. By exploiting the inherent efficiency benefits of the phasor-based dehomogenisation procedure, on-the-fly dehomogenisation to a single-scale structure is obtained. The presented code includes procedures for structural verification of the final dehomogenised structure by comparison to the multi-scale solution. The code is introduced in terms of the underlying theory and its major components, including examples and potential extensions, and can be downloaded from \href{https://github.com/peterdorffler/deHomTop808.git}{Github}.
  
\end{abstract}

\vspace{0.35cm}

\begin{multicols}{2}

  \section{Introduction}
Topology optimisation is an established and systematic tool in engineering design and research. The premise is to obtain optimised structural layouts within a given physical design domain, according to specific objectives and constraints. This methodology is particularly valuable in industry applications due to its limited requirement for prior knowledge of the design. Topology optimisation has been applied to numerous fields of study, varying from elastic problems to photonics. Educational dissemination of topology optimisation methods has been achieved through publications of complete interactive apps or software, originating with the \texttt{top99} code by \citet{Sigmund2001}. Most of these publications focus on smaller-scale problems, whereas \citet{Aage2015} provides a large-scale topology optimisation framework using PETSc. An extensive review of educational and publicly available codes is presented in \citet{WangXSZhang2021}.  

Giga-scale topology optimisation was shown by \citet{Aage2017} and \citet{Baandrup2020} to reveal important aspects of structural design, as novel high-performing structures with bridging length scales were found. These frameworks require a considerable amount of computational resources to achieve these results, due to the discretisation needed to realise the minimum length scale. An alternative approach is to relax the length scale discretisation using a periodic homogenised design representation. Homogenisation-based topology optimisation constitutes the foundation of the research field, first introduced in the seminal paper by \citet{BendsoeKikuchi1988}. Due to the multi-scale nature of the optimised structures, obtaining a corresponding realisation on a single length-scale is not trivial, and has likely been a contributing factor as to why this original method stayed passive until recent years. 

The single-scale reconstruction of homogenised results involves approximating the conformality and constant periodicity of the multi-scale structure on a finite length scale, as first established in \citet{Pantz2008} with the introduction of the projection-based dehomogenisation approach, which was later improved by \citet{Groen2018}. Numerous other frameworks have been proposed, based on a variety of different concepts. \citet{Wu2021b} presented an approach using meshing techniques and \citet{Stutz2022} developed a streamline tracing based procedure. A pattern-generating neural network was employed in \citet{Elingaard2022}, whereas \citet{Garnier2022} utilised reaction-diffusion-driven pattern formation. Along similar lines, \citet{Woldseth2023} recently proposed a phasor-based dehomogenisation heuristic, based on noise functions from computer graphics. \citet{Senhora2022} utilised spinodal metamaterials, with inherent similarities in parameterisation, to the phasor noise approach. The dehomogenisation capabilities have also been extended to incorporate multiple loads~\citep{Jensen2022} and manage 3D domains~\citep{Geoffroy-Donders2020,Groen2020,Jensen2023a}.

By considering a homogenised solution in a layer-wise manner, the oriented periodic multi-scale microtructures are projected to finite periodic laminates with smoothly varying orientations. To ensure structural integrity on a finite length-scale, the dehomogenised design must be separated from the underlying homogenisation assumptions, reflected by the orientation conformality objective of the projection. There is an inherent conflict between the microscale assumption of perfect bonding and the macroscale discontinuities arising from evaluating the microstructure parameters at the centre of each macroscale element~\citep{Kumar2020}. Bonding on the macroscale is ensured by the smoothness of the conformal-like mapping, but requires distortion of the microstructure geometry~\citep{Allaire2019}. The dehomogenised design should converge towards the correct microstructure shape for increasingly fine length-scales. To this end, the proportions of the microstructure should be preserved in the realisation, which facilitates requirements of spatially coherent layer-spacing. The concepts of best approximating conformality and obtain uniform layer-spacing causes conflict for any non-trivial orientation field, and the heuristic nature of different dehomogenisation approaches emerges from measures aiming to obtain a balanced trade-off between these incompatible requirements.

\begin{figure*}[t]
    \centering
    \makebox[\textwidth][c]{
        \begin{subfigure}{90mm}
            \includegraphics[width=90mm]{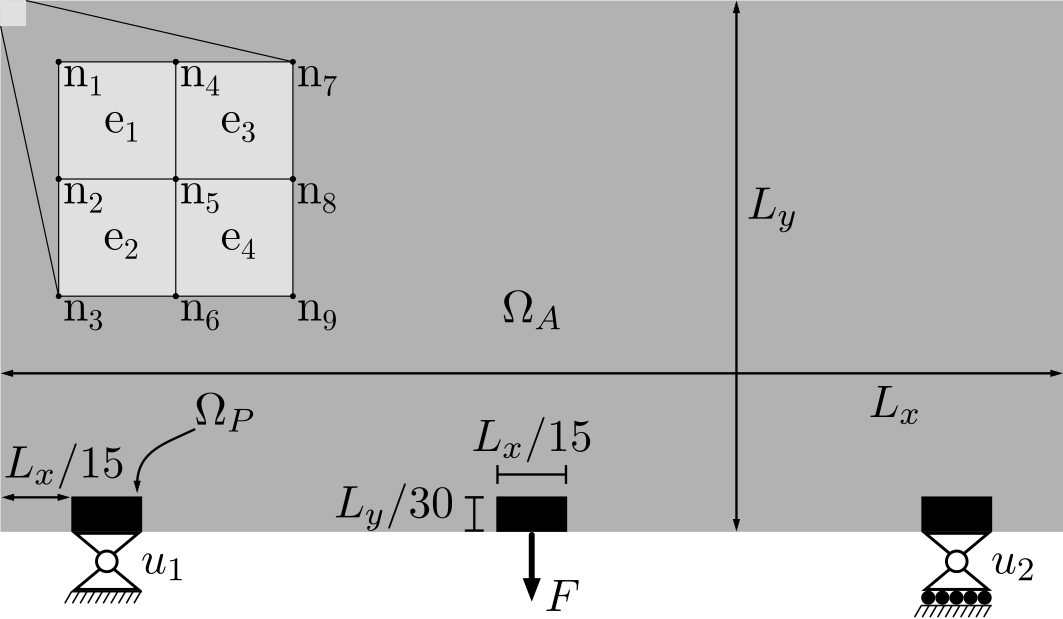}
            \caption{}
            \label{fig:model}
        \end{subfigure}\hspace{5mm} %
        \begin{subfigure}{90mm}
            \includegraphics[width=90mm]{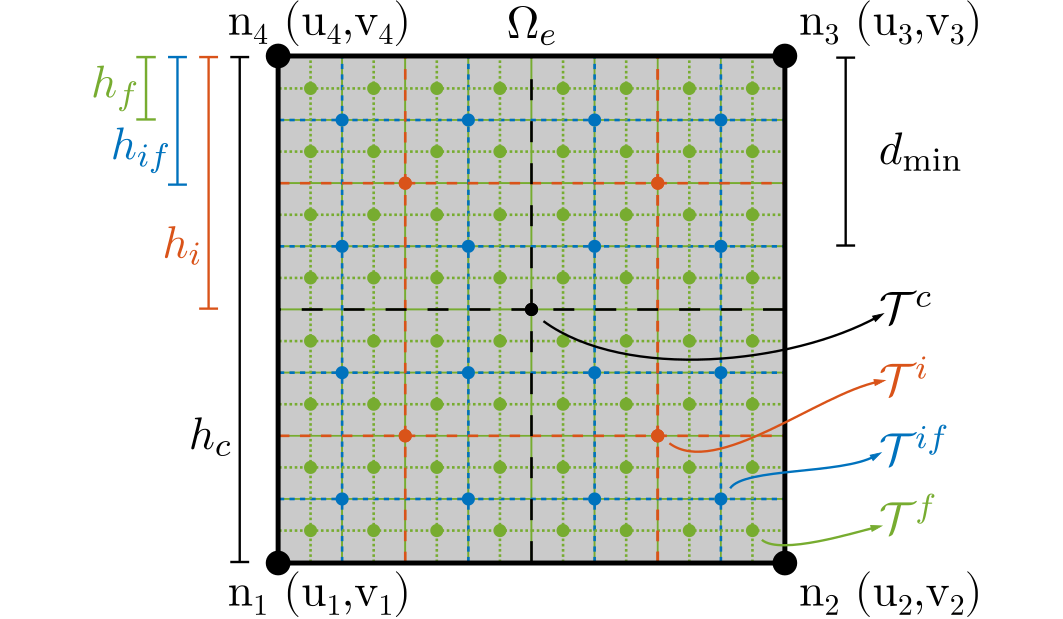}
            \caption{}
            \label{fig:FEmodel}
        \end{subfigure}
    }
    \vspace{5mm}\\
    \makebox[\textwidth][c]{
        \begin{subfigure}{185mm}
            \includegraphics[width=185mm]{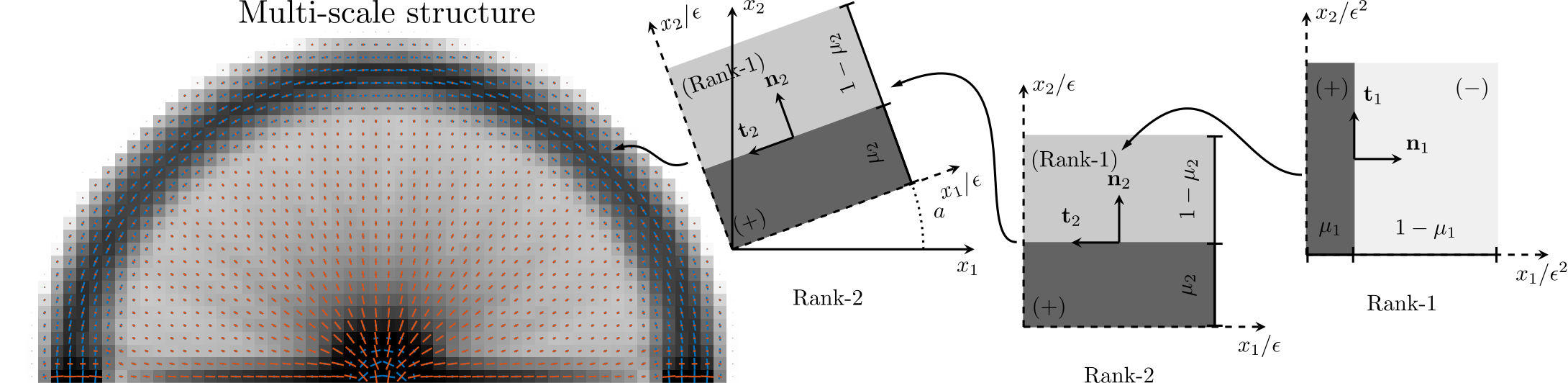}
            \caption{}
            \label{fig:rankModel}
        \end{subfigure}
    }
    \vspace{5mm}\\
    \makebox[\textwidth][c]{
        \begin{subfigure}{185mm}
            \includegraphics[width=185mm]{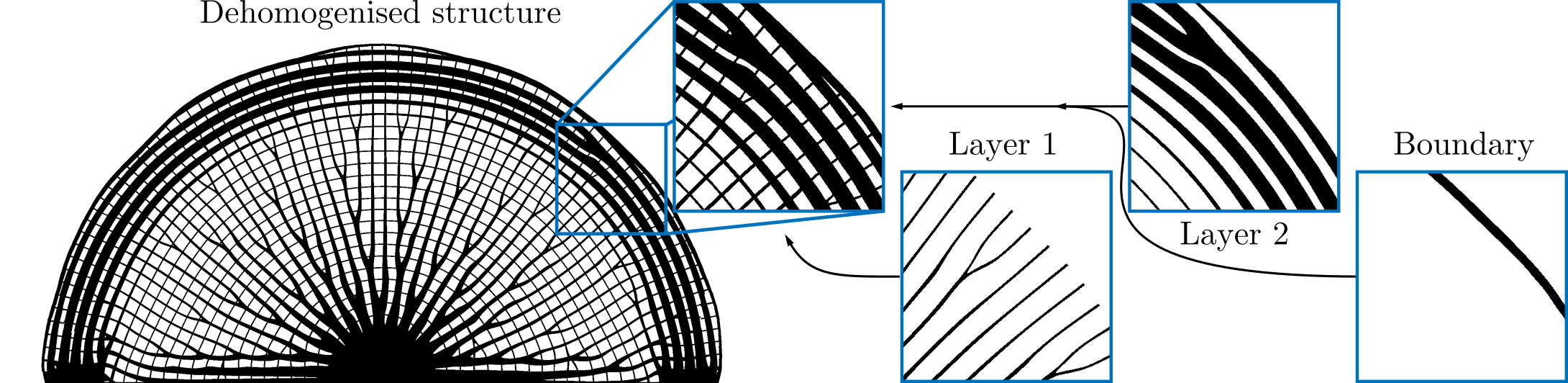}
            \caption{}
            \label{fig:dehomModel}
        \end{subfigure}
    }%
    \caption{
        An overview of the example model and the multi-scale topology optimisation and dehomogenisation  principles. The model of the baseline problem being optimised, as well as the element and node counting conventions are illustrated in (a). The relation between the FE-grid, FE-element and dehomogenisation grids is illustrated in (b). (c) illustrates the nature of the multi-scale structure being optimised, with element-wise Rank-2 material, and (d) how the multi-scale structure relates to the realised single-scale structure obtained by phasor-based dehomogenisation.
    }
    \label{fig:test_model}
\end{figure*}

Nevertheless, dehomogenisation approaches have been proven to produce large-scale topology optimisation results from coarse parameterisations of multi-scale structures. The obtained structural performance is consistently within 10\% of the conventional large-scale approaches, while the computational cost is reduced by several orders of magnitudes, even allowing for solving on conventional workstation PCs~\citep{Groen2018,Jensen2022,Jensen2023a,Woldseth2023}.

The phasor noise-based dehomogenisation procedure \citep{Woldseth2023} has proven to provide a highly efficient and well-performing alternative to the more conventional approaches. By exploiting the compact representation and real-time rendering properties of procedural noise functions~\citep{Tricard2020}, many of the computational restrictions of existing methods are overcome, achieving advantageous scalability properties facilitating potential for on-the-fly dehomogenisation. The dehomogenisation implementation is, however, nontrivial and involves a number of computer graphics aspects and heuristic adaptions.  
Motivated by these observations, an integrated and accessible dehomogenisation and homogenisation-based topology optimisation framework is presented.

The \deHomTop code presents an integrated workflow combining minimum compliance multi-scale topology optimisation utilising a stiffness-optimal material parameterisation with dehomogenisation to a fine-scale physica structural design. The code constitutes a comprehensive design tool, aimed at giving everyday structural engineers an efficient and adaptable large-scale topology optimisation framework, that can run on simple computer hardware. The \deHomTop code is capable of,
\begin{itemize}[leftmargin=*]
    \item[1.] Maximising stiffness of a single-load case scenario subjected to a constraint on the maximum allowed material use.
    \item[2.] Performing on-the-fly dehomogenisation of the stiffness-optimised structure to a given minimum length scale.
    \item[3.] Verify the dehomogenised fine-scale structure for stiffness and volume fraction performance. 
\end{itemize}

A multi-scale composite material, the orthogonal Rank-2 laminate, is used as the design material due to its stiffness optimal properties for single-load problem cases. The Rank-2 material is, however, parameterised on two different length scales, which makes the realisation of the microstructure challenging. Instead, a single-scale translation of this parameterisation is introduced, with a corresponding mapping to its multi-scale counterpart, to obtain the Rank-2 material properties. This single-scale parameterisation is subsequently utilised to dehomogenise the Rank-2 structure by the phasor-based approach. With the single-scale parameterisation, minimum length-scale control is introduced, increasing the relevancy for manufacturing-oriented design applications.
For improved readability and compactness, the \deHomTop code consists of a new formulation and implementation of the homogenisation-based optimisation procedure, as well as a simplified version of the original phasor-based dehomogenisation procedure \citep{Woldseth2023}. A visual overview of the code capabilities is seen in \autoref{fig:test_model}.
The code presented builds on the concept of the \texttt{top88} code \citep{Andreassen2011} and the \texttt{top250} code \citep{Ferrari2021} utilising the modified OC routine. 
The code was developed and tested using MATLAB, version R2023b, including MATLAB Image Processing Toolbox\footnote[1]{The code can also be executed without the MATLAB Image Processing Toolbox, but the behaviour may change, which is discussed in \autoref{sec:howto}.} and is tested on Linux kernel 6.6 Pop!\_OS 22.04 LTS, Mac OS Sonoma 14.4 and Windows 11 operative systems.

To provide fundamentals for how \deHomTop is operated, this paper starts with an introduction of the executable and its various inputs in \autoref{sec:howto}. For understanding the mathematical models and methodologies utilised in this code, fundamental theory is presented for the multi-scale topology optimisation problem in \autoref{sec:TO} and the phasor-based dehomogenisation in \autoref{sec:dehom}. \autoref{sec:code} provides a more detailed description of the \deHomTop code, in terms of how the code is organised, the different parameters, and how it relates to the theory in \autoref{sec:TO}- \SecNum{sec:dehom}. Examples and performance considerations are presented in \autoref{sec:ex}. Potential extensions to the code are presented in \autoref{sec:extension}, and final remarks are given in \autoref{sec:end}. The \deHomTop Matlab code including additional models can be downloaded from \href{https://github.com/peterdorffler/deHomTop808.git}{Github}.



  \begin{table*}[b]
    \centering
    \caption{Input and output arguments of \deHomTopns.}
    \begin{tabular}{@{}llll@{}}
        \toprule
        Argument & Type & Domain & Description \\ \midrule
        \texttt{nelX} and \texttt{nelY} & \texttt{int} & $\mathbb{N}$ & \begin{tabular}[c]{@{}l@{}} Define the physical dimensions and discretisation of the rectangular \\ computational domain.  \end{tabular} \\
        \texttt{volFrac} & \texttt{double} & $[0,1]$ & \begin{tabular}[c]{@{}l@{}} Specifies the allowed material volume fraction within the\\ computational domain.  \end{tabular} \\
        \texttt{rMin} & \texttt{double} & $\mathbb{R}_{>0}$ & \begin{tabular}[c]{@{}l@{}} Determines the filter radius for the multi-scale material.  \end{tabular} \\
        \texttt{wMin} and \texttt{wMax} & \texttt{double} & $(0,1]$ & \begin{tabular}[c]{@{}l@{}} Define the minimum and maximum relative bounds for the thickness\\ of the laminated material. Note that \texttt{wMin} $\leq$ \texttt{wMax}.  \end{tabular} \\
        \texttt{dMin} & \texttt{double} & $\mathbb{R}_{>0}$ & \begin{tabular}[c]{@{}l@{}} Sets the physical minimum length scale of the dehomogenised design.  \end{tabular} \\
        \texttt{deHomFrq} & \texttt{int} & $\mathbb{Z}_{\geq 0}$ & \begin{tabular}[c]{@{}l@{}} Frequency for on-the-fly dehomogenisation during optimisation.  \end{tabular} \\
        \texttt{eval} & \texttt{logical} & \{0,\,1\}& \begin{tabular}[c]{@{}l@{}}  Whether to perform post evaluation of dehomogenised structure.  \end{tabular} \\ \midrule
        \texttt{rhoPhys} & \texttt{double} & $[0,1]^{n \times m}$ & Represents the relative element densities of the dehomogenised result. \\
        \texttt{TO} & \texttt{struct} & - & Structure array containing information about the multi-scale structure.
        \\ 
        $\ $\rotatebox[origin=c]{180}{$\Lsh$} \texttt{TO.w} &   \texttt{double} & $ [0,1]^{k \times l} $ & \begin{tabular}[c]{@{}l@{}}  Multi-scale thickness field, ordered column-wise $k$-sized fields for $l$ layers.\\ The default number of layers is $l=2$. \end{tabular}\\ 
        $\ $\rotatebox[origin=c]{180}{$\Lsh$} \texttt{TO.N} &  \texttt{double} &   $\mathbb{R} ^{k \times 2 \times l}$ & \begin{tabular}[c]{@{}l@{}} Multi-scale layer normal fields, ordered component-column-wise $k$-sized\\ fields for $l$ layers. \end{tabular}\\ 
        $\ $\rotatebox[origin=c]{180}{$\Lsh$} \texttt{TO.f} & \texttt{double} & $[0,1]$ & Multi-scale resulting volume fraction. \\ 
        $\ $\rotatebox[origin=c]{180}{$\Lsh$} \texttt{TO.J} & \texttt{double} & $\mathbb{R}_{\geq 0}$  & Multi-scale resulting volume compliance. \\ \bottomrule
    \end{tabular}
    \label{tab:input_args}
\end{table*}

\section{Getting Started with \deHomTop} \label{sec:howto}
The Matlab code is executed with the following function call,
\begin{lstlisting}[numbers=none,deletekeywords={eval},xleftmargin=1em,framexleftmargin=0em]
[rhoPhys,TO] = deHomTop808(nelX, nelY, volFrac, rMin, wMin, wMax, dMin, deHomFrq, eval, TO)
\end{lstlisting}
The first nine arguments are required to run the code, while the tenth argument, \texttt{TO}, represents a previously obtained multi-scale result, facilitating direct dehomogenisation of the multi-scale structure encoded in \texttt{TO}. The input and output arguments are detailed in \autoref{tab:input_args}.

The code executes in three phases; the first phase is the initialisation of the multi-scale optimisation problem and the phasor-based dehomogenisation problem; the second phase performs the multi-scale topology optimisation, from which the optimised solution is obtained. Depending on \texttt{deHomFrq}, on-the-fly dehomogenisation can also be performed in this phase. The final phase is constituted by the post dehomogenisation of the optimised result, and if \texttt{eval} is enabled, the dehomogenised structure is evaluated for structural compliance. If \texttt{TO} is provided as input, the second phase is skipped.


In each optimisation iteration the following line is printed
\begin{lstlisting}[numbers=none,columns=fullflexible,xleftmargin=1em,framexleftmargin=0em]
Itr:   - Obj: - J: - S: - Vol: - (ph: -) ch: - Time: - (ph: -)
\end{lstlisting}
indicating the iteration count \verb|Itr|, current objective \verb|Obj|, corresponding compliance value \verb|J|, weighted indicator field volume fraction \verb|S|, the design volume fraction \verb|Vol|, maximal iterative change \verb|ch| and computational time \verb|Time|. \verb|(ph: -)| indicates the corresponding phasor-based dehomogenisation result, and is only assigned positive values in iterations where dehomogenisation is executed, which is controlled by the \texttt{deHomFreq} value. During the optimisation iteration, the multi-scale and dehomogenised structures are plotted. From the post-dehomogenisation phase, the volume fraction, structural compliance and weighted compliance errors are evaluated with respect to the multi-scale result, and printed to the user.
\begin{Figure}
    \centering
            \includegraphics[width=\linewidth]{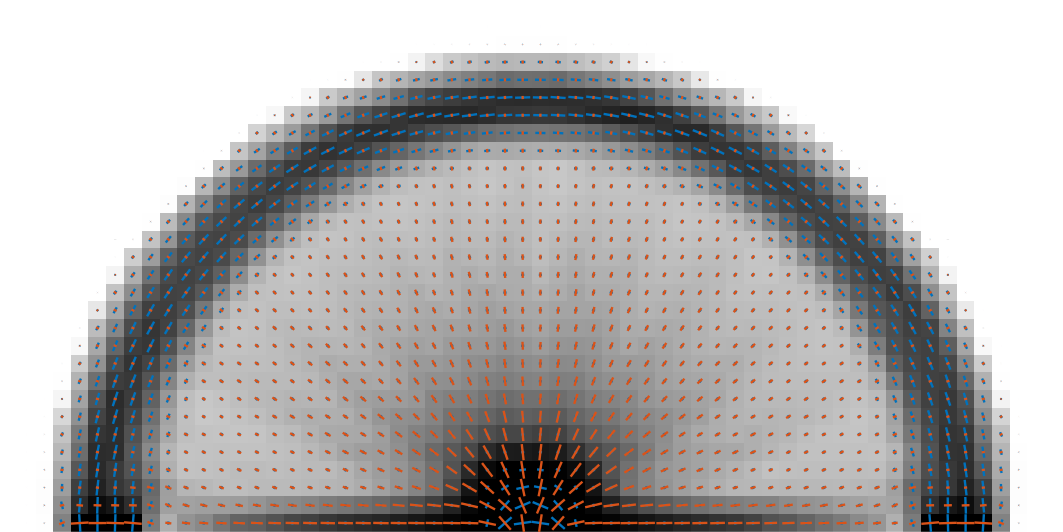}
   \captionof{figure}{Final multi-scale structure from running the \deHomTop code the first time, with lines indicating the Rank-2 orientations and layer widths. The greyscale indicates relative density.}
    \label{fig:firstRun01}
\end{Figure}

The default problem is a single-load bridge problem with two distributed simple supports (\autoref{fig:model}). The bridge is subjected to surface traction, where supports and reactions are applied to solid passive areas of the computational domain. Running the code with
\begin{lstlisting}[numbers=none,columns=fullflexible,xleftmargin=1em,framexleftmargin=0em]
[rhoPhys0,TO] = deHomTop808(60,30,0.3,2,0.1,1,0.2,10,true)
\end{lstlisting}
will run all three phases of the code sequentially. The resulting multi-scale structure from the second phase is presented in \autoref{fig:firstRun01}. After the dehomogenisation in the third phase, the obtained single-scale structure (\autoref{fig:firstRun02}) is compared to the multi-scale target in terms of the volume fraction error, $\epsilon_f=(f^{s}-f^{o})/f^{o}$, with $f^{o}$ and $f^{s}$ denoting the optimised (\autoref{eq:volfrac}) and dehomogenised (\autoref{eq:volfrac_dehom}) volume fractions, respectively. If $\texttt{eval}=\texttt{true}$, the combined volume-weighted compliance error \texttt{wt err} is also evaluated, corresponding to     $\epsilon_\mathcal{S}=(\mathcal{S}^{s}-\mathcal{S}^{o})/\mathcal{S}^{o}$ with the volume-weighted compliance $\mathcal{S}^p=J^p f^p, \quad p\in \{o,s\}$ and $J$ denoting the compliance (\autoref{eq:comp}).
The dehomogenised and multi-scale structures are evaluated on different mesh sizes, meaning that, due to \textit{h}-convergence effects, a negative bias towards the multi-scale compliance is to be expected. 

A summary of these results is printed in the following format

\begin{lstlisting}[numbers=none,columns=fullflexible,deletekeywords={length,conv},commentstyle=\color{black},xleftmargin=1em,framexleftmargin=0em]
Multi-scale structure, intermediate design: J: 10.244 Vol: 0.300 Total time: 31.680 

Dehomogenisation to single-scale structure, with minimum length-scale: 0.200...
    Vol: 0.322 err: 7.34% Time: 0.840 (total: 32.555)

Analysing dehomogenisation result...
    J: 10.449 err: 2.00% wt err: 9.49% Time: 9.857 (total:  42.413)
    Evalulated on 1200x600 grid (x20 scaled), results subjected to h-conv. effects
\end{lstlisting}
The \texttt{TO} struct allows for direct dehomogenisation of the optimised structure at a different minimum feature size, e.g. $\texttt{dmin} = \texttt{dmin}/2$, by changing the function call to
\begin{lstlisting}[numbers=none,columns=fullflexible,xleftmargin=1em,framexleftmargin=0em]
rhoPhys1 = deHomTop808(60,30,0.3,2,0.1,1,0.1,10,true,TO)
\end{lstlisting}
 resulting in \autoref{fig:firstRun03}. This second step can be repeated for different length scales as desired. The results from the two example code executions are presented in \autoref{tab:firstRun}, where the results have been obtained on an HP 840 Elitebook running Pop!\_OS 22.04 LTS with 32 GiB of memory and an 11th-gen i7-1185G7 processor.

\begin{Figure}
    \centering
            \includegraphics[width=\linewidth]{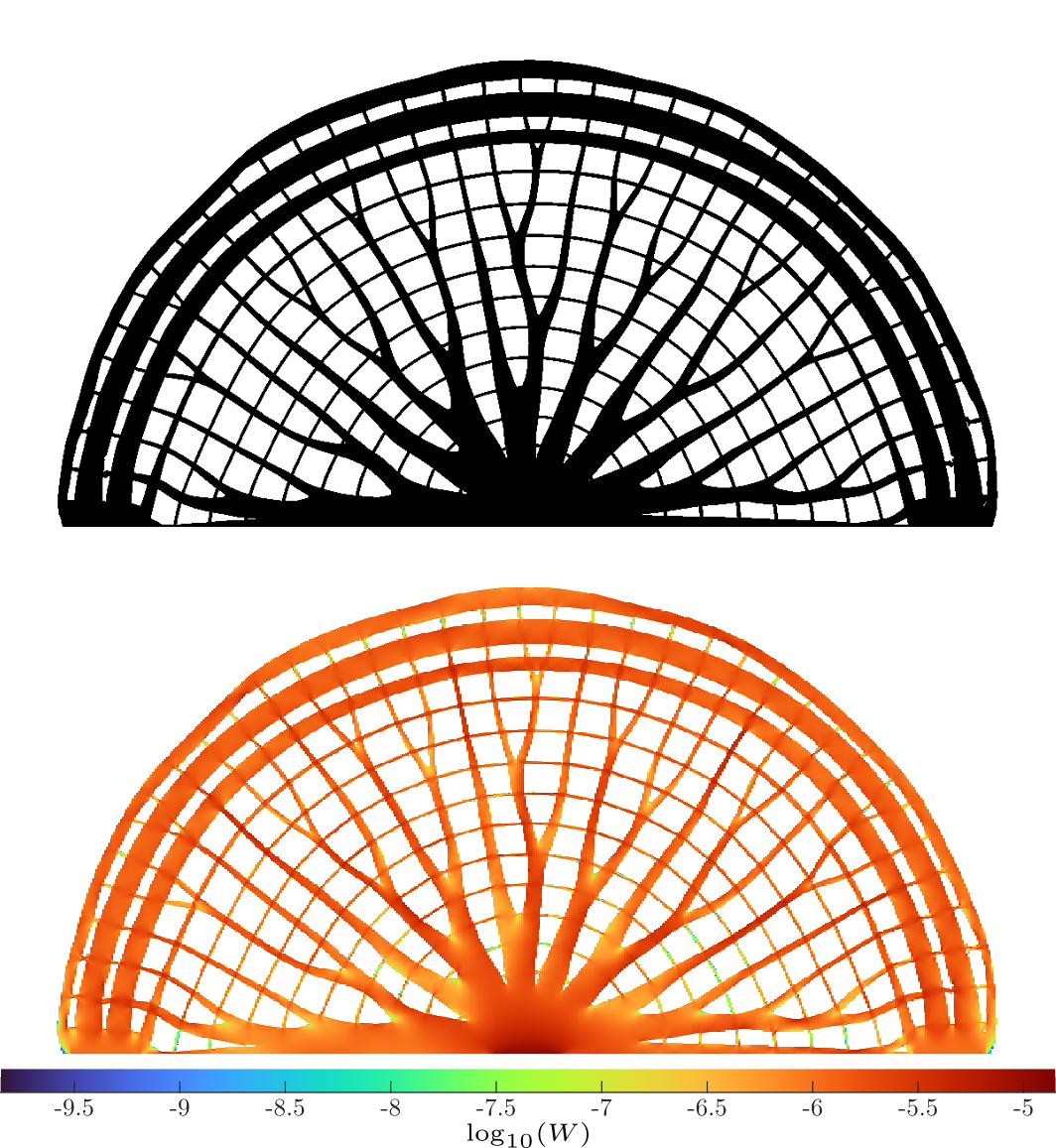}
   \captionof{figure}{Final single-scale structure obtained by post dehomogenisation with \texttt{dmin=0.2} of the multi-scale structure in \autoref{fig:firstRun01} (top). Plot of the local energy of the post-dehomogenised structure (bottom).}
    \label{fig:firstRun02}
\end{Figure}

\begin{Table}
    \centering
   \captionof{table}{Optimisation (TO) and dehomogenisation (Dehom) result from running the highlighted example cases of \deHomTopns, in terms of compliance ($J$), volume fraction ($f$), the relative volume fraction error wrt. the optimisation result ($\epsilon_f$) and the volume fraction weighted compliance relative error wrt. the optimisation result ($\epsilon_\mathcal{S}$).}
    \begin{tabular}{@{}lccc@{}}
    \toprule
    & TO & Dehom \#1 & Dehom \#2 \\ \midrule
    Grid   & 60$\times$30 & 1200$\times$600 & 2400$\times$1200\\[1.1ex]
    $J$     & 10.24&10.45&10.78\\
    $f$   & 0.300&0.322&0.309\\[1.1ex]
    $\epsilon_f$ & -&7.34\% & 3.04\% \\
    $\epsilon_\mathcal{S}$ & - & 9.49\%& 8.46\%\\[1.1ex]
    Time [s]   & 27.64 & 0.66 & 1.61 \\
    Analysis time [s] &-&9.32 & 51.35 \\ \bottomrule
    \end{tabular}
    \label{tab:firstRun}
\end{Table}

If the MATLAB Image Processing Toolbox is not installed on the user's system, the function \texttt{removeIslands2D()} (lines 762-765) cannot be executed as this function uses a MATLAB connected component labeling implementation. To get around this, the user can disable this function by commenting out lines 763 and 764; disabling this functionality may result in subpar structural performances. Alternatively, the user can source a substitute connected component labelling implementation.

\begin{Figure}
    \centering
            \includegraphics[width=\linewidth]{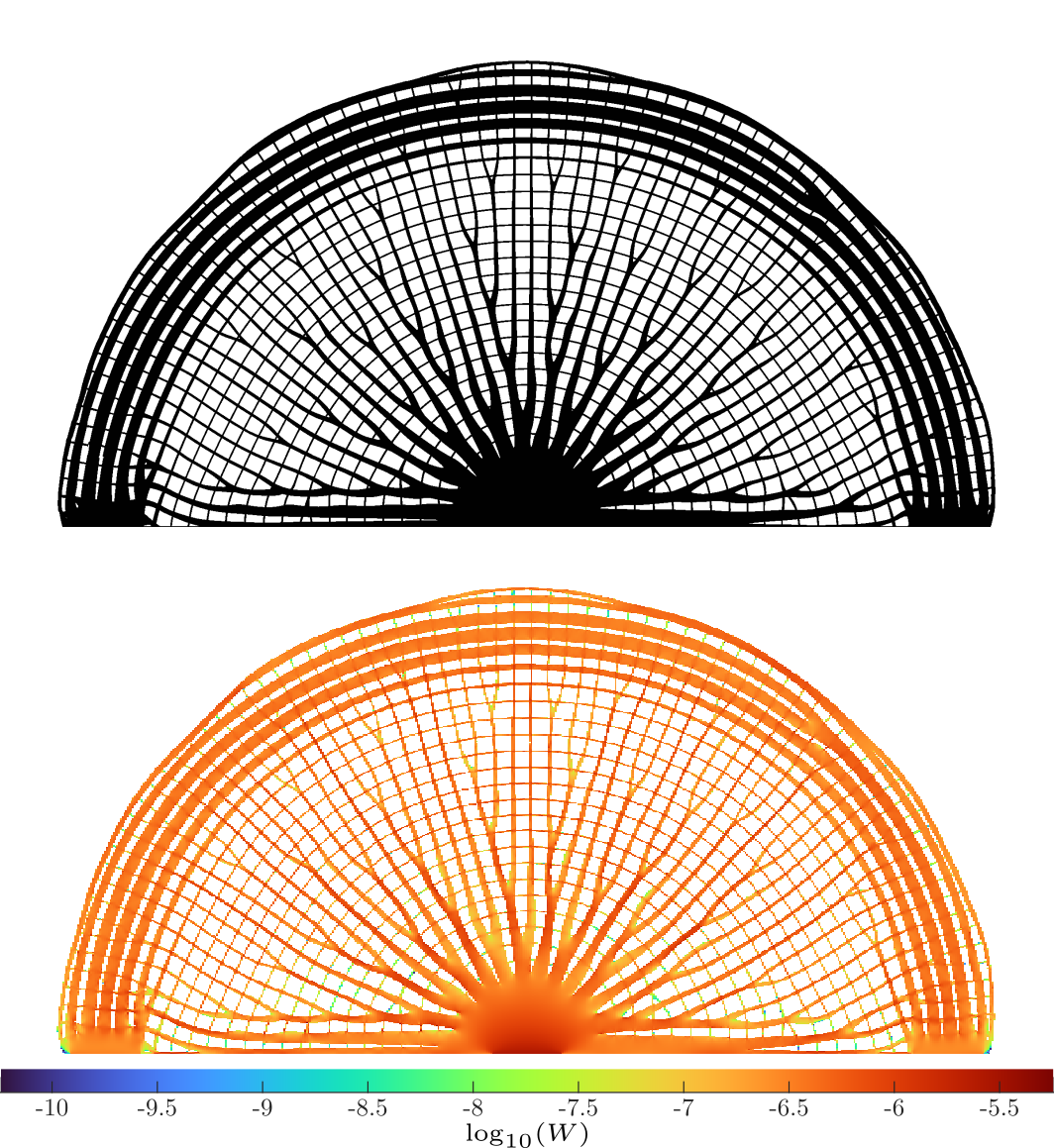}
   \captionof{figure}{
        Result from the second featured call to the \deHomTop code, with the optimised multi-scale structure from the first run as input. Dehomogenisation and analysis are performed for a halved minimal feature size, \texttt{dmin=0.1}, compared to \autoref{fig:firstRun01}. The dehomogenised structure (top) as well as the local energy distribution (bottom) of this single-scale structure is plotted.
    }
    \label{fig:firstRun03}
\end{Figure}


\section{Setup and formulations}\label{sec:setup}

A structural rectangular domain, $\Omega$, with dimensions $L \in L_x \times L_y$ is considered. $\Omega$ is partitioned into a design domain, $\Omega_A \subseteq \Omega$, and a solid passive domain, $\Omega_P \subseteq \Omega$. The domain is subjected to surface traction $\mathbf{f}$ at Neumann boundaries, $\Gamma_N \in \Omega_P$, and fixed displacements are applied to Dirichlet boundaries, $\Gamma_D \in \Omega$. $\Omega$ is discretised on four different grids, $\mathcal{T}^{c}$, $\mathcal{T}^{i}$, $\mathcal{T}^{if}$, and $\mathcal{T}^{f}$, with the element sizes, $h^{c}$, $h^{i}$, $h^{if}$, and $h^{f}$, respectively, such that $h^{c} \leq h^{i} \leq h^{if} \leq h^{f}$ (\autoref{fig:FEmodel}). The multi-scale optimisation problem is solved on $\mathcal{T}^{c}$ while the final single-scale structure is projected to $\mathcal{T}^{f}$. $\mathcal{T}^{i}$ and $\mathcal{T}^{if}$ are intermediate grids, utilised in the dehomogenisation procedure, to limit the computational cost of smoothing and interpolating the structure.

\subsection{Homogenisation-based topology optimisation}\label{sec:TO}

A multi-scale optimisation problem, formulated on both a macro and a micro scale, is considered. On the macro scale $\Omega$ is discretised using $N_e$ equal-sized quadrilateral elements $e\in \mathcal{T}^{c}$. Let $\Omega_e\subseteq\Omega$ denote the domain spanned by the element   $e\in\mathcal{T}^{c}$. On the microscale $\Omega_e$ consists of a constant periodic laminate, the so-called Rank-2 microstructure. The Rank-2 microstructure consists of two periodic orthogonally oriented lamellar materials, parameterised on two different length scales, $\bm{y}_i = \bm{x}/\epsilon^i,\; \epsilon_i\rightarrow0$, for $i\in\{1,2\}$ (\autoref{fig:rankModel}). On the smallest length scale, $\bm{y}_1$, the lamellar material has a stiff phase, $(+)$,  and a compliant phase, $(-)$, with Youngs moduli $E^{(+)}$ and $E^{(-)}$, respectively, and identical Poisson's ratio, $\nu = 1/3$. On the second length scale, $\bm{y}_2$, the lamellar material is similarly structured, but the compliant phase is described by the effective stiffness of the $\bm{y}_1$ lamellar. The stiff phases are parameterised by their relative widths $\mu_i \in [0,1]$, for $i \in \{1,2\}$.

Assuming that the compliant phase is void, $E^{(-)} = 0$, will make the constitutive model singular, and thus non-positive definite. To mitigate this, an isotropic background material, with the stiffness $E_{\min} = 10^{-9}$, is introduced to approximate void and simplify the expression. The stiffness of the solid phase is set to unit stiffness, $E^{(+)} = 1$. Following the description of a Rank-N microstructure from \citet{Jensen2022} with Cartesian aligned laminates, the constitutive matrix is defined in \autoref{eq:cmat}.
\begin{equation}\label{eq:cmat}
    \begin{split}
    \mathbf{C}^H (\mu_1,\mu_2) = \frac{E_{\min}}{1 - \nu^2} 
    \begin{bmatrix}
        1 & \nu & 0 \\ 
        \nu & 1 & 0 \\ 
        0 & 0 & \frac{1-2}{\nu} 
    \end{bmatrix}
    \\
    + \frac{E^{(+)}}{1 - \mu_2 + \mu_1\mu_2(1 - \nu^2)} \\
    \begin{bmatrix}
        \mu_1 & \mu_1 \mu_2 \nu & 0 \\ 
        \mu_1 \mu_2 \nu & \mu_2(1 - \mu_2 + \mu_1 \mu_2) & 0 \\ 
        0 & 0 & 0 
    \end{bmatrix}
\end{split}
\end{equation}
The Rank-2 material is rotated with respect to $\bm{x}$, using the standard planar rotation (\autoref{eq:rotcmat}), where the basis-transformation matrix is given in \autoref{eq:rottmat} with $c = \cos(a)$ and $s = \sin(a)$, and $a \in [-4\pi,4\pi]$ being the orientation angle. 
\begin{equation}\label{eq:rotcmat}
    \widetilde{\mathbf{C}}^H (\mu_1,\mu_2,a) = \mathbf{T}^{\intercal}(a) \, \mathbf{C}^H (\mu_1,\mu_2) \, \mathbf{T}(a)
\end{equation}
\begin{equation}\label{eq:rottmat}
    \mathbf{T}(a) = 
    \begin{bmatrix}
        c^2 & s^2 & c\,s \\
        s^2 & c^2 & -c\,s \\
        -2c\,s & 2c\,s & c^2 - s^2 
    \end{bmatrix}
\end{equation}

The relative widths, $\mu_i$, are defined on two different scales, making a single-scale interpretation challenging. Instead, the interpretation of relative widths from single-scale, $\Bar{w}_i$, to multi-scale is significantly more straightforward to obtain~\citep{JensenThesis}. 
To this end, $\mu_i$ is obtained from $\Bar{w}_i$ by \autoref{eq:w2mu}-\autoref{eq:homrho}, where $p_i$ is the stiffness contribution from each layer and $\rho_N$ the relative density of a lamellar with $N$ layers.
\begin{equation}\label{eq:w2mu}
    \mu_i = \frac{ p_i(\bar{w})  \rho_2(\bar{w}) }{1 - \rho_{n-1}(\mu)}
\end{equation}
\begin{equation}\label{eq:w2mu_p}
    p_i = \frac{\bar{w}_i}{\sum_{j=1}^{N} \bar{w}_j}
\end{equation}
\begin{equation}\label{eq:homrho}
    \rho_N(\xi) = 1 - \prod_{i=1}^{N} (1 - \xi_i),\quad \xi\in\{\mu,\; \Bar{w}\}
\end{equation}


$\Bar{w}_i$ is a physical relative width, which depends on two design variables; the relative width $w_i \in [w_{\min},w_{\max}]$, where $0 \leq w_{\min} \leq w_{\max} \leq 1$, and an indicator value $s_i \in [0,1]$. The indicator determines if the material is active, $s=1$, or void, $s=0$~\citep{Giele2021a}. $w$ and $s$ are filtered to obtain $\Tilde{w}$ and $\Tilde{s}$ (\autoref{eq:filter}), where $k$ is a kernel based on the neighbourhood $\Omega^k_e \in \Omega$ defined from the filter radii $r_{\min}$ and $\alpha_{\text{rsc}}\,r_{\min}$, for $\Tilde{w}$ and $\Tilde{s}$, respectively. $\alpha_{\text{rsc}} \geq 1$ is a scaling parameter for the indicator filter radius.
\begin{equation}\label{eq:filter}
    \Tilde{\xi}_e = \frac{\sum_{i \in \Omega^k_e} k_i \xi_i }{\sum_{i \in \Omega^k_e} k_i}, \quad \xi\in\{w,\; s\}
\end{equation}

$\Tilde{s}$ is projected using the three-field robust approach~\citep{Wang2011} to $\Bar{s}^m$, for $m \in \{e,i,d\}$, representing the eroded, intermediate and dilated fields, respectively (\autoref{eq:proj}). $\eta^m \in [0,1]$ are projection thresholds and $\beta \in (0, \infty ]$ is the projection sharpness. To ensure minimum length-scale on the physical relative widths, the indicator and widths are convoluted (\autoref{eq:ip2}).
\begin{equation}\label{eq:proj}
    \Bar{s}_e^m = \frac{\tanh (\beta \eta) \, \tanh( \beta ( \Tilde{s}_e - \eta ) )}{ \tanh (\beta \eta^m) \, \tanh( \beta ( 1 - \eta^m ) ) }
\end{equation}

\begin{equation}\label{eq:ip2}
    \Bar{w}_i^m = \Tilde{w}_i \, \Bar{s}^m
\end{equation}

Let $\mathbf{w}_i \in \mathbb{R}^{N_e}$, $\mathbf{s}\in \mathbb{R}^{N_e}$ and $\mathbf{a}\in \mathbb{R}^{N_e}$ be vectors representing the design on $\Omega$, with restrictions $\mathbf{w}_i=1$ and $\mathbf{s}=1$ in $\Omega_P$, for solid passive material. Further, let $\bm{\kappa}$ denote the combined set of design variables, such that $\bm{\kappa} = \{ \mathbf{w}_1,\mathbf{w}_2,\mathbf{s},\mathbf{a} \}$. The optimisation problem of the multi-scale minimum compliance problem is stated in (\autoref{eq:TopOptObj})-(\autoref{eq:TopOptBox}).
\begin{eqnarray}\label{eq:theTopOptProb}
     & \displaystyle \min_{\bm{\kappa}}  & : \Phi (\bm{\kappa},\mathbf{u}) = \gamma_{0} J+ \gamma_{S} \frac{1}{V_{\Omega}} \int_{\Omega} \bar{\mathbf{s}}^d \mathrm{d}\Omega\label{eq:TopOptObj}   \\
    & \textrm{s.t.}                     & : \mathbf{K}( \bm{\mu}^e_1,\bm{\mu}^e_2,\mathbf{a} ) \mathbf{u} = \mathbf{f},    \label{eq:TopOptSys} \\
     &  						            & : f^o \leq f^*, \label{eq:TopOptCon} \\
    &                                   & : \underline{\bm{\kappa}} \leq  \bm{\kappa}   \leq  \overline{\bm{\kappa}} \label{eq:TopOptBox} 
\end{eqnarray}
The optimisation objective (\autoref{eq:TopOptObj}) is to minimise a weighted sum of the structural compliance (\autoref{eq:comp}), 
and a non-uniqueness penalty term~\citep{Giele2021a,Jensen2022}, with weight factors $\gamma_0$ and $\gamma_S$ respectively. $\gamma_{0}$ is the volume-weighted solid compliance of $\Omega$, and $\gamma_{S} = 1/20$ scales the penalty term.
\begin{equation} \label{eq:comp}
    J = \langle \mathbf{f},\mathbf{u} \rangle
\end{equation}

The objective is subjected to an upper bound constraint (\autoref{eq:TopOptCon}) on the overall volume (\autoref{eq:volfrac}), with $V_{\Omega}$ denoting the volume of $\Omega$.
\begin{equation} \label{eq:volfrac}
    f^o = \frac{1}{V_{\Omega}} \int_{\Omega} \rho_2(\{\bar{\mathbf{w}}^d_1,\bar{\mathbf{w}}^d_2\}) \mathrm{d}\Omega 
\end{equation} 

The displacements, $\mathbf{u}$, are obtained from the linear elastic problem~(\autoref{eq:TopOptSys}) considering the eroded multi-scale interpretation of $\mathbf{w}_i$ (\autoref{eq:w2mu}).

The design variables are subjected to box constraints based on their defined ranges (\autoref{eq:TopOptBox}). 
The optimisation problem is solved in a nested form, where in each iteration of the procedure, the design is updated using the OC-resembling update rule from~\cite{Ferrari2021}. Sensitivities with respect to the objective and volume fraction constraint are computed using the discrete adjoint method.

    \subsection{Phasor-based dehomogenisation}\label{sec:dehom}


    To provide a rudimentary understanding of the subprocedures of the phasor-based dehomogenisation procedure, the fundamental mathematical definitions and heuristic details are introduced. For the purpose of dehomogenising Rank-2 material structures, the procedure assumes a set of frame fields $\mathcal{F} = \{  \{\mathbf{n}_1 (a)\}, \{\mathbf{n}_2 (a)\} \}$ and a corresponding set of relative thicknesses $ \mathcal{W} = \{  \{\bar{\mathbf{w}}^i_1\}, \{\bar{\mathbf{w}}^i_2\} \}$, describing the homogenised solution to be realised, are given.

    Let $\mathcal{K}_i$ denote a finite set of phasor kernels such that, for lamination layer $i\in\{1,2\}$, $\mathcal{E}_i=\{e{\in} \mathcal{T}^c: \Bar{w}_{ie}{\in}[w_{min},\,1)\}$ denotes the set of intermediate density kernels and \mbox{$\zeta:\mathcal{E}\rightarrow \mathcal{K}_i$} is an injective mapping between each intermediate density element, and a phasor kernel $\zeta(e)\in\mathcal{K}_i$. The phasor kernel $\zeta(e)\in \mathcal{K}_i$ inherits a spatial location $\mathbf{\mathring{x}}_e\in \mathcal{T}^c$, at the centre of element $e$, and a unit vector $\mathbf{n}_{ie}=(n_{ie}^x,\;n_{ie}^y)^{\intercal}$ corresponding to the lamination orientation normal of element $e$. The corresponding signal emitted from kernel $\zeta(e)$, sampled at spatial location $\mathbf{x}\in \mathcal{T}^{i}$, is denoted $\mathcal{G}_e(\mathbf{x})$ (\autoref{eq:phasor_signal}).
    \begin{equation}\label{eq:phasor_signal}
        \begin{split}
\mathcal{G}_e(\mathbf{x})=\mathrm{exp}\left({\displaystyle-\tilde{\beta}\Delta^{\mathbf{r}}_e(\mathbf{x})}\right)\qquad\qquad\qquad\qquad\quad\;\\
\mathrm{exp}\left({\displaystyle2i \pi \omega \textbf{n}_{ie}\cdot\left(\mathbf{x}{-}\mathbf{\mathring{x}}_e\right){+}i\varphi_e}\right), \mathbf{x}\in \mathcal{T}^{i}
        \end{split}
    \end{equation}

This signal is the product of two exponential functions, where the first controls the spatial magnitude in terms of a Gaussian and the second defines the periodic oriented complex oscillator originating from the kernel centre. The span and shape of the Gaussian are determined by the signal bandwidth $\tilde{\beta}\in \mathbb{R}_{>0}$ and the distance transformation $\Delta_e^\mathbf{r}: \mathbb{R}^2\rightarrow \mathbb{R}_{\geq 0}$. The oscillator originating at the kernel centre is a complex wave field of a specified frequency, with magnitude $\omega\in\mathbb{R}$, orientation $\mathbf{n}_{ie}$ and phase shift $\varphi_e\in[-\pi,\pi]$. The phase shift controls the position of the wave in relation to the kernel centre, such that $\mathcal{G}_e(\mathbf{\mathring{x}}_e)=\mathrm{exp}(i\varphi_e)$.

In accordance with \cite{Woldseth2023}, an anisotropic distance measure is considered (\autoref{eq:2d_anisotropic_distance}). This defintion of the distance measure changes the shape of the Gaussian kernel from an isotropic circle, for $\mathbf{r}{=}(1,1)^\intercal$, to an anisotropic ellipse with major axis oriented perpendicular to the kernel normal $\mathbf{n}_{ie}$, for $r_2>r_1$.
\begin{equation}\label{eq:2d_anisotropic_distance}
        \Delta_e^{\mathbf{r}}(\mathbf{x})=
        \left\|\begin{pmatrix} \sqrt{r_1}n_{ie}^y &{-} {\sqrt{r_1}}n_{ie}^x\\ {\sqrt{r_2}}n_{ie}^x & \sqrt{r_2}n_{ie}^y\end{pmatrix}(\mathbf{\mathring{x}}_e{-}\mathbf{x})\right\|_2^2, \mathbf{r}{\in} \mathbb{R}^2_{{>}0}
\end{equation}

\begin{figure*}[htb]
    \centering
    \begin{minipage}[b]{0.2\textwidth}
\begin{center}
    \includegraphics[width=\linewidth]{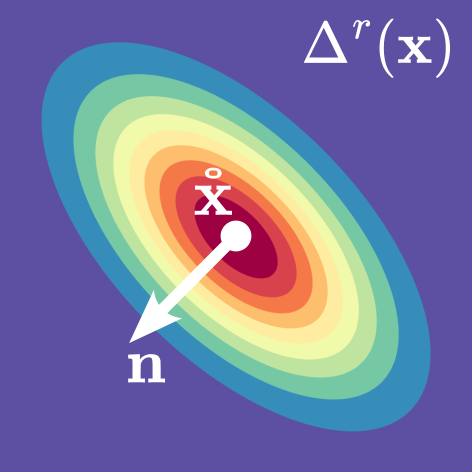}
    (a) Gaussian
    \end{center}
\end{minipage}
\hfill
\begin{minipage}[b]{0.26\textwidth}
\begin{center}
    \includegraphics[width=\linewidth]{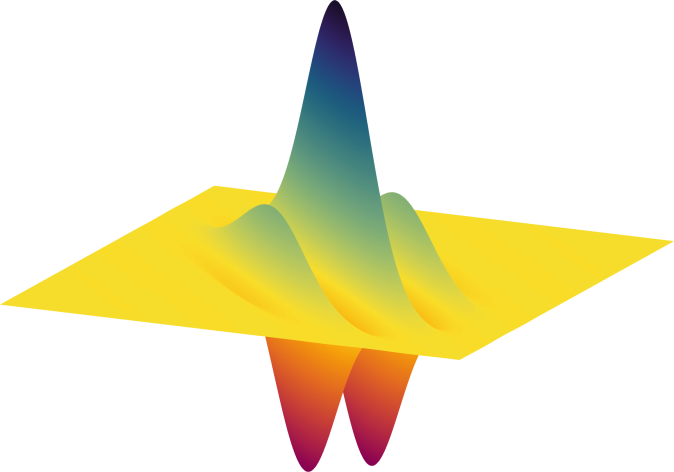}
    \vspace{1mm}
    (b) $\Re\left(\mathcal{G}\right)$
    \end{center}
\end{minipage}
\hfill
\begin{minipage}[b]{0.26\textwidth}
\begin{center}
    \includegraphics[width=\linewidth]{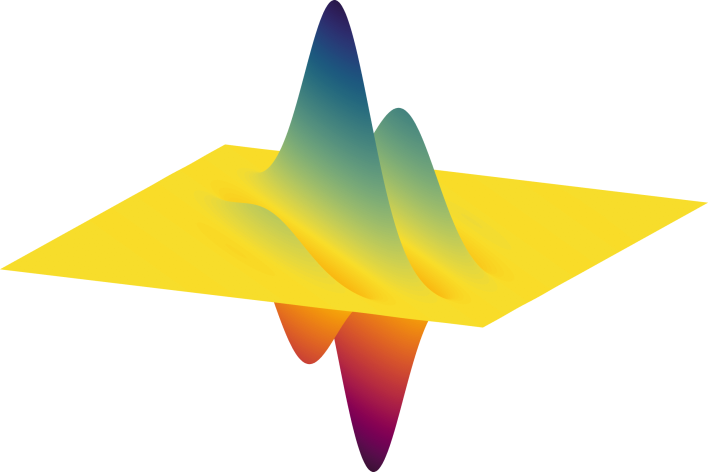}
        \vspace{1mm}
    (c) $\Im\left(\mathcal{G}\right)$
    \end{center}
\end{minipage}
\hfill
\begin{minipage}[b]{0.24\textwidth}
 \begin{center}
    \includegraphics[width=\linewidth]{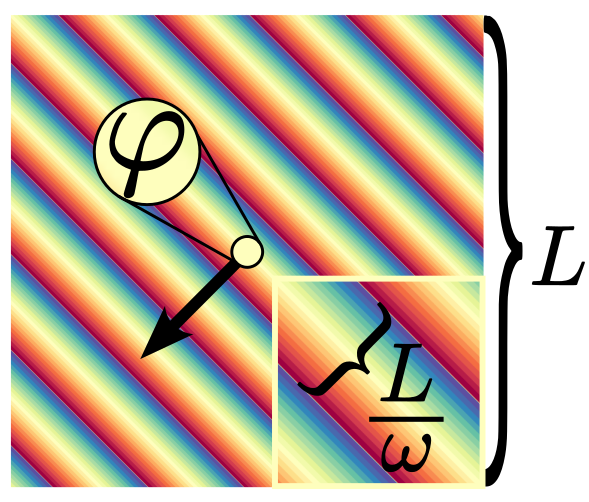}
    (d) $\text{Arg}\left(\mathcal{G}\right)$
    \end{center}
\end{minipage}
    \caption{Illustrating the relation between the definition of a single phasor kernel and its correspoding spatial signal. (a) presents the anisotropic contour-lines of the Gaussian weights about the kernel centre $\mathbf{\mathring{x}}$ with normal $\mathbf{n}$ and degree of anisotrpy $\mathbf{r}=(1/2,2)^\intercal$. The weighted complex signal emitted from this kernel, for some fixed frequency $\omega$, is separated into its real and imaginary parts in (b) and (c) respectively. (d) presents the real-valued argument of this complex signal, where the weight-induced magnitude variations in the complex field are normalised to form a perfectly contrasting oriented periodic wave. The kernel phase shift $\varphi$ controls the value at the kernel centre, and the wavelength $L/\omega$ is controlled by the input frequency $\omega$ and the span $L$ of the domain sampled.}
    \label{fig:phasor_kernel_concept}
\end{figure*}


    \autoref{fig:phasor_kernel_concept} illustrates the relation between the phasor kernel defining parameters and the resulting kernel signal. When sampling the combined signal from a set of multiple phasor kernels, additional mesures are needed to control the blended transitions in the low-magnitude regions of the complex phasor field, to ensure smooth transitions in the real-valued oriented wave field.

    Given the interpolated unit orientation vector $\mathbf{n}(\mathbf{x})$ of a sampling point $\mathbf{x}\in \mathcal{T}^{i}$, a sampling filter is introduced, in the form of a kernel $\mathcal{A}_e(\mathbf{x})$ with bandwidth $\tilde{\alpha}\in \mathbb{R}_{>0}$, to regularise the signal emitted from $\zeta(e)$ (\autoref{eq:sampling_filter_kernel}).
    \begin{equation}\label{eq:sampling_filter_kernel}
        \begin{split}
            \mathcal{A}_e(\mathbf{x})=\\
            \mathrm{exp}{\left(\dfrac{\tilde{\beta}^2\Delta_e^\mathbf{r}(\mathbf{x})-\pi^2\|\Lambda_j(\mathbf{x})\|_2^2+2i\tilde{\alpha}\Lambda_e(\mathbf{x})\cdot(\mathbf{x}{-}\mathbf{\mathring{x}}_e)}{\tilde{\alpha}+\beta}\right)}
        \end{split}
    \end{equation}
    The term $\Lambda_e(\mathbf{x})=\omega(\mathbf{n}_{ie}-\mathbf{n}(\mathbf{x}))$ is a similarity weight between the orientation of the kernel $\zeta(e)$ being sampled and the interpolated orientation at the sampling point $\mathbf{x}$, such that for $\Lambda_e(\mathbf{x})\rightarrow0$ these orientations are identical and the magnitude of the kernel impact is maximised.
    The filter kernel is integrated directly in the sampling of the phasor field, which is obtained for sampling point $\mathbf{x}\in \mathcal{T}^{i}$ by computing the complex sum of all kernel signals at this spatial location (\autoref{eq:combined_signals}).

    \begin{equation}\label{eq:combined_signals}
        \mathcal{G}(\mathbf{x})=\displaystyle\sum_{\zeta(e)\in \mathcal{K}_i}\mathcal{A}_e(\mathbf{x})\mathcal{G}_e(\mathbf{x}),\; \mathbf{x}\in \mathcal{T}^{i}
    \end{equation}


\begin{figure*}[htb]
    \centering
    \begin{minipage}[t]{0.175\textwidth}
\begin{center}
    \includegraphics[width=\linewidth]{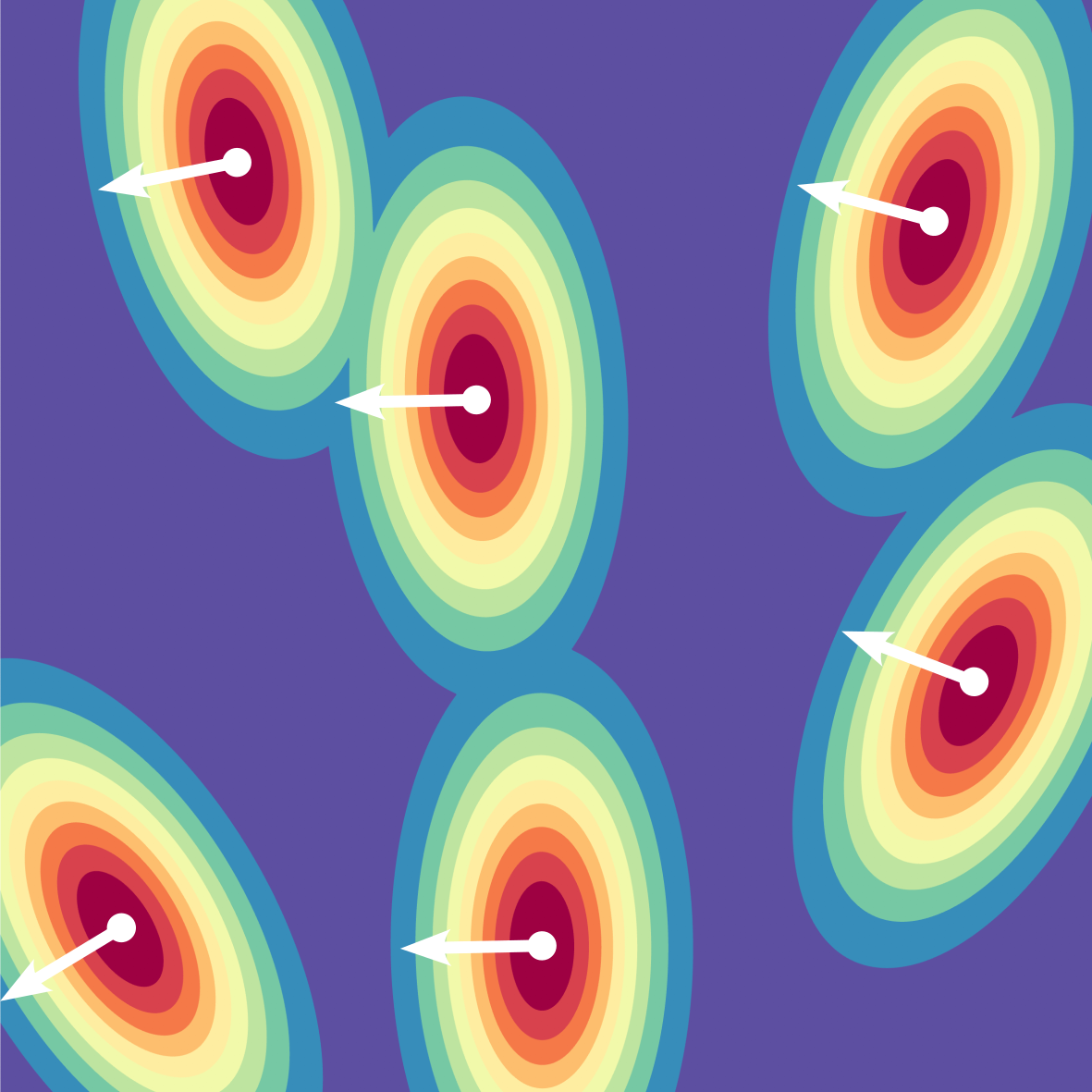}
    (a) Phasor kernels
    \end{center}
\end{minipage}
\hfill
\begin{minipage}[t]{0.395\textwidth}
\begin{center}
    \includegraphics[width=\linewidth]{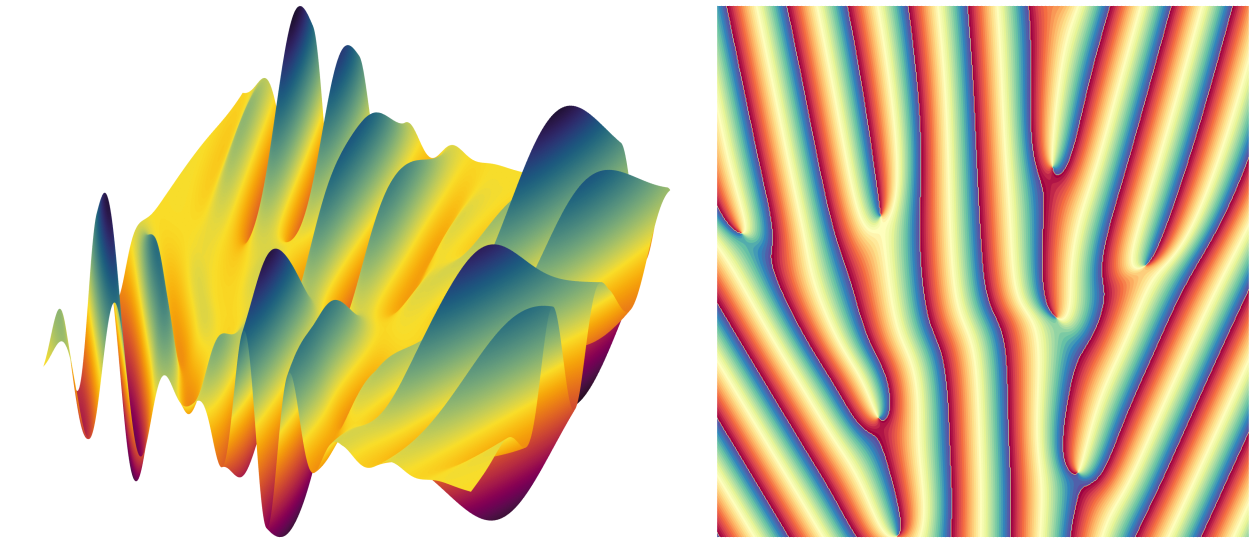}
    \vspace{1mm}
    (b) Unaligned phases: $\Re(\mathcal{G}(\mathbf{x}))$ and $\phi(\mathbf{x})$
    \end{center}
\end{minipage}
\hfill
\begin{minipage}[t]{0.395\textwidth}
\begin{center}
    \includegraphics[width=\linewidth]{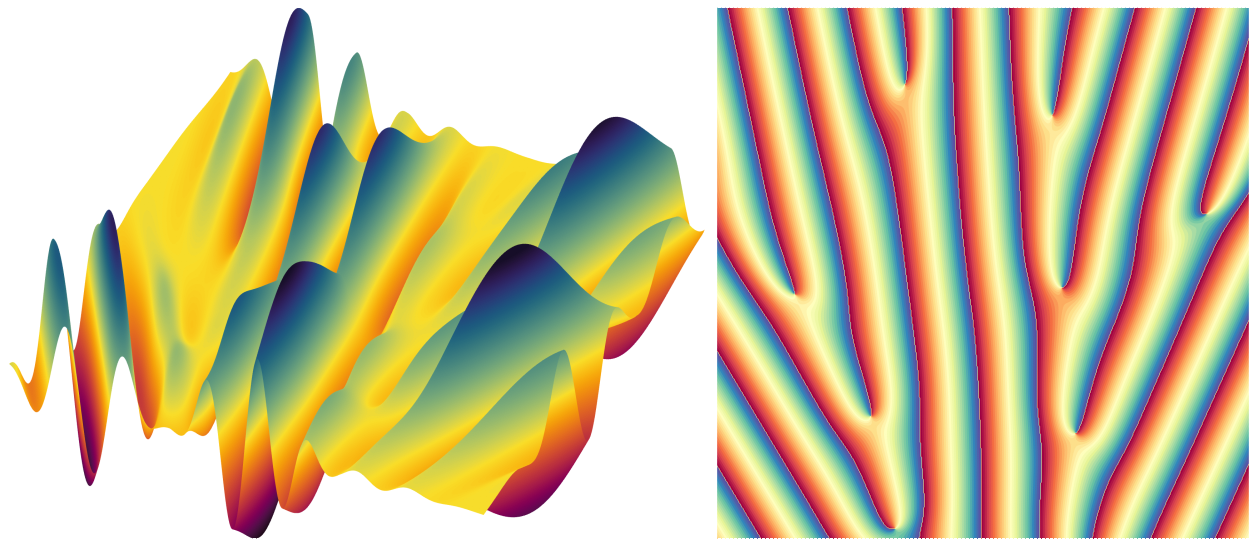}
        \vspace{1mm}
    (c) Aligned phases: $\Re(\mathcal{G}(\mathbf{x}))$ and $\phi(\mathbf{x})$
    \end{center}
\end{minipage}

    \caption{A sparse set of phasor kernels (a) are sampled to a phasor field with $\Re(\mathcal{G}(\mathbf{x}))$ and $\phi(\mathbf{x})$ shown in (b) illustrating the variance in the spatial magnitude of the blended signals (left) and the corresponding isolated phase (right). Misaligned kernel phase shifts result in unnecessary local curvature in (b), which is corrected for by phase alignment to obtain the combined signal illustrated in (c). }
    \label{fig:kernel_blend_and_align}
\end{figure*}


The relation of the kernel phase shift and the location of the emitted wave signal at the kernel centre, may cause the signal from neighbouring kernels to be out-of-phase, causing destructive inference in the sampled phasor field (\cite{Tricard_orientable_2020}). To this end, phase alignment is performed to obtain a set $\varphi_e,\:\zeta(e)\in \mathcal{K}_i$ of phase shifts ensuring spatial coherency between the separate kernel wavefronts. The phase alignment procedure is a fixed-iteration heuristic, where the phase shift of each kernel $\zeta(e)\in \mathcal{K}_i$ is updated sequentially in each iteration, based on a weighted contribution from the nearby kernels. Similarly to \cite{Woldseth2023}, an anisotropic neighbourhood definition is utilised to define the subset $\mathcal{N}_e\subseteq\mathcal{K}_i$ of nearby kernels for kernel $\zeta(e)\in \mathcal{K}_i$ (\autoref{eq:align_neighbourhood}).
\begin{equation}\label{eq:align_neighbourhood}
    \mathcal{N}_e=\left\{\zeta(f)\in \mathcal{K}_i:\; \Delta_e^\mathbf{r}(\mathbf{\mathring{x}}_f)\leq R^2_e\right\}
\end{equation}
Based on the definitions of the neighbouring kernels in the current iterate, the phase shift $\varphi_{e}$ of kernel $\zeta(e)\in \mathcal{K}_i$ is the real-valued argument of the combined signal from its neighbours (\autoref{eq:phase_alignment_update}-\autoref{eq:phase_alignment_update2}). To account for the inherent $\pi$-rotational invariance of the microstructure representation in this update, the corrected phase shift $\tilde{\varphi}_{fe}$ and orientation $\tilde{\mathbf{n}}_{fe}$ contributions from kernel $\zeta(f)\in \mathcal{N}_e$, relative to the kernel $\zeta(e)\in \mathcal{K}_i$ being updated, are determined according to \autoref{eq:opposing_orient_correct}-\autoref{eq:opposing_orient_correct2}.


\begin{align}\label{eq:phase_alignment_update}
        \varphi_e=\text{Arg}\left(\sum_{\zeta(f)\in \mathcal{N}_e} \mathrm{exp}\left({-\dfrac{1}{R_e}\|\mathbf{\mathring{x}}_e-\mathbf{\mathring{x}}_f\|_2^2}\right)\tilde{\mathcal{G}}_e^f\right)\\\label{eq:phase_alignment_update2}
       \tilde{\mathcal{G}}_e^f= \mathrm{exp}\left({\displaystyle2i\pi \omega \mathbf{\tilde{n}}_{fe}\cdot(\mathbf{\mathring{x}}_e-\mathbf{\mathring{x}}_f)+i\tilde{\varphi}_{fe}}\right)
\end{align}

\begin{align}
   &\tilde{\varphi}_{fe}=\begin{cases}
        \pi-\varphi_f & \textit{if } \mathbf{n}_{ie}\cdot \mathbf{n}_{if}<0 \\
        \varphi_i     & \textit{otherwise}
    \end{cases}\label{eq:opposing_orient_correct}\\
    &\mathbf{\tilde{n}}_{fe}=\begin{cases}
        -\mathbf{n}_{if} & \textit{if } \mathbf{n}_{ie}\cdot \mathbf{n}_{if}<0 \\
        \mathbf{n}_{if}  & \textit{otherwise}\label{eq:opposing_orient_correct2}
    \end{cases}
\end{align}

    
After alignment, $\mathcal{G}(\mathbf{x})$ can be sampled from the kernels $\zeta({e})\in \mathcal{K}_i$ on a discretised grid $\mathbf{x}\in\mathcal{T}^i$ to obtain the complex phasor field (\autoref{fig:kernel_blend_and_align}). Cubic interpolation of $\mathcal{G}(\mathbf{x})$ from $\mathcal{T}^i$ to $\mathcal{T}^{if}$ ensures sufficient detail is obtained at the finer resolution. From this, $\phi(\mathbf{x})=\text{Arg}(\mathcal{G}(\mathbf{x}))$ is the sawtooth wave-field obtained by isolating the phase of $\mathcal{G}(\mathbf{x})$ from its magnitude. $\psi(\mathbf{x})=\sin(\phi(\mathbf{x}))$ denotes the phasor sine-wave, and ${\tau}(\mathbf{x})=\arcsin(\psi(\mathbf{x}))/\pi+0.5$ the normalised triangular wave-conversion corresponding to a level-set function of the dehomogenised design. 


\begin{figure*}[htb]
    \centering
    \begin{minipage}[t]{0.32\textwidth}
\begin{center}
    \includegraphics[width=\linewidth]{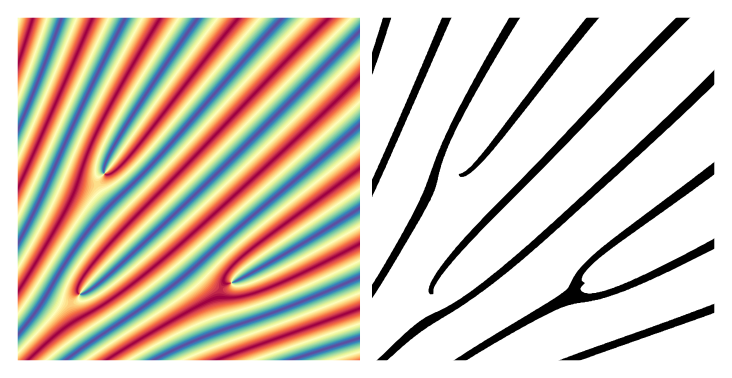}
    (a) Disconnected branches $(\tau(\mathbf{x}))$
    \end{center}
\end{minipage}
\hfill
\begin{minipage}[t]{0.32\textwidth}
\begin{center}
    \includegraphics[width=\linewidth]{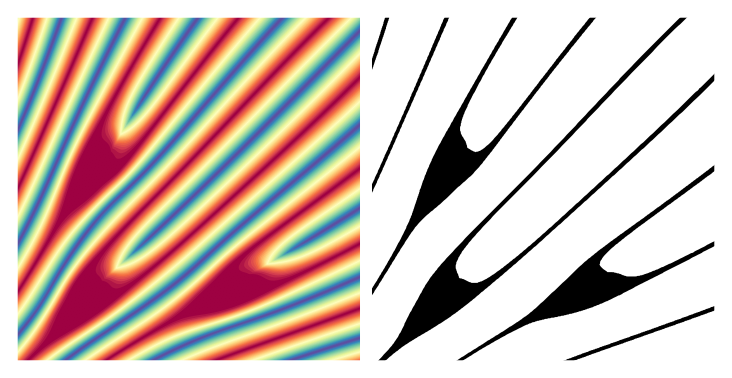}
    \vspace{1mm}
    (b) Solidified branches $(\hat{\tau}(\mathbf{x}))$
    \end{center}
\end{minipage}
\hfill
\begin{minipage}[t]{0.32\textwidth}
\begin{center}
    \includegraphics[width=\linewidth]{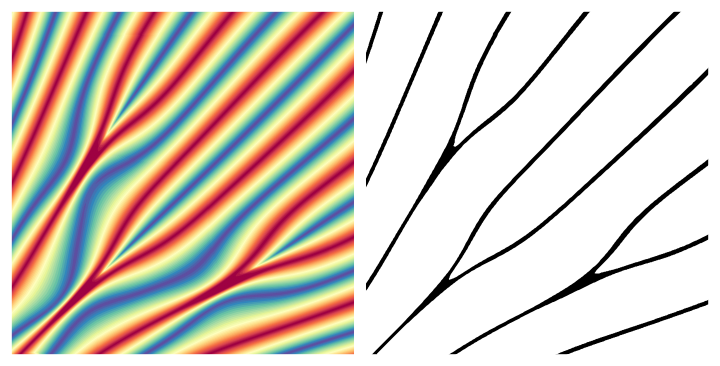}
        \vspace{1mm}
    (c) Pinched branches $(\tilde{\tau}(\mathbf{x}))$
    \end{center}
\end{minipage}

    \caption{Post-processing of the wave-field obtained from phasor sampling (a) is needed to ensure connectivity by solidication (b) and adhere to the desried relative thickness (c)}
    \label{fig:branch_closure}
\end{figure*}


To best adhere to the orientations $\mathbf{n}_e,\; e\in \mathcal{T}^c$ and an imposed constant periodicity $\omega$, for a non-trivial orientation field, the sampled phasor field $\mathcal{G}(\mathbf{x}),\; \mathbf{x}\in \mathcal{T}^{if}$ must contain point-singularities, which imposes branch-like bifurcations in the wave-field translations. Therefore, thresholding the triangular wave-field $\tau(\mathbf{x})$ may result in a solid-void design with disconnected structural members at these bifurcation locations (\autoref{fig:branch_closure}(a)), which are subjected to a post-processing procedure to improve structural integrity. 

The convenient relation between singularities in the complex phasor field and the branching disconnections in the corresponding real-valued conversions, allows for identifying a discrete set of points $\mathcal{B}\subset\mathcal{T}^{if}$ by identifying where the magnitude of the phasor field vanishes to \mbox{zero (\autoref{eq:branching_point_set})}.
\begin{equation}\label{eq:branching_point_set}
    \mathcal{B}=\{\mathbf{x}\in\mathcal{T}^{if}\;:\;|{\mathcal{G}}(\mathbf{x})|<\epsilon,\;\epsilon\rightarrow 0\}
\end{equation}

 Depending on the global phase of the phasor sine-wave $\psi(\mathbf{x})$, relative to a bifurcation location, the degree of disconnection of the branches may vary. Branches with a minimal degree of disconnection will require less invasive efforts to ensure connectivity, compared to those with maximal degree of disconnection. $\varrho(\boldsymbol{\gamma})\in [0,1]$ is introduced as a measure describing the degree of disconnection of a branching point $\boldsymbol{\gamma}\in\mathcal{B}$, and is computed based on the average normalised phasor sine-wave value within a circular neighbourhood $\mathcal{M}$ (\autoref{eq:disconnection_degree2}) with diameter equal to one wave-length $\lambda$ at the current periodicity $\omega$ (\autoref{eq:disconnection_degree}). 


\begin{align}\label{eq:disconnection_degree}
    \varrho(\boldsymbol{\gamma})=1-\dfrac{1}{|\mathcal{M}(0.5\lambda)|}\displaystyle\sum_{\mathbf{x}\in \mathcal{M}(0.5\lambda)}\frac{\psi(\mathbf{x})+1}{2}\\\label{eq:disconnection_degree2}
    \mathcal{M}_{\boldsymbol{\gamma}}(s)=\{\mathbf{x}\in\mathcal{T}^i\;:\;\|\boldsymbol{\gamma}-\mathbf{x}\|_2^2\leq s^2\},\; \forall \boldsymbol{\gamma}\in \mathcal{B}
\end{align}

    
To determine the unit vector $\boldsymbol{\vec{\gamma}}$ indicating the closure direction for the branch about the branching point $\boldsymbol{\gamma}\in \mathcal{B}$, the degrees of disconnection at the locations corresponding to moving $\pm\frac{1}{3}$-wavelenghts along the layer-normal $\mathbf{n}(\boldsymbol{\gamma})$ at the branching point $\boldsymbol{\gamma}$ are compared to determine the best fit (\autoref{eq:closure_direction}). Keeping the control-points within $\pm\frac{1}{3}\lambda$ of the identified bifurcation point limits potential interference of neighbouring waves on the closure direction selection.

\begin{equation}\label{eq:closure_direction}
    \vec{\boldsymbol{\gamma}} = %
    \begin{cases}
        \mathbf{n}(\boldsymbol{\gamma}) & \text{if } \varrho\left(\boldsymbol{\gamma}{+}\frac{\lambda}{3}\mathbf{n}(\boldsymbol{\gamma})\right)\leq\varrho\left(\boldsymbol{\gamma}{-}\frac{\lambda}{3}\mathbf{n}(\boldsymbol{\gamma})\right)\\
        - \mathbf{n}(\boldsymbol{\gamma}) & \text{otherwise}
    \end{cases}
\end{equation}

From the obtained closure direction and degree of disconnection of $\boldsymbol{\gamma}\in \mathcal{B}$ a perturbed branch centre $\boldsymbol{\dot{\gamma}}$ and a closure control-point $\boldsymbol{\ddot{\gamma}}$ are identified (\autoref{eq:new_centre}).  
\begin{equation}\label{eq:new_centre}
\boldsymbol{\dot{\gamma}}=\boldsymbol{\gamma}+\frac{\lambda}{3}\varrho(\boldsymbol{\gamma})\vec{\boldsymbol{\gamma}}\quad \land \quad \boldsymbol{\ddot{\gamma}}=\boldsymbol{\gamma}+\frac{2\lambda}{3}\varrho(\boldsymbol{\gamma})\vec{\boldsymbol{\gamma}}
\end{equation}

The point $\boldsymbol{\dot{\gamma}}$ is chosen as the centre location for ensuring connection by branch solidification. This is obtained by introducing a modified Gaussian $\hat{\Pi}_{\boldsymbol{\gamma}}(\mathbf{x})$ about this centre, and from the union $\hat{\Pi}(\mathbf{x})$ a density weighted localised phaseshift $\hat{\pi}(\boldsymbol{x})$ is introduced. A normalised triangular wave $\hat{\tau}(\mathbf{x})$ with solidified branch-shaped connections (\autoref{fig:branch_closure}(b)) can then be obtained by the union of the $\pm\hat{\pi}(\boldsymbol{x})$ phase-shifted sine waves $\hat{\psi}(\mathbf{x})$ (\autoref{eq:solidified_triangular}-\autoref{eq:branch_local_gaussian}). 
\begin{align}
    &\hat{\tau}(\mathbf{x})=\frac{1}{\pi}\arcsin(\hat{\psi}(\mathbf{x}))+0.5\label{eq:solidified_triangular}\\
    &\hat{\psi}(\mathbf{x})=\max\{\sin(\phi(\mathbf{x})+\hat{\pi}(\mathbf{x})),\;\sin(\phi(\mathbf{x})-\hat{\pi}(\mathbf{x}))\}\label{eq:solidified_triangular}\\
    &\hat{\pi}(\mathbf{x})=\hat{\Pi}(\mathbf{x})\pi\left(1-\tau(\mathbf{x})\right)\label{eq:phase_shift_mag}\\
    &\hat{\Pi}(\mathbf{x})=\displaystyle\max_{\boldsymbol{\gamma}\in\mathcal{B}}\hat{\Pi}_{\boldsymbol{\gamma}}(\mathbf{x})\label{eq:region_based_shift}\\
    &\hat{\Pi}_{\boldsymbol{\gamma}}(\mathbf{x})={\text{smoothstep}\left(\Pi_{\boldsymbol{\gamma}}(\mathbf{x})\right)}=3\Pi_{\boldsymbol{\gamma}}(\mathbf{x})^2-2\Pi_{\boldsymbol{\gamma}}(\mathbf{x})^3\label{eq:smoothstep_local_gaussian}\\
   &\Pi_{\boldsymbol{\gamma}}(\mathbf{x})=\mathrm{e}^{-4{\omega^2}\Delta^\mathbf{s}_{\boldsymbol{\dot{\gamma}}}(\mathbf{x})
        {\left(1-0.5\psi(\mathbf{x})\right)}},\; \mathbf{s}=\left(\dfrac{1}{2\pi},\;1\right)^\intercal\label{eq:branch_local_gaussian}
\end{align}

This branch solidification ensures connection of $\hat{\tau}(\mathbf{x})$, but also introduces excessive material in the branching region. To reduce the amount of material in the branches while maintaining appropriate branch-shapes, the localised pinch-procedure is applied (\cite{Woldseth2023}). This is an iterative procedure, where for each iterate $k\in\{1,...,k_{max}\}$ the material of a branch is reduced and its shape corrected by utilising the control points $\boldsymbol{\gamma}_k$ and $\tilde{\boldsymbol{\gamma}}_k$ to move the branch centre from $\boldsymbol{\dot{\gamma}}$ towards $\boldsymbol{\ddot{\gamma}}$, with step-magnitude determined by the degree of disconnection (\autoref{eq:pinch_control_points}).
\begin{equation}\label{eq:pinch_control_points}
\begin{split}
\boldsymbol{\gamma}_k&=(1-d_k(\boldsymbol{\gamma}))\boldsymbol{\dot{\gamma}}+d_k(\boldsymbol{\gamma})\boldsymbol{\ddot{\gamma}}\quad \land \\
\boldsymbol{\tilde{\gamma}}_k&= (1-d_k^2(\boldsymbol{\gamma}))\boldsymbol{\dot{\gamma}}+d_k^2(\boldsymbol{\gamma})\boldsymbol{\ddot{\gamma}},\\ 
d_k(\boldsymbol{\gamma})&=\dfrac{1-\varrho(\boldsymbol{\gamma})}{2}(k-1)
\end{split}
\end{equation}

The pinch operator is directed by the gradients of the union $\tilde{\Pi}_k(\mathbf{x})$ of anisotropic Gaussians constructed for each $\boldsymbol{\gamma}\in \mathcal{B}$, centred at the correponding current iterate location $\boldsymbol{\gamma}_k$ (\autoref{eq:pinch_deriv_field}).
\begin{equation}\label{eq:pinch_deriv_field}
\tilde{\Pi}_k(\mathbf{x})=\max_{\gamma\in\mathcal{B}}\mathrm{e}^{-{\omega^2}\Delta^{\mathbf{\tilde{s}}}_{\boldsymbol{{\gamma}_k}}(\mathbf{x})},\; \mathbf{\tilde{s}}=\left(\frac{1}{2\pi},1\right)^\intercal
\end{equation}

To localise the pinch impact and improve the resulting shape, a union $\bar{\Pi}_{k}(\mathbf{x})$ of modified Gaussians, controlled by the interpolated orientation $\mathbf{n}(\mathbf{x})$ and relative thickness $w(\mathbf{x})$ at the sampling point $\mathbf{x}\in\mathcal{T}^{if}$, centred at the control points $\boldsymbol{\tilde{\gamma}}_k$ is introduced (\autoref{eq:pinch_localisation}). 

\begin{equation}
\begin{split}
    \bar{\Pi}_{k}(\mathbf{x})&=\max_{\gamma\in\mathcal{B}}\mathrm{e}^{-{2\omega^2}\Delta_{\boldsymbol{\tilde{\gamma}_k}}^\mathbf{\tilde{s}_k}(\mathbf{x})-\tilde{\delta}_{\boldsymbol{\tilde{\gamma}}_k}},\\ 
    \mathbf{\tilde{s}}_k&=\left(\frac{3}{4\pi}\left(1-\frac{k-1}{2}\right),1\right)^\intercal, \\ 
    \tilde{\delta}_{\boldsymbol{\tilde{\gamma}}_k}&=\dfrac{2\omega}{(2-w({\mathbf{x}}))}|\mathbf{n}(\mathbf{x})\cdot(\mathbf{x}-\boldsymbol{\gamma}_k)|\label{eq:pinch_localisation}
\end{split}
\end{equation}

Based on the introduced pinch control measures for an iterate $k\in\{1,...,k_{max}\}$, the corresponding transformation of the triangular wave-field from $\tilde{\tau}_{k-1}(\mathbf{x})$ to $\tilde{\tau}_{k}(\mathbf{x})$ at iteration $k$ is governed by the combined pinch perturbation $\mathcal{P}_k(\mathbf{x})$ and a perturbation magnitude scaling factor $\delta(\mathbf{x})$ (\autoref{eq:stepmag_dit}-\autoref{eq:update_rho}). 
\begin{align}
   & \mathcal{P}_k(\mathbf{x})=\delta(\mathbf{x})\bar{\Pi}_k(\mathbf{x})\nabla\tilde{\Pi}_k(\mathbf{x}),\quad\delta(\mathbf{x})=\dfrac{\lambda}{3}(1-w(\mathbf{x}))\label{eq:stepmag_dit}\\
    &\tilde{\tau}_k(\mathbf{x})=\tilde{\tau}_{k-1}\left(\mathbf{x}-{\mathcal{P}}_k(\mathbf{x})\right),\quad \tilde{\tau}_0(\mathbf{x})=\hat{\tau}(\mathbf{x})\label{eq:update_rho}
\end{align}
After completing $k_{max}$ iterations, the shape and material usage of the branch connections are effectively improved. Given the connected triangular wave-field $\tilde{\tau}_i$ for lamination layer $i\in \{1,2\}$, the corresponding thickness-adhering solid-void projection $\rho^s_i$ is obtained by \autoref{eq:layer_dens} (\autoref{fig:branch_closure}(c)).
\begin{equation}\label{eq:layer_dens}
    \rho^s_i=\mathcal{H}\left(\mathbf{w}_i-\tilde{\tau}_i\right),\;i\in\{1,2\}
\end{equation}
The combined dehomogenised structure, represented by the density-field $\rho^s$, is then obtained by the union of the lamination layers within the specified structural domain $\mathcal{I}\subseteq\Omega$ (\autoref{eq:layer_union}), and the corresponding dehomogenised volume fraction is computed according to \autoref{eq:volfrac_dehom}.
\begin{equation}\label{eq:layer_union}
    \rho^s=\left(\displaystyle\bigcup_{i\in\{1,2\}}\rho^s_i\right)\bigcap \mathcal{I}
\end{equation}

\begin{equation} \label{eq:volfrac_dehom}
    f^s = \frac{1}{V_{\Omega}} \int_{\Omega} \rho^s \mathrm{d}\Omega 
\end{equation} 




  \begin{figure*}
    \centering
    \makebox[\textwidth][c]{
        
\tikzstyle{startstop} = [rectangle, rounded corners, 
minimum width=2.5cm, 
minimum height=0.5cm,
text width=2.5cm, 
text centered, 
draw=black, 
fill=mycolor5!50]

\tikzstyle{io} = [trapezium, 
trapezium stretches=true, 
trapezium left angle=70, 
trapezium right angle=110, 
text width=3.75cm, 
minimum width=3.75cm, 
minimum height=0.5cm, text centered, 
draw=black, fill=blue!30]

\tikzstyle{process} = [rectangle, 
minimum width=2.5cm, 
minimum height=0.5cm, 
text centered, 
text width=2.5cm, 
draw=black, 
fill=mycolor4!50]

\tikzstyle{decription} = [rectangle, 
minimum width=4.5cm, 
minimum height=0.5cm, 
text centered, 
text width=4.5cm, 
draw=black,
fill=white]

\tikzstyle{decision} = [diamond, 
minimum width=1cm, 
minimum height=0.5cm, 
text centered, 
text width=2cm,
aspect=1.75,
draw=black, 
fill=mycolor3!50]

\tikzstyle{arrow} = [thick,->,>=stealth]

\begin{tikzpicture}[node distance=1.5cm]

\node (in1) [startstop] {Input model parameters};
\node (pro0) [process, below of=in1, yshift=0.15cm] {Prepair TO and Dehom};
\node (dec2) [decision, below of=pro0, yshift=-0.1cm] {TO data?};
\node (pro1) [process, below of=dec2, yshift=-0.75cm] {Obtain multi-scale design};
\node (pro2) [process, below of=pro1, yshift=-0.5cm] {FE-analysis};
\node (pro3) [process, below of=pro2, yshift=-0.25cm] {Obtain object functions and gradients};
\node (pro5) [process, below of=pro3, yshift=-0.3cm] {Update deisgn};
\node (pro6) [process, below of=pro5, yshift=0.5cm] {Dehomogenise};
\node (pro7) [process, below of=pro6, yshift=0.6cm] {Display design};
\node (dec1) [decision, below of=pro7, yshift=0.cm] {Converged?};
\node (pro10) [process, below of=dec1, yshift=-0.cm] {Store TO data};
\node (pro8) [process, below of=pro10, yshift=-0.1cm] {Post dehomogenise};
\node (pro9) [startstop, below of=pro8, yshift=-0.cm] {FE-analysis of Dehom design};

\node at ($(pro1.north west)+(-0.65,0.01)$) [thick,draw=none,anchor=south west] (toloop) {\textbf{TO loop}};

\draw [arrow] (dec1) -- node[anchor=north] {No} ([xshift=-0.15cm] dec1.west) |- (pro1);
\draw [arrow] (dec1) -- node[anchor=east] {Yes} (dec1.south) -- (pro10);

\draw [arrow] (dec2) -- node[anchor=south] {Yes} ([xshift=-1.cm] dec2.west) |- (pro8);
\draw [arrow] (dec2) -- node[anchor=east] {No} (dec2.south) -- (pro1);

\draw [arrow] (in1) -- (pro0);
\draw [arrow] (pro0) -- (dec2);
\draw [arrow] (pro1) -- (pro2);
\draw [arrow] (pro2) -- (pro3);
\draw [arrow] (pro3) -- (pro5);
\draw [arrow] (pro5) -- (pro6);
\draw [arrow] (pro6) -- (pro7);
\draw [arrow] (pro7) -- (dec1);
\draw [arrow] (pro10) -- (pro8);
\draw [arrow] (pro8) -- (pro9);


\node (pro0D) [decription, right of=pro0, xshift=2.75cm] {Get FE model, filters, initial design, and grids, $\mathcal{T}^{c}$, $\mathcal{T}^{i}$, $\mathcal{T}^{if}$, and $\mathcal{T}^{f}$}; \draw [dashed] (pro0) -- (pro0D);

\node (dec2D) [decription, right of=dec2, xshift=2.75cm] {TO result, $\Bar{w}$ and $N$, must be provided prior}; \draw [dashed] (dec2) -- (dec2D);

\node (pro1D) [decription, right of=pro1, xshift=2.75cm] {%
\begin{itemize}[nosep,leftmargin=*,noitemsep,nolistsep]
    \item Filter and projection, $w \rightarrow \Tilde{w}$ and $s \rightarrow \Tilde{s} \rightarrow \bar{s}$
    \item Indicator, $\bar{w} = \Tilde{w}\,\Bar{s}$
    \item Multi-scale, $\bar{w} \rightarrow \mu$
\end{itemize}%
}; \draw [dashed] (pro1) -- (pro1D);

\node (pro2D) [decription, right of=pro2, xshift=2.75cm] {\begin{itemize}[nosep,leftmargin=*,noitemsep,nolistsep]
    \item Assemble $\mathbf{K}$ with $\mathbf{C}(\mu,a)$
    \item Solve $\mathbf{K}\, \mathbf{u} = \mathbf{F}$
    \item Compute $\partial \mathbf{K} / \partial (\mu, a)$
\end{itemize}}; \draw [dashed] (pro2) -- (pro2D);

\node (pro3D) [decription, right of=pro3, xshift=2.75cm] {\begin{itemize}[nosep,leftmargin=*,noitemsep,nolistsep]
    \item Compute $\mathcal{J}$, $\mathcal{P}$ and $f$
    \item Compute $\partial\mathcal{J}/ \partial (w,s,a)$, $\partial\mathcal{P}/\partial s$ and $\partial f / \partial (w,s)$
\end{itemize}}; \draw [dashed] (pro3) -- (pro3D);

\node (pro5D) [decription, right of=pro5, xshift=2.75cm] {\begin{itemize}[nosep,leftmargin=*,noitemsep,nolistsep]
    \item Get $x_0 \leftarrow (w,a,s)$
    \item Call $x_{1} =\hspace{-0.5mm} \texttt{OC}(x_0,\partial \Phi, g,\partial g)$
    \item Update $x_{1} \rightarrow (w,a,s)$
\end{itemize}
}; \draw [dashed] (pro5) -- (pro5D);

\node (dec1D) [decription, right of=dec1, xshift=2.75cm] {Is $ |x_0 - x_{1}| < \texttt{stopCrit}$  and $\texttt{loop} > \texttt{loopMax}$?}; \draw [dashed] (dec1) -- (dec1D);

\node (pro10D) [decription, right of=pro10, xshift=2.75cm] {\begin{itemize}[nosep,leftmargin=*,noitemsep,nolistsep]
    \item Post-analysis intermediate design.
    \item Store design and result.
\end{itemize} 
}; \draw [dashed] (pro10) -- (pro10D);


\node (pro9D) [decription, right of=pro9, xshift=2.75cm] {\begin{itemize}[nosep,leftmargin=*,noitemsep,nolistsep]
    \item Assemble $\mathbf{K}$ with $\mathbf{C}(\rho)$ on $\mathcal{T}^f$
    \item Solve $\mathbf{K}\, \mathbf{u} = \mathbf{F}$
    \item Plot evaluated design
\end{itemize}}; \draw [dashed] (pro9) -- (pro9D);

\node (dpro0) [startstop, right of=pro0, xshift=7.5cm,yshift=-1.25cm] {Process input design};
\node (dpro4) [process, below of=dpro0, yshift=-0.9cm] {Construct domain boundary};
\node (dpro1) [process, below of=dpro4, yshift=-0.9cm] {Phase alignment};
\node (dpro2) [process, below of=dpro1, yshift=-0.6cm] {Phasor sampling};
\node (dpro5) [process, below of=dpro2, yshift=-1.45cm] {Close branches};
\node (dpro6) [process, below of=dpro5, yshift=-1.25cm] {Interpolate lamination layers};
\node (dpro7) [startstop, below of=dpro6, yshift=-0.85cm] {Project and combine final design};

\draw [arrow] (dpro0) -- (dpro4);
\draw [arrow] (dpro4) -- (dpro1);
\draw [arrow] (dpro1) -- (dpro2);
\draw [arrow] (dpro2) -- (dpro5);
\draw [arrow] (dpro5) -- (dpro6);
\draw [arrow] (dpro6) -- (dpro7);


\node (dpro0D) [decription, right of=dpro0, xshift=2.75cm] {
\begin{itemize}[nosep,leftmargin=*,noitemsep,nolistsep]
    \item Find active kernels on $\mathcal{T}^c$
    \item Interpolate design $\mathcal{T}^c{\rightarrow} \mathcal{T}^i$
\end{itemize}
}; \draw [dashed] (dpro0) -- (dpro0D);

\node (dpro1D) [decription, right of=dpro1, xshift=2.75cm] {
\begin{itemize}[nosep,leftmargin=*,noitemsep,nolistsep]
    \item Construct $\mathcal{N}(\mathcal{F},\mathcal{T}^c,\mathcal{W})$
    \item Obtain  $\varphi(\mathcal{F},\mathcal{T}^c,\mathcal{W})$
\end{itemize}
}; \draw [dashed] (dpro1) -- (dpro1D);

\node (dpro2D) [decription, right of=dpro2, xshift=2.75cm] {
\begin{itemize}[nosep,leftmargin=*,noitemsep,nolistsep]
    \item Localised sampling of $\mathcal{G}(\mathcal{F},\phi,\mathcal{T}^i)$ from $\mathcal{T}^c\rightarrow\mathcal{T}^{i}$
    \item Interpolate $\mathcal{G}$ to $\mathcal{T}^i\rightarrow\mathcal{T}^{if}$
    \item Get real-valued sawtooth wave $\Tilde{\phi} = \text{Arg}(\mathcal{G})$
\end{itemize}
}; \draw [dashed] (dpro2) -- (dpro2D);

\node (dpro4D) [decription, right of=dpro4, xshift=2.75cm] {
\begin{itemize}[nosep,leftmargin=*,noitemsep,nolistsep]
    \item Get boundary kernels and align orientations on $\mathcal{T}^c$
    \item Sample boundary phasor wave on $\mathcal{T}^{i}$
    \item Cut indicator field and thickness threshold boundary on $\mathcal{T}^{f}$
\end{itemize}
}; \draw [dashed] (dpro4) -- (dpro4D);

\node (dpro5D) [decription, right of=dpro5, xshift=2.75cm] {
\begin{itemize}[nosep,leftmargin=*,noitemsep,nolistsep]
    \item Locate singularity maxima in $\mathcal{G}$ on $\mathcal{T}^{i}\rightarrow\mathcal{T}^{if}$
    \item Solidify branches by local phase shift of phasor sine-wave on $\mathcal{T}^{if}$
    \item Locally pinch branches on triangular wave-field
\end{itemize}
}; \draw [dashed] (dpro5) -- (dpro5D);

\node (dpro6D) [decription, right of=dpro6, xshift=2.75cm] {
\begin{itemize}[nosep,leftmargin=*,noitemsep,nolistsep]
    \item Interpolate triangular wave-fields $\mathcal{T}^{if} \rightarrow \mathcal{T}^{f}$
    \item Cut according to indicator field
\end{itemize}
}; \draw [dashed] (dpro6) -- (dpro6D);

\node (dpro7D) [decription, right of=dpro7, xshift=2.75cm] {
\begin{itemize}[nosep,leftmargin=*,noitemsep,nolistsep]
    \item Threshold triangle wave wrt. $\mathcal{W}$ on $\mathcal{T}^f$
    \item Boolean union of lamination layers and boundary to obtain $\rho$
\end{itemize}
}; \draw [dashed] (dpro7) -- (dpro7D);


\node at ($(dpro0.north west)+(-0.25,0.15)$) [thick,draw=none,anchor=south west] (dehom) {\textbf{Phasor-based dehomogenisation}};


\begin{scope}[on background layer]
  \draw[thick,draw=none,fill=bycolor7!40] ($(pro1.north west)+(-0.75,0.6)$) rectangle ($(pro10.south east)+(5.4,-0.6)$);

    \coordinate (A) at ($(pro6.east)+(5.666,0.0)$);
    \draw [dashed,thick] (pro8) -|  (A);
    \draw [dashed,thick] (pro6) --  (A);
    \draw [dashed,thick] (A) |-  (dehom);

    \draw[thick,draw=none,fill=mycolor6!25] ($(dpro0.north west)+(-0.30,0.80)$) rectangle ($(dpro7.south east)+(5.5,-0.75)$);
  
\end{scope}


\end{tikzpicture}

    }%
    \caption{%
    Overview of the \deHomTop %
    structure, with emphasis on the topology optimisation (TO) and dehomogenisation (Dehom) components and how they interact.}
    \label{fig:code_flowchart}
\end{figure*}
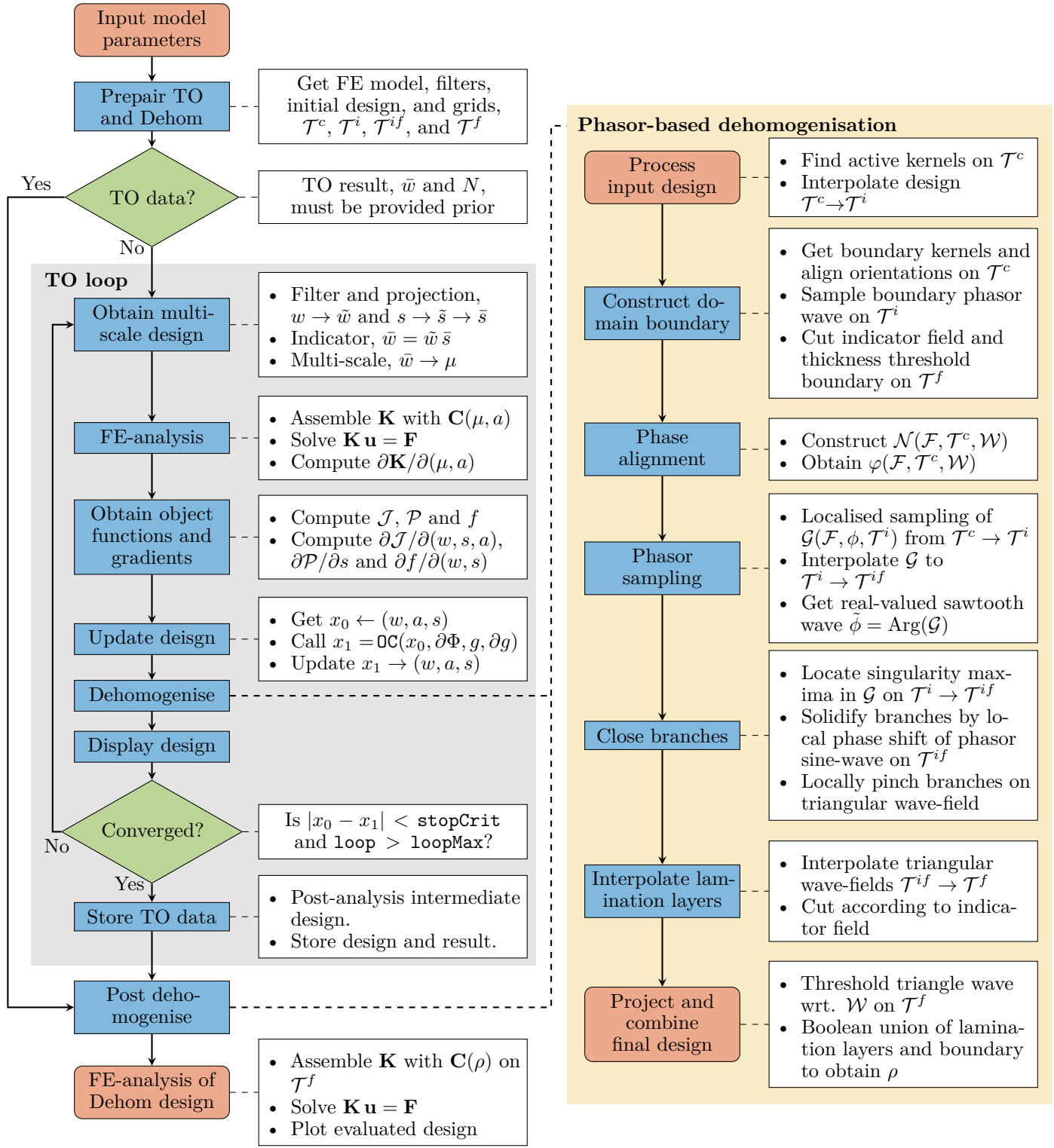


\section{General structure of the code}\label{sec:code}

The general structure of the code is given in the flowchart in \autoref{fig:code_flowchart}.
The main code spans from line 1 to line 182 and is divided into three phases; initialisation, optimisaiton and dehomogenisation. 

\subsection{Initialisation phase}

The initialisation phase (lines 4-56) begins by specifying the material properties (line 6) and building the FE-model (lines 8-12). Given the provided mesh size and scale, the FE-model of the bridge problem in \autoref{fig:model} is obtained from the function \texttt{prepFEA()}.
This FE-model contains information on the FE-analysis and grid $\mathcal{T}^c$, based on 4-noded bilinear Q4 elements, where nodes and elements are counted in y-major order (\autoref{fig:model}). The local node count is anti-clockwise, starting from the southwest corner of the elements (\autoref{fig:FEmodel}).
A crucial part of correct FE-analysis and successful multi-scale modelling is applying forces as surface tractions, instead of point forces, on solid elements to avoid stress raisers. In \texttt{prepFEA()}, this is achieved by introducing a passive block count, relative to $\mathcal{T}^c$ (lines 184-198).
\begin{lstlisting}[stepnumber=1,firstnumber=184,columns=fullflexible]
function [passiveElms,edofMat,iK,jK,U,F,freeDofs,Kp] = prepFEA(nelX,nelY,sc) (*@\Suppressnumber@*) 
... (*@\Reactivatenumber{192}@*)
[px,py] = deal(ceil(nelX/(15*sc))*sc,ceil(nelY/(30*sc))*sc);               
forceElmsB1 = floor(nelY*nelX*0.5) - (0:(ceil(0.5*px)-1))*nelY;            
forceElmsB2 = nelX*nelY - forceElmsB1 + nelY;
forceElms = union(forceElmsB1,forceElmsB2);
iF = edofMat(forceElms,:);                                                 
sF = repmat(-1*[0 1 0 1 0 0 0 0],[numel(forceElms),1]); 
F = sparse(iF(:),ones(size(iF(:))),sF(:),numDof,1); F = F/sum(abs(F(:)));
\end{lstlisting}
The surface traction is defined by first identifying elements subjected to a force  (\texttt{forceElms}) and  determining the triplet force values (\texttt{sF}) based on the defined node counting convention $\mathtt{edofMat_e = [u_1, v_1, u_2, v_2, u_3, v_3, u_4, v_4]}$ (lines 192-197). The constructed sparse force vector is scaled to unit magnitude (line 198). 
As promoted in \cite{Sigmund2022}, and to ensure well-behaved FE-analysis, the bridge problem's simply supported boundary conditions are implemented as multipoint constraints using the Penalty method \citep{Felippa2004} to achieve mesh independent point supports. Based on the degrees of freedom at $u_1$ (\texttt{fixedDofsB1}) and $u_2$ (\texttt{fixedDofsB2}), a sparse penalty block matrix (\texttt{Kp}) is constructed with a penalty weight of $10^4$ (lines 199-208). 
\begin{lstlisting}[stepnumber=1,firstnumber=199,columns=fullflexible]
[iB1,jB1] = meshgrid(fixedDofsB1); [iB2,jB2] = meshgrid(fixedDofsB2);      
Kp = sparse([iB1(:);iB2(:)], [jB1(:);jB2(:)],1e4,numDof,numDof);    
fixedDofs = fixedDofsB1(1)-1;   (*@\Suppressnumber@*)      
... (*@\Reactivatenumber{208}@*)
end
\end{lstlisting}
The imposed penalty weight is relatively low to account for the stiffness of the void ($E_{\min} = 10^{-9}$), as a higher weight would result in an ill-conditioned system. Another approach is to implement the multipoint constraints with Lagrange multipliers, but this results in a non-positive definite system, which would increase the complexity of solving the system.
The last boundary condition is a single constrained node in the $y$-direction, to prevent rigid body translation (line 201).
The final part of \texttt{prepFEA()} locates the passive design elements \texttt{passiveElms}, outputted as \texttt{pasE} in the main code. Based on the passive design, the active design variables are found (lines 8-10). The combined set of design variables is accessed as $\bm{\kappa} = \{ \mathbf{w}_1,\mathbf{w}_2,\mathbf{s},\mathbf{a} \}$, where $\bm{\kappa}$ is denoted \texttt{x} in the code.

On lines 14-21 the filtering \autoref{eq:filter} and robust projection \autoref{eq:proj} are initialised. The filter is based on a convolution filter, with approximate Gaussian kernels, \texttt{hS} and \texttt{hW}, for the indicator and width fields, respectively. The kernel size of the indicator field filter radius is based on the scalar \texttt{rsc} and the input width filter radius \texttt{rMin}.

On lines 23-28 the starting guess for orientation design variables are found from the principle stress directions of an isotropic solid domain. The isotropic constitutive elements of an upper-triangular matrix are obtained from the function \texttt{getIsotropicElasticity()}, which is passed to the FE-analysis, \texttt{doFEA()}. The constitutive elements are mirrored to a square matrix for computing the stresses. Due to angle invariants, the orientation field is corrected to $[0,2\pi]$ for any negative angles. 
The scaling factor for compliance (\texttt{cScale}) is further obtained.

On lines 30-33 the starting guess for the indicator and width fields are defined. All elements are initially considered active, while the width is found from (\autoref{eq:homrho}) and bounded by (\autoref{eq:TopOptBox}). Elements with passive design values are further enforced. 

The dehomogenisation routine is initialised on lines 35-39. From the function \texttt{prepPhasor()} the grids $\mathcal{T}^{i}$, $\mathcal{T}^{if}$, and $\mathcal{T}^{f}$, as well as distance parameters are computed. From this, the element-wise maximum neighborhood candidate sets are found in \texttt{prepPhasorKernels()}. Another FE-model is created using \texttt{prepFEA()}, on $\mathcal{T}^{f}$, to obtain passive design information for the dehomogenisation.

On lines 41-47 the figures for plotting during optimisation are created, using a tiled layout. Finally, on lines 49-56 the initialisation phase is completed with initialising the combined design vector \texttt{x}, the box constraints \texttt{xmin0} and \texttt{xmax0} (\autoref{eq:TopOptBox}), and the move limits \texttt{xmove}. The indicator penalty scale \texttt{sWeight} (\autoref{eq:TopOptObj}) is set. If an optimisation result is provided through the \texttt{TO} argument, the maximum number of design iterations is changed from 300 to 0, thus skipping the optimisation loop.

\subsection{Optimisation phase}

The optimisation routine is defined from line 58 to line 149, where the first part is an optimisation loop and the final 10 lines constitute the post-evaluation step. 

The optimisation loop starts by checking stopping criteria and the iteration count, whereafter the time, loop and projections parameters are updated on lines 61 to 66. 
Hereafter, from lines 68 to 73 the single-scale design is obtained. First the active design variables of $w$ and $s$ are filtered (\autoref{eq:filter}). Subsequently, $\tilde{s}$ is projected using the three-field approach (\autoref{eq:proj}), and finally multiplied with $\tilde{w}$, to get the single-scale eroded physical relative widths (\autoref{eq:ip2}).  

The multi-scale design is obtained on lines 75-76, where the relative densities (\autoref{eq:homrho}) are obtained. From the eroded single-scale physical relative widths, the multi-scale widths, $\mu$, are obtained by the function \texttt{getMuFromW()} (line 210), with \autoref{eq:w2mu} is combined with \autoref{eq:w2mu_p}. \texttt{getMuFromW()} further computed the sensitivities of \autoref{eq:w2mu} with respect to $w$.

\begin{lstlisting}[stepnumber=1,firstnumber=210,columns=fullflexible]
function [mu,dmudw] = getMuFromW(rho,w) (*@\Suppressnumber@*) 
... (*@\Reactivatenumber{213}@*)
wSum = w(:,1)+w(:,2);        
mu(:,1) = w(:,1)./wSum.*rho; 
mu(:,2) = w(:,2).*rho./(wSum.*(1 - mu(:,1)));    (*@\Suppressnumber@*) 
... (*@\Reactivatenumber{224}@*)
end
\end{lstlisting}

From the multi-scale design, the FE-analysis is commenced starting on line 78 by obtaining the Rank-2 elastic properties and their sensitivities with respect to $\mu$ and $a$ by the function \texttt{getConstitutiveRank2()} (line 226). 
The constitutive matrix (\autoref{eq:cmat}) is obtained followed by the rotation (\autoref{eq:rotcmat}). To efficiently perform this construction by vectorisation, the element matrices are stored as an $\mathbb{R}^{3 \times 3 \times N_e}$-array, where the page-wise matrix operations (\texttt{pagemtimes()}, \texttt{pagectranspose()}) are utilised. However, due to symmetry, only the lower triangles of the constitutive matrices are considered, which is extracted utilising the two functions \texttt{TRI()} and \texttt{T2D()}.  
\begin{lstlisting}[stepnumber=1,firstnumber=226,columns=fullflexible]
function [CT,dCTdmu1,dCTdmu2,dCTda] = getConstitutiveRank2(a,mu,E,nu,EMin) (*@\Suppressnumber@*) 
... (*@\Reactivatenumber{230}@*)
[C,dCdmu1,dCdmu2] = deal(zeros(3,3,numel(a)));
T2D = @(A) reshape(permute(A,[3,1,2]),[],9);                               
TRI = @(A) reshape(A(:,[1,2,3,5,6,9]),1,[],6);    (*@\Suppressnumber@*)                            
... (*@\Reactivatenumber{234}@*)
denominator = (1 - mu2+mu1.*mu2*(1-nu^2));                                 
C(1,1,:) = E./denominator.*(mu1)+EMin/(1-nu^2);    (*@\Suppressnumber@*)                           
... (*@\Reactivatenumber{240}@*)
T = [c.^2 s.^2 c.*s; s.^2 c.^2 -c.*s; -2*c.*s 2*c.*s c.^2-s.^2];           
Tt = pagectranspose(T);                                                    
CT = TRI(T2D(pagemtimes(pagemtimes(Tt,C),T)));   (*@\Suppressnumber@*)                            
... (*@\Reactivatenumber{258}@*)
end
\end{lstlisting} 
The FE-analysis is completed (line 79) by the function call to \verb|doFEA()|, where the stiffness matrix is assembled from the lower triangles of the constitutive matrices (lines 267-271). This is achieved by utilising the page-wise matrix operation for the triplet stiffness values. The penalty term, \texttt{Kp}, is scaled and added to the system. The displacement state field, \texttt{U}, is solved using Cholesky factorisation.
\begin{lstlisting}[stepnumber=1,firstnumber=267,columns=fullflexible]
function U = doFEA(F,U,C,iK,jK,freeDofs,KE0,Kp)
K = sparse(iK,jK,reshape(sum(pagemtimes(KE0,C),3),[],1));                  
K = K + Kp*max(diag(K));                                                   
U(freeDofs)=decomposition(K(freeDofs,freeDofs),'chol','lower') \F(freeDofs);
end
\end{lstlisting}

On lines 81-88 the compliance objective function (\autoref{eq:TopOptObj}) and sensitivities are computed. 
The sensitivities wrt. the state field are precomputed on lines 82-83 to get the element-wise contributions. 
The stiffness matrix sensitivities are obtained similarly to how the stiffness matrix is assembled (line 74), which can be used to obtain the element-wise matrix-vector products for the adjoint sensitivities (lines 87-88). 
\begin{lstlisting}[stepnumber=1,firstnumber=74,columns=fullflexible]
dKdMu1 = sum(pagemtimes(KE0,dCdmu1),3);
\end{lstlisting}

The indicator field volume penalty function and sensitivities (\autoref{eq:TopOptObj}) are computed on lines 90-91.

On lines 93-95, the volume fraction constraint function and sensitivities are computed by the mean of the relative densities (\autoref{eq:homrho}). The dilated volume fraction is also updated for the three-field robust approach. 

As the final part of the sensitivity analysis the chain and product rule sensitivities are computed from design domain modifications, i.e., multi-scale transformation, indicator field, projection and filtering (lines 97-106). 


The design is updated on lines 108-118 with the modified OC routine from the \texttt{top250} code~\citep{Ferrari2021}.

The on-the-fly dehomogenisation routine is called on lines 120-126, which will be discussed in detail in \autoref{sec:dehom_code_struct}.

Finally, the results are printed and plotted on lines 128-135. To plot efficiently during optimisation, only the iterative data is updated in the already initialised plots.

Breaking from the optimisation loop, the optimised intermediate result is post-evaluated on lines 138-148 if the \texttt{TO} data is not already provided, otherwise the \texttt{TO} data will be stored as the final part of the post-evaluation section.

\subsection{Dehomogenisation phase}

The final phase of the main code is divided into two sections; first, on lines 150-159, the multi-scale structure is dehomogenised by the \verb|phasorDehomogenise()| procedure (\autoref{sec:dehom_code_struct}), where the resulting single-scale volume fraction is evaluated wrt. the multi-scale structure. 

If the \texttt{eval} input is \texttt{true}, the dehomogenised structure is evaluated for structural performance on lines 161-181. As the dehomogenised structure is single-scale, isotropic elastic properties are assigned to each element, to compute the displacement field, from which the compliance is obtained and evaluated. 
To give a visual evaluation of the structure, the strain energy density is computed on lines 169-172.
\begin{lstlisting}[stepnumber=1,firstnumber=169,columns=fullflexible]
strain = reshape(BMatrixMid*U(edofMat'),3,1,[]);                       
stress = pagemtimes(reshape(permute(C(:,:,[1,2,3,2,4,5,3,5,6]),...     
    [3,1,2]),3,3,[]),strain);
W0 = 0.5*squeeze(sum(stress.*strain,1));                               
\end{lstlisting}
The stress and element stress fields are again found using the page-wise matrix operation. Finally, the strain energy density is plotted with log-10 scaling to enhance the visual energy contrast in the plot.

\subsection{Dehomogenisation structure}\label{sec:dehom_code_struct}
The main Matlab routine for the phasor-based  dehomogenisation procedure, outlined in \autoref{fig:code_flowchart}, is implemented in the function \verb|phasorDehomogenise()| (line 356).
\begin{lstlisting}[stepnumber=1,firstnumber=356,columns=fullflexible]
function rho = phasorDehomogenise(grid,wMin,w,N,alignItr,align)
\end{lstlisting}
From the homogenised design this routine requires the minimum relative thickness \verb|wMin|, layerwise relative thicknesses \verb|w| and lamination normals \verb|N| on $\mathcal{T}^c$ as inputs. The dehomogenisation procedure specifications are provided by the number of phase alignment iterations \verb|alignItr|, the \verb|grid| structure, obtained by the \verb|prepPhasor()| function (line 273), and the \verb|align| structure, obtained by the \verb|prepPhasorKernels()| function (line 340). These structures contain precomputed sizes, parameters and index-sets utilised throughout the dehomogenisation process, that can be reused when dehomogenising designs with shared minimum feature size and coarse mesh representation.

To enable the sequential projections from $\mathcal{T}^c$ to the intermediate and fine grids, $\mathcal{T}^{a},\;a\in\{i\;,if,\;f\}$, the homogenised information on the coarse grid is interpolated to obtain smoothed design representations on the finer grids (line 361-364).
\begin{lstlisting}[stepnumber=1,firstnumber=361,columns=fullflexible]
indicator = getIndicator(wMin,w,grid,nLayer); 
[Nx,Ny] = filterVectorField(grid,nLayer,actKernels,Nx,Ny);
[iNx,iNy,ifNx,ifNy,grid] = interpOrientations(grid,Nx,Ny,nLayer);
[w_i,w_if,w_f] = interpThickness(grid,nLayer,w,wMin,indicator.coarse);
\end{lstlisting}
\verb|getIndicator()| interpolates layer-wise smoothed material indicator fields from $\mathcal{T}^c$ to $\mathcal{T}^i$. \verb|filterVectorField()| and \verb|interpolateVectorFields()| corrects for localised artefacts in the orientation fields on $\mathcal{T}^c$ and interpolates to $\mathcal{T}^{i}$ and $\mathcal{T}^{if}$ while accounting for spurious behaviour caused by interpolation of periodic data. The layer-wise relative thicknesses are interpolated in \verb|interpThickness()|.

The procedure \verb|addBoundary()| (line 365) generates a synthetic outer layer by constructing a set of phasor kernels along the structural boundary on $\mathcal{T}^c$ and samples a single wave outlining the structural body on $\mathcal{T}^{i}$, which is utilised for indicator field smoothing and establishing a varying thickness boundary, according to local layer thicknesses, on $\mathcal{T}^f$. Additionally, the layer-wise active phasor kernels aligned with the boundary orientations are identified within this procedure. Some of the local orientation and phase-shift adaptions from \citet{Woldseth2023} have been removed to focus on the fundamental elements of this procedure. 

\begin{lstlisting}[
stepnumber=1,firstnumber=365,columns=fullflexible]
[domain,boundary,align] = addBoundary(grid,w_f,align,wMin,indicator,Nx,Ny);
\end{lstlisting}

\verb|phaseAlignment()| and \verb|phasorSampling()| comprise the main components of the phasor noise methodology, by performing the phase alignment procedure (\autoref{eq:phase_alignment_update}-\autoref{eq:opposing_orient_correct}) on $\mathcal{T}^c$ (line 367) and the phasor signal sampling (\autoref{eq:combined_signals}) to $\mathcal{T}^{i}$ (line 369), respectively. \verb|non_solid| is computed as the layer-wise intermediate density fields on intermediate grid $\mathcal{T}^{i}$ (line 368).
\begin{lstlisting}[
stepnumber=1,firstnumber=367,columns=fullflexible]
align = phaseAlignment(grid,Nx,-Ny,alignItr, actKernels,align,nLayer,w); 
non_solid = (w_i < 1-1e-3).*indicator.intermed;
phasor_i = phasorSampling(grid,Nx,-Ny,iNx,-iNy,actKernels,...  
    non_solid>=0.01,align,nLayer);
\end{lstlisting}

 The branch connection procedure is applied layer-wise by \verb|closeBranches()| (line 385) given the set of inputs (line 379-384). \verb|inp1| is the sampled complex phasor field on $\mathcal{T}^{i}$ and \verb|inp3| its counterpart on $\mathcal{T}^{if}$, obtained by cubic interpolation (line 372-376). The branch point localisation is performed on $\mathcal{T}^{i}$ and \verb|inp2| gives the region of interest for this search, based on the indicator for intermediate layer densities. \verb|inp4|, \verb|inp5| and \verb|inp6| provide the interpolated thicknesses and orientation unit vectors on $\mathcal{T}^{if}$, to be utilised for closing the branches. 
\begin{lstlisting}[stepnumber=1,firstnumber=379,columns=fullflexible]
    inp1 = reshape(phasor_i(:,r),grid.isize(1:2));    
    inp2 = reshape(non_solid(:,r)>=0.5,grid.isize(1:2));
    inp3 = reshape(phasor_if(:,r),grid.ifsize(1:2)); 
    inp4 = reshape(w_if(:,r),grid.ifsize(1:2)); 
    inp5 = reshape(ifNx(:,r),grid.ifsize(1:2)); 
    inp6 = reshape(ifNy(:,r),grid.ifsize(1:2)); 
    triangular_if(:,r) = closeBranches(grid,inp1,inp2,inp3,inp4,inp5,inp6);
\end{lstlisting}
The final dehomogenised design is assembled by combining the thresholded individual lamination layers and the structural boundary (line 388-395). \verb|rho| represents the finalised structure, which is initialised as the approximately solid regions within the smoothed domain, as these regions are not sampled from the phasor kernels defined by the intermediate density elements. Linear interpolation is applied to obtain the connected wave-field on $\mathcal{T}^f$ (line 391), which is thresholded to solid-void based on the relative thicknesses (line 392). After combining the lamination layers, the smoothed structural indicator removes potential artefacts outside the structural domain, and the varying thickness boundary is added (line 395). Potential floating structural members are removed by the connected component operation \texttt{removeIslands2D()}.
\begin{lstlisting}[stepnumber=1,firstnumber=387,columns=fullflexible]
rho = reshape(max(w_f,[],2) >= 0.99,grid.fsize(1:2)); 
for r = 1:nLayer  
    grid.intp_if.Values(:) = (triangular_if(:,r)+1)/2; 
    triangular_if_f = grid.intp_if({grid.y_f',grid.x_f'});
    thresholded =  triangular_if_f(:) >= 1-w_f(:,r); 
    rho(:) = rho(:)+thresholded;
end
rho = removeIslands2D(min(rho.*domain+boundary,1),0.1);
\end{lstlisting}
These descriptions provide an overview of the functionalities in \deHomTop and how they collectively contribute to obtaining a dehomogenised design. 

\subsubsection{Phase-alignment}
The implementation of the phase alignment routine is performed on $\mathcal{T}^c$, given the orientations and indexes of active kernels, for all lamination layers. 
\begin{lstlisting}[stepnumber=1,firstnumber=690,columns=fullflexible]
function aCand = phaseAlignment(grid,dirx,diry,alignItr,...
                            active,aCand,nLayer,w)
\end{lstlisting}
The first part of the code computes and stores precomputed information (line 693-717) which is reused in the iterative alignment process (line 720-733). For each layer, the phase shifts are initialised to zero for any kernel not identified as being sufficiently aligned with the domain boundary from \verb|addBoundary()|. For boundary-aligned kernels, the phase shifts are initialised to $-\pi/2$ (line 719), and the kernel alignment order is determined based on the minimal distance to these boundary-aligned kernels (line 723). In each alignment iteration, the phase shifts of the active kernels are updated sequentially, according to this pre-computed ordering. The update for a single kernel (\autoref{eq:phase_alignment_update}), is computed based on the weighted average contribution \verb|avg_shift| (line 729) from its alignment neighbourhood \verb|idNj| (line 728), by computing the corresponding real-valued argument (line 730) to obtain the updated phase shift.

\begin{lstlisting}[
    stepnumber=1,firstnumber=719, columns=fullflexible]
pshift = 0*aCand.idbnd; pshift(aCand.idbnd(:)) = -pi/2;
for k = 1:nLayer
    [orientk,idnk,weightk] = deal(orient(:,:,k),indx(:,:,k),weight(:,:,k));
    aind = idM(active(:,k));
    [~,ordid] = sortrows([aCand.bdist(aind,k),w(aind,k)],[1,-2]);
    for iter = 1:alignItr 
        for j = aind(ordid)
            idnj = idnk(:,j); wj = weightk(idnj,j);
            ortj = orientk(idnj,j); ortj2 = pi*(1-ortj)/2; 
            idNj = idN(j,idnj);
            avg_shift = sum(wj.*exp(1i*(ortj.*pshift(idNj,k)+ortj2)));
            pshift(j,k) = atan2(imag(avg_shift),real(avg_shift));
\end{lstlisting}

\verb|idN| is a global index-set, which for each kernel covers the maximal span of the neighbourhood $\mathcal{N}_j$ (\autoref{eq:align_neighbourhood}) regardless of the kernel orientation. \verb|idnj| indicates, for the considered kernel orientation and active kernel set, the local index of current neighbours \verb|idNj|. As only the phase shifts are updated in each iteration, more efficient alignment can be obtained by precomputing the remainder of the neighbour contributions \verb|wj| (\autoref{eq:phase_alignment_update}), as well as the orientation coherency correction \verb|ortj|. 

\subsubsection{Phasor field sampling}
The phasor field is sampled on the intermediate grid $\mathcal{T}^{i}$, in a layer-wise manner, by summation of the phasor kernel signal responses at locations within the specified material indicator field. To sample the kernels, their corresponding orientations and aligned phase-shifts are required on $\mathcal{T}^c$, and for the integrated filter, the interpolated orientations on $\mathcal{T}^{i}$.
\begin{lstlisting}[stepnumber=1,firstnumber=737,columns=fullflexible]
function phasor = phasorSampling(grid,Dx,Dy,Dxx,Dxy,...
                            active,indicator_i,align,nLayer)
\end{lstlisting}

The computation of \verb|s| corresponds to the anisotropic distance (\autoref{eq:2d_anisotropic_distance}) from the sampling points, within a maximal cover of the intended sampling span, to the kernel being sampled (line 747-748). Gaussian weight cut-off is utilised to localise the sampling of the signal (line 749). $\Lambda_j(\mathbf{x})$ and $\|\Lambda_j(\mathbf{x})\|_2^2$ utilised in the sampling filter (\autoref{eq:sampling_filter_kernel}) are computed for the localised selection (lines 751-753). \verb|wfilt_j| is computed as the corresponding filtered signal response of the current kernel (line 754-756) and added to the sum of signal responses (\autoref{eq:combined_signals}) stored in the \verb|phasor| array (line 757). 

\begin{lstlisting}[stepnumber=1,firstnumber=746]  
for k=idK(active(j,:))
    [xdv,ydv] = deal(xdist*Dx(j,k), ydist*Dy(j,k));
    s = grid.rx*(xdist*Dy(j,k)-ydist*Dx(j,k)).^2+ grid.ry*(xdv+ydv).^2;
    idcutoff = s < grid.cutoff; idxy = idXY(idcutoff(:));distance cut-off
    [xdv,ydv,s] = deal(xdv(idcutoff), ydv(idcutoff), s(idcutoff));
    d_jx = grid.omega*(Dx(j,k)-Dxx(idxy,k));
    d_jy = grid.omega*(Dy(j,k)-Dxy(idxy,k));
    d_jsq = (d_jx.^2+d_jy.^2);
    wfilt_j = exp(-s*grid.babpainv- pi^2*d_jsq*grid.bpainv + ... 
         2*1i*(pi*grid.omega*(xdv+ydv)+ (d_jx.*xdist(idcutoff) + ...
         d_jy.*ydist(idcutoff))*grid.abpainv) +1i*pshift(j,k));
    phasor(idxy,k) = phasor(idxy,k)+(wfilt_j); 
end
\end{lstlisting}

\subsubsection{Branch closure}
The branch closure procedure, with function call introduced in \autoref{sec:dehom_code_struct}, consists of three main steps, namely the branch point localisation (\autoref{eq:branching_point_set}), branch solidification (\autoref{eq:solidified_triangular}-\autoref{eq:branch_local_gaussian}) and pinch procedure (\autoref{eq:pinch_deriv_field}-\autoref{eq:update_rho}).
\begin{lstlisting}[firstnumber=398,columns=fullflexible]
function triangular_if = closeBranches(grid,phasor_i, bregion,phasor_if,w_if,ifNx,ifNy)
\end{lstlisting}

The set $\mathcal{B}$ of branch points is determined from the sampled phasor field $\mathcal{G}(\mathbf{x})$ for  $\mathbf{x}\in\mathcal{T}^{i}$ (\autoref{eq:combined_signals}), and subsequently projected to $\mathcal{T}^{if}$ by the \verb|locateBranchinPoints()| procedure (line 401). 
\begin{lstlisting}[firstnumber=401,stepnumber=1,columns=fullflexible]
[gamma_x,gamma_y,numBranch] = locateBranchingPoints(phasor_i,bregion,grid);
\end{lstlisting}

The degree of connection \verb|condeg| (\autoref{eq:disconnection_degree}) is computed based on the circular neighbourhood extracted from the precomputed \verb|grid.discomp| (lines 407-416).
\begin{lstlisting}[firstnumber=416,
    stepnumber=1]  
condeg = (sum(sif_pad(idofMat).*circle(:)',2) /sum(circle(:))+1)/2;
\end{lstlisting}
A simplified computation approximating the closure direction (\autoref{eq:closure_direction}) based on single-point values is utilised to determine the closure control points (lines 418-433). 

The main procedure for performing branch solidification (\autoref{eq:solidified_triangular}-\autoref{eq:branch_local_gaussian}) is executed sequentially for each branching point (lines 450-458). For each branching point, the phasor sine-wave \verb|sine_if| is locally updated based on the weighted phase shift \verb|pihat| induced by the current branch point.
\begin{lstlisting}[stepnumber=1,firstnumber=450,columns=fullflexible]  
for K=1:numBranch(*@\Suppressnumber@*) 
    ...(*@\Reactivatenumber{453}@*)
    Pihat =exp(-2*sigma_inv*(wx*xd.^2+yd.^2).* (1-0.5*triangular_if(idL)));
    Pihat2 = Pihat.*Pihat; Pihat = 3*Pihat2-2*Pihat2.*Pihat;
    pihat = pi*Pihat.*(1-(2/pi*asin(sine_if(idL))+1)/2);
    sine_if(idL) = max(sine_if(idL), ...
        max(sin(atan_if(idL)-pihat),sin(atan_if(idL)+pihat))); 
end
\end{lstlisting}

After branch closure is ensured for all branching points on $\mathcal{T}^{if}$, the pinch procedure is locally applied to each branching region on the converted triangular wave-field, by performing $k_{max}=3$ pinching steps. The anisotropic Gaussian with gradients specifying the pinch direction \verb|Pi_pinch_k| (\autoref{eq:pinch_deriv_field}) is located about the centre specified by the current pinch step locations \verb|ixx| and \verb|iyy| corresponding to $\gamma_{k}$. \verb|vx_k| and \verb|vy_k| are the directional gradients of this Gaussian determined by a derivative filter and scaled according to the largest gradient magnitudes. 

The modified Gaussian for localising the pinch magnitude \verb|Pi_local_k| (\autoref{eq:pinch_localisation}) is approximated by the oriented distances \verb|xdm| and \verb|ydm| to the iterated point $\tilde{\gamma}_{k}$.
\begin{lstlisting}[stepnumber=1,firstnumber=486,columns=fullflexible]
for k=1:3 
    step = min((1-condeg(K))/2*(k-1),1); 
    ixx = round(step*ix2+(1-step)*ix)-min(xlims)+1;
    iyy = round(step*iy2+(1-step)*iy)-min(ylims)+1; (*@\Suppressnumber@*) 
        ...(*@\Reactivatenumber{494}@*)  
    Pi_pinch_k = exp(-0.5*sigma_inv*(wx*xdx3.^2+ydx3.^2)); 
    vx_k = sum(Pi_pinch_k(idofMat(:,[1:3,7:9])).*kx(:)',2); 
    vy_k= sum(Pi_pinch_k(idofMat(:,[1,3,4,6,7,9])).*ky(:)',2);(*@\Suppressnumber@*) 
        ...(*@\Reactivatenumber{510}@*)  
    step = step.^2; 
    xdm =step.*xd2+(1-step).*xd; ydm =step.*yd2+(1-step).*yd; 
    Pi_local_k = exp((-sigma_inv*(wx*(1+(k-1)/2)*xdm.^2+... 
            ydm.^2)-c_weight.*abs(ydm)));
    pw = ws./3.*Pi_local_k.*(1-w_patch);
   triangular_if(xlims,ylims) = pinchPatch( ...
    triangular_if(xlims,ylims),pw.*vx,pw.*vy,idy,idx,ny,nx);
end
\end{lstlisting}
The combined pinch magnitudes and directions \verb|pw.*vx| and \verb|pw.*vy| (\autoref{eq:stepmag_dit}) are then passed to the \verb|pinchPatch| procedure performing the corresponding interpolation (\autoref{eq:update_rho}). After all branching point regions have been subjected to the iterative pinch procedure, the triangular wave-field with connected branches is obtained. 

\subsubsection{Structural boundary}
Due to the realisation of a single-scale structure at finite periodicity, the global phase shift of the lamination layers with respect to the boundary of the structural domain may cause non-load carrying members towards the outer periphery of the structure. \citet{Woldseth2023} exemplified why the addition of a varying thickness structural boundary is highly beneficial for the performance of the dehomogenised structure. A simplified version of the phasor-based boundary from this original work is implemented in \deHomTop in the function \verb|addBoundary()| (line 544), where the structural boundary is constructed and additionally utilised to smooth the upscaled structural indicator field to reduce staircase artefacts from the underlying coarse mesh. This structural boundary is constructed as an artificial lamination layer defined by kernels on $\mathcal{T}^c$ which are sampled as phasor noise on $\mathcal{T}^{i}$ and finally modified and projected to $\mathcal{T}^f$ to obtain the desired results.
\begin{lstlisting}[stepnumber=1,firstnumber=544,columns=fullflexible]
function [indicator_f,shellwave,align] = ...
    addBoundary(grid,w, align,wmin,indicator,Nx,Ny)
\end{lstlisting}
A filtered structural indicator field is first obtained by combining the individual layer indicator-fields, thresholding and applying a Gaussian filter with zero padding (lines 547-552). This filtered field is subsequently subjected to a Gaussian derivative filter to obtain the derivative components \verb|vx| and \verb|vy| (lines 553-555). The angle of orientation \verb|vdir| and derivative magnitude \verb|vmag| are derived from these components (line 556) and the kernels within the structural boundary region \verb|potential| are identified based on the derivative magnitude, as this measure tends to zero away from the boundary region (line 557). 

\begin{lstlisting}[stepnumber=1,firstnumber=554,columns=fullflexible]
vx = reshape(sum(idsp(idofMat).*kx(:)',2),grid.csize(1:2));
vy = reshape(sum(idsp(idofMat).*ky(:)',2),grid.csize(1:2)); 
[vdir,vmag] = deal(atan2(vy,vx), vx.^2+vy.^2);     
[potential,bvec] = deal(vmag>1e-3, [cos(vdir(:)),sin(vdir(:))]);    
\end{lstlisting}
The boundary orientations \verb|bvec| of the boundary region kernels are aligned in a similar manner to the phase alignment procedure (line 559-571). Instead of the phase shift, the angle of the orientation vectors are aligned, and minor modifications are imposed to reduce alignment between orientations with a relative angular difference approaching $\pi/2$, to better adapt the boundary to smaller holes within the structure. After alignment, the orientations are upscaled by complex interpolation to $\mathcal{T}^{i}$ (lines 573-575), to be used for the phasor sampling filter.

\begin{lstlisting}[stepnumber=1,firstnumber=577,columns=fullflexible]
idbnd = potential & vmag >= 0.25*max(vmag(:)); 
new_omg = min(1./(8*grid.h_c),grid.omega/2); omg_rat = new_omg/grid.omega;  (*@\Suppressnumber@*) 
     ...(*@\Reactivatenumber{582}@*)
atmp = align; atmp.pshift = pi*((1-idsfr(:))*0.5+1/3); 
bphasor = phasorSampling(grid,bvec(:,1),bvec(:,2), ...
    cos(vec_a(:)),sin(vec_a(:)),idbnd(:),bregion(:),atmp,1);
\end{lstlisting}
The boundary phasor kernels \verb|idbnd| are identified (line 577), prescribed a periodicity \verb|new_omg| (line 578) and assigned individual kernel phase-shifts (line 582). The phase-shift is dependant upon the filtered indicator field value at the kernel grid point to control the location of the signal wave relative to the structural domain boundary. The complex phasor noise signal \verb|bphasor| is sampled using \verb|phasorSampling()| directly from these pre-defined phase shifts (line 583-584).

The layer-wise identification of kernels that are aligned with the boundary kernels is based on the element-wise dotproducts between the sets (lines 586-588). These identified kernels are utilised in the initialisation of kernel phase shifts in \verb|phaseAlignment()|, and the layer-wise minimal distance to these kernels (lines 589-592) determine the phase alignment order. 

\begin{lstlisting}[stepnumber=1,firstnumber=586,columns=fullflexible]
ddot = Nx.*bvec(:,1)+Ny.*bvec(:,2); adot = abs(ddot); nd = max(adot,[],2);(*@\Suppressnumber@*) 
     ...(*@\Reactivatenumber{588}@*)
align.idbnd = align.idbnd & abs(nd-adot)==0 & adot>0.95*max(nd);
for r = 1:size(w,2)   
    align.bdist(:,r) = min([(grid.X_c(:)-grid.X_c(align.idbnd(:,r))').^2+...
          (grid.Y_c(:)-grid.Y_c(align.idbnd(:,r))').^2, 0*idbnd(:)+1],[],2);
end
\end{lstlisting}

From the sampled phasor boundary-wave the indicator cut-field \verb|cutfi| is constructed in a similar manner to \citet{Woldseth2023} on $\mathcal{T}^{i}$ and projected to $\mathcal{T}^{f}$ before thresholding to obtain the smoothed structural domain \verb|indicator_f| (lines 592-600). 

\begin{lstlisting}[stepnumber=1,firstnumber=595,columns=fullflexible]
bsawtooth = atan2(imag(bphasor),real(bphasor)); 
positive = (sin(bsawtooth) > 1e-1).* sqrt(indicator_i.*(1-indicator_i)); 
cutfi = 10*positive.*sin(bsawtooth+pi/2)+indicator_i;(*@\Suppressnumber@*) 
     ...(*@\Reactivatenumber{602}@*)
indicator_f = indicator_f > 0.5;  
\end{lstlisting}

The boundary-wave is projected to $\mathcal{T}^f$ and thresholded according to the localised maximal lamination layer thickness, scaled according to the periodicity ratio between the structural boundary and the structural infill (lines 604-608). 
\begin{lstlisting}[stepnumber=1,firstnumber=604,columns=fullflexible]
Inter_i.Values(:) = bphasor(:);
bphasor_f = Inter_i({grid.y_f,grid.x_f});
shellwave = 2/pi*asin(sin(atan2(imag(bphasor_f),real(bphasor_f))));
wmaxs = max(w,[],2); shellth = 2*omg_rat*min(max(wmaxs(:),wmin),0.99);
shellwave(:) = indicator_f(:).*max(shellwave(:),outline(:)) >= 1-shellth;
\end{lstlisting}
The structural domain indicator field is subjected to a final clean-up removing potential isolated artefacts utilising \texttt{removeIslands2D()} (line 609). 




\section{Examples}\label{sec:ex}




The following section presents a collection of tests and application examples for the \deHomTop code, facilitating discussions of its performance and limitations.

\subsection{Length-scale convergence}
The \deHomTop base-case results in a single-scale structure with volume-weighted compliance narrowly within 10\% of the multi-scale solution. This design is evaluated on a relatively large minimal length-scale, meaning there is expected solution quality loss due to the level of abstraction from the infinitely periodic homogenisation assumptions. \autoref{tab:firstRun} reveals the improvement in solution quality obtained when decreasing the minimum feature size. This is related to the crucial property of a well-defined dehomogenisation procedure, in that the performance convergence towards the homogenised solution as the minimal length scale approaches zero. The resolution of the mesh utilised for optimisation also has a significant effect on the solution quality, affecting the level of detail captured in the multi-scale solution.

Therefore, this first set of numerical experiments is focused on testing the length-scale convergence behaviour of dehomogenised designs for the baseline bridge example, realised with decreasing \verb|dmin|, for a select set of different optimisation grid resolutions.

\autoref{fig:conv_vf} and \autoref{fig:conv_s} illustrate how the relative error of the obtained single-scale designs, compared to the multi-scale optimised solution, depends on the minimum length-scale \verb|dmin|, for the volume fraction and volume-weighted compliance, respectively. The grid resolution utilised for optimisation is indicated by \verb|nelX|, and \verb|nelY=0.5*nelX| for each case; hence results are effected by \textit{h}-convergence. \verb|dmin| is relative to the grid resolution, such that for \verb|nelX=180| the dehomogenised result obtained for the length-scale given \verb|dmin=0.2| corresponds to \autoref{fig:basic_180_02}, which has a significantly higher level of detail than the base case run in \autoref{fig:firstRun01}.

\begin{Figure}
    \centering
%
%

%
\begin{tikzpicture}

    \begin{axis}[%
            width=1.0\linewidth,
            height=0.8\linewidth,
            xmin=0.0,
            xmax=0.4,
            ymin=-0.002,
            ymax=0.35,
            xlabel={$d_{\text{min}}$},
            ylabel={$\epsilon_f$},
            xtick={0,0.05,0.1,0.15,0.2,0.25,0.3,0.35,0.4},
            xticklabels={$0$,$0.05$,$0.1$,$0.15$,$0.2$,$0.25$,$0.3$,$0.35$,$0.4$},
            axis background/.style={fill=white},
            legend style={at={(0.03,0.97)}, anchor=north west, legend cell align=left, align=left, draw=white!15!black},
            legend columns=2,
            xmajorgrids,
            ymajorgrids,
        ]

        \addlegendimage{empty legend}
        \addlegendentry{}
        \addlegendimage{empty legend}
        \addlegendentry{\makebox[0pt][l]{\hspace{-11mm}nelX}}

        \addplot[color=mycolor1,mark=*,line width=1pt] table[row sep=crcr] {
                x	y\\
                0.400000 0.327182 \\
                0.300000 0.277796 \\
                0.200000 0.132385 \\
                0.100000 0.064802 \\
                0.050000 0.019415 \\
            };
        \addlegendentry{30}

        \addplot[color=mycolor2,mark=*,line width=1pt] table[row sep=crcr] {
                x	y\\
                0.400000 0.141144 \\
                0.300000 0.110254 \\
                0.200000 0.067669 \\
                0.100000 0.025103 \\
                0.050000 -0.001336 \\
            };
        \addlegendentry{60}

        \addplot[color=mycolor3,mark=*,line width=1pt] table[row sep=crcr] {
                x	y\\
                0.400000 0.080923 \\
                0.300000 0.070136 \\
                0.200000 0.035844 \\
                0.100000 0.008933 \\
                0.050000 -0.001087 \\
            };
        \addlegendentry{120}

        \addplot[color=mycolor4,mark=*,line width=1pt] table[row sep=crcr] {
                x	y\\
                0.400000 0.063437 \\
                0.300000 0.053464 \\
                0.200000 0.029101 \\
                0.100000 0.007152 \\
                0.050000 0.000125 \\
            };
        \addlegendentry{180}

        \addplot[color=mycolor5,mark=*,line width=1pt] table[row sep=crcr] {
                x	y\\
                0.400000 0.052730 \\
                0.300000 0.045873 \\
                0.200000 0.022268 \\
                0.100000 0.006323 \\
                0.050000 NaN \\
            };
        \addlegendentry{240}

        \addplot[color=mycolor6,mark=*,line width=1pt] table[row sep=crcr] {
                x	y\\
                0.400000 0.050323 \\
                0.300000 0.035546 \\
                0.200000 0.019358 \\
                0.100000 0.005227 \\
                0.050000 NaN \\
            };
        \addlegendentry{300}


    \end{axis}

\end{tikzpicture}%
    
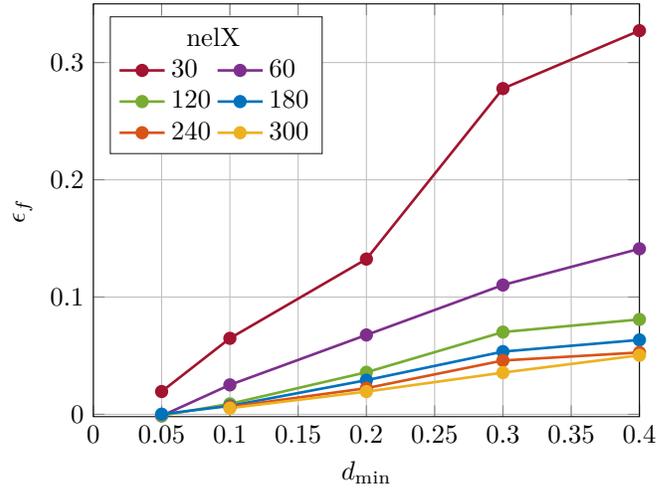
\captionof{figure}{Convergence of volume fraction error ($\epsilon_f$) for different grid discretisations wrt. length-scale.}
    \label{fig:conv_vf}
\end{Figure}

\begin{Figure}
    \centering
%
%

%
\begin{tikzpicture}

    \begin{axis}[%
            width=0.95\linewidth,
            height=0.8\linewidth,
            xmin=0.0,
            xmax=0.4,
            ymin=0.05,
            ymax=0.35,
            ytick={0.0631,0.1,0.1585,0.2512},
            yticklabels={-$1.2$,$-1.0$,$-0.8$,   $-0.6$},
            xtick={0,0.05,0.1,0.15,0.2,0.25,0.3,0.35,0.4},
            xticklabels={$0$,$0.05$,$0.1$,$0.15$,$0.2$,$0.25$,$0.3$,$0.35$,$0.4$},
            xlabel={$d_{\text{min}}$},
            ylabel={$\log_{10}(\epsilon_\mathcal{S})$},
            ymode=log,
            log basis y={10},
            axis background/.style={fill=white},
            legend columns=2,
            legend style={at={(0.03,0.97)}, anchor=north west, legend cell align=left, align=left, draw=white!15!black},
            xmajorgrids,
            ymajorgrids,
        ]


        \addlegendimage{empty legend}
        \addlegendentry{}
        \addlegendimage{empty legend}
        \addlegendentry{\makebox[0pt][l]{\hspace{-11mm}nelX}}

        \addplot[color=mycolor1,mark=*,line width=1pt] table[row sep=crcr] {
                x	y\\
                0.400000 0.303828 \\
                0.300000 0.240320 \\
                0.200000 0.159324 \\
                0.100000 0.116634 \\
                0.050000 0.106477 \\
            };
        \addlegendentry{30}

        \addplot[color=mycolor2,mark=*,line width=1pt] table[row sep=crcr] {
                x	y\\
                0.400000 0.206395 \\
                0.300000 0.108788 \\
                0.200000 0.093521 \\
                0.100000 0.085531 \\
                0.050000 0.069209 \\
            };
        \addlegendentry{60}

        \addplot[color=mycolor3,mark=*,line width=1pt] table[row sep=crcr] {
                x	y\\
                0.400000 0.100229 \\
                0.300000 0.100347 \\
                0.200000 0.083332 \\
                0.100000 0.074261 \\
                0.050000 0.058267 \\
            };
        \addlegendentry{120}

        \addplot[color=mycolor4,mark=*,line width=1pt] table[row sep=crcr] {
                x	y\\
                0.400000 0.148810 \\
                0.300000 0.109999 \\
                0.200000 0.084909 \\
                0.100000 0.075571 \\
                0.050000 0.058289 \\
            };
        \addlegendentry{180}

        \addplot[color=mycolor5,mark=*,line width=1pt] table[row sep=crcr] {
                x	y\\
                0.400000 0.158256 \\
                0.300000 0.119768 \\
                0.200000 0.090021 \\
                0.100000 0.075113 \\
            };
        \addlegendentry{240}

        \addplot[color=mycolor6,mark=*,line width=1pt] table[row sep=crcr] {
                x	y\\
                0.400000 0.186449 \\
                0.300000 0.146977 \\
                0.200000 0.099374 \\
                0.100000 0.078002 \\
            };
        \addlegendentry{300}


        \draw [name path=f,color=bycolor7,dashed,thin,opacity=0.05] (0.0,0.1) -- (0.4,0.1);
        \path[name path=axis0] (axis cs:0,0.0001) -- (axis cs:1,0.0001);

        \addplot [
            thick,
            color=bycolor7,
            fill=bycolor7,
            fill opacity=0.2
        ]
        fill between[
                of=f and axis0,
                soft clip={domain=0:1},
            ];
    \end{axis}

\end{tikzpicture}%
    \captionof{figure}{Convergence of log-scaled volume-weighted compliance error ($\log_{10}(\epsilon_\mathcal{S})$) for different grid discretisations wrt. length-scale.}
    \label{fig:conv_s}
\end{Figure}

Both the volume fraction ($\epsilon_f$) and volume-weighted compliance ($\epsilon_\mathcal{S}$) errors exhibit the desired convergence behaviour as \verb|dmin| is reduced, across all grid resolutions. It is, however, evident that the coarsest grid-resolution suffers from loss of detail, causing the structural performance to be diminished. For the remaining grid resolutions the volume-weighted compliance error converges well within a 10\% deviance from the multi-scale solution. 

The convergence behaviours of the different grid-resolutions additionally reveal a limitation of dehomogenisation to a relatively large minimum length scale. This leads to an excessive downsampling of the multi-scale details to obtain the single-scale projection, meaning that the level of detail in the optimised solution cannot be sufficiently captured. 

\begin{Figure}
    \centering
    \includegraphics[width=\linewidth]{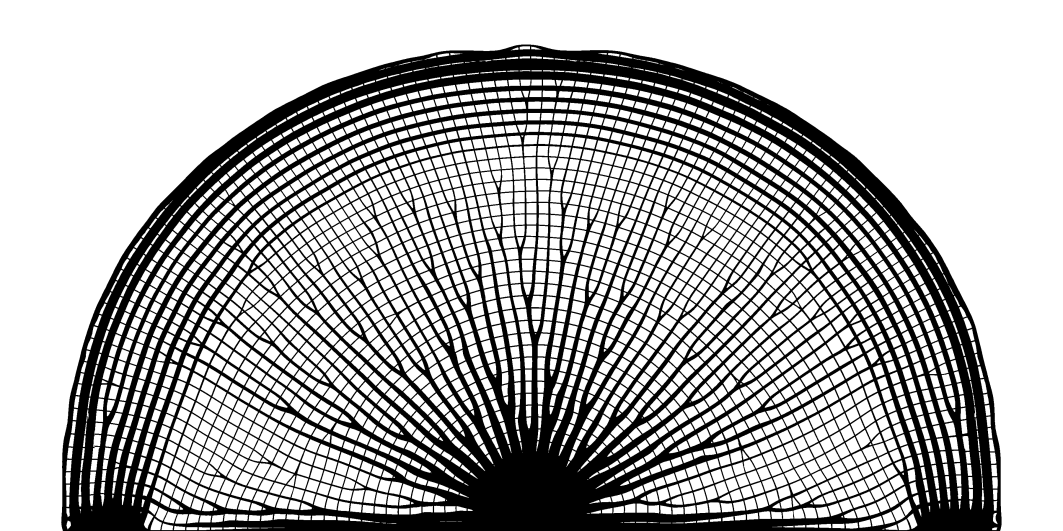}
    \captionof{figure}{Bridge model with \texttt{nelX} = 180 and \texttt{dmin} = 0.2}
    \label{fig:basic_180_02}
\end{Figure}

\subsection{Benchmark: Michell cantilever}
The performance of the \deHomTop code is benchmarked to the first proposed dehomogenisation for manufacturable designs \citep{Groen2018} by considering the Michell cantilever. The FE-model of the Michell cantilever can be included in the code by replacing \texttt{prepFEA()} with \texttt{prepFEA\_cant()}. 
The Poisson's ratio is set to $\nu = 0.3$ (line 6) in accordance with ~\cite{Groen2018}.
\begin{lstlisting}[firstnumber=6,columns=fullflexible]
[E,EMin,nu] = deal(1,1e-9,0.3); 
\end{lstlisting}
The \deHomTop code is executed with,
\begin{lstlisting}[numbers=none,columns=fullflexible,xleftmargin=1em,framexleftmargin=0em]
deHomTop808(80,40,0.5,1.5,0.1,1,0.2,0,true);
\end{lstlisting}
where $d_{\min} = 0.2$ and $w_{\min} = 0.1$ corresponds to a mapped cell size and feature size of $40 h_f$ and $2 h_f$ \citep{Groen2018}, respectively. To best replicate the numerical setup in \cite{Groen2018}, by ensuring consistent mapping grid resolutions, the minimum thickness for \deHomTop is increased to $w_{\min} = 0.1$ compared to $w_{\min} = 0.05$ in \cite{Groen2018}.

%
\begin{Figure}
    \centering
    \includegraphics[width=\linewidth]{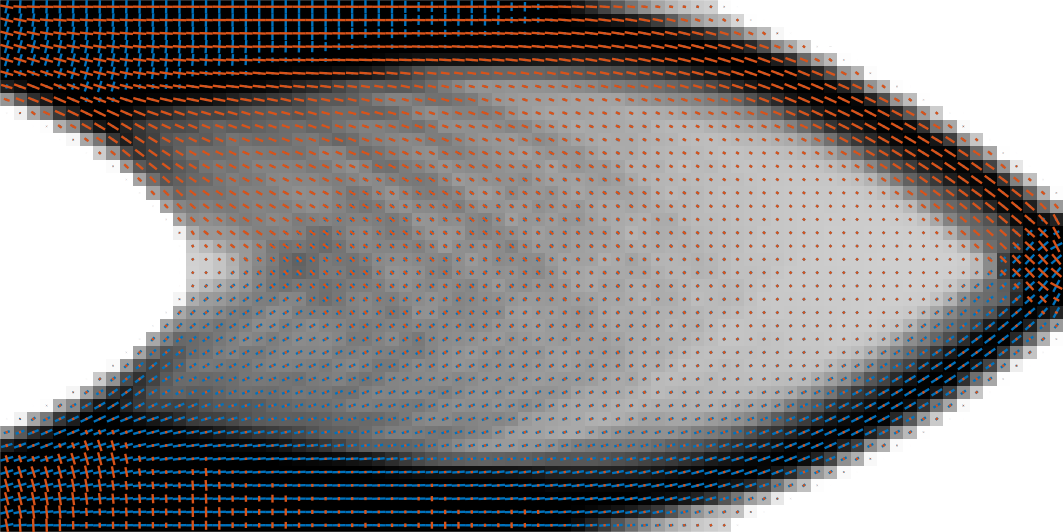}
    \captionof{figure}{Multi-scale structure of the Michell cantilever obtained with \deHomTopns.}
    \label{fig:cant_jeroen_hom}
\end{Figure}

\begin{Table}
\centering
    \captionof{table}{Optimisation benchmark results of the Michell cantilever. For \deHomTop the time includes plotting.}

    \begin{tabular}{@{}lcc@{}}
        \toprule
                                            & \deHomTopns    & \citeauthor{Groen2018} \\ \midrule
        $J$                                 & 58.10 & 58.31   \\
        $f$                                 & 0.500 & 0.500   \\ 
        Time [s]                            & 15.32 & 199.50 \\ \bottomrule
    \end{tabular}

    \label{tab:BenchmarkTO}
\end{Table}

The comparative results of the optimisation are seen in \autoref{tab:BenchmarkTO}, and for the multi-scale structure in \autoref{fig:cant_jeroen_hom}. A similar compliance is achieved by the two frameworks, closely corresponding to the result provided in \cite{Sigmund2016} of $J=56.73$, even with a different design domain regularisation for \deHomTop compared to \cite{Groen2018,Sigmund2016}. The key observation is the computational time comparison, where \deHomTop is an order of magnitude faster than \cite{Groen2018}. Due to the timings being recorded on different hardware, an exact comparison is discouraged, but the order of magnitude is expected to be representative as the linear systems are of the same size, and both codes are executed in Matlab.


The dehomogenisation result is seen in \autoref{tab:BenchmarkDEHOM}, and the single-scale structure is seen in \autoref{fig:cant_jeroen_dehom}. The \deHomTop result is very close to the optimised solution. However, the dehomogenisation result of \cite{Groen2018} is performing slightly better, which is to be expected for this test case, as discussed in \cite{Woldseth2023}. Nevertheless, the computational time is almost two orders of magnitude faster than \cite{Groen2018}, which underlines both the efficiency of phasor-based dehomogenisation and the \deHomTop implementation in general.
\begin{Table}
\centering
    \captionof{table}{Dehomogenisation benchmark results of the Michell cantilever. For \deHomTop the time includes plotting.}
    \begin{tabular}{@{}lcc@{}}
        \toprule
                            & \deHomTopns     & \citeauthor{Groen2018} \\ \midrule
        $J$                   & 59.52        & 59.55 \\
        $f$                   & 0.513         & 0.500 \\
        $\epsilon_\mathcal{S}$         & 5.13\%        & 2.13\% \\
        Time [s]            & 1.26         & 126.20 \\
        Time (analysis) [s] & 17.72        & - \\ \bottomrule
    \end{tabular}
    \label{tab:BenchmarkDEHOM}
\end{Table}
\begin{Figure}
    \centering
    \includegraphics[width=\linewidth]{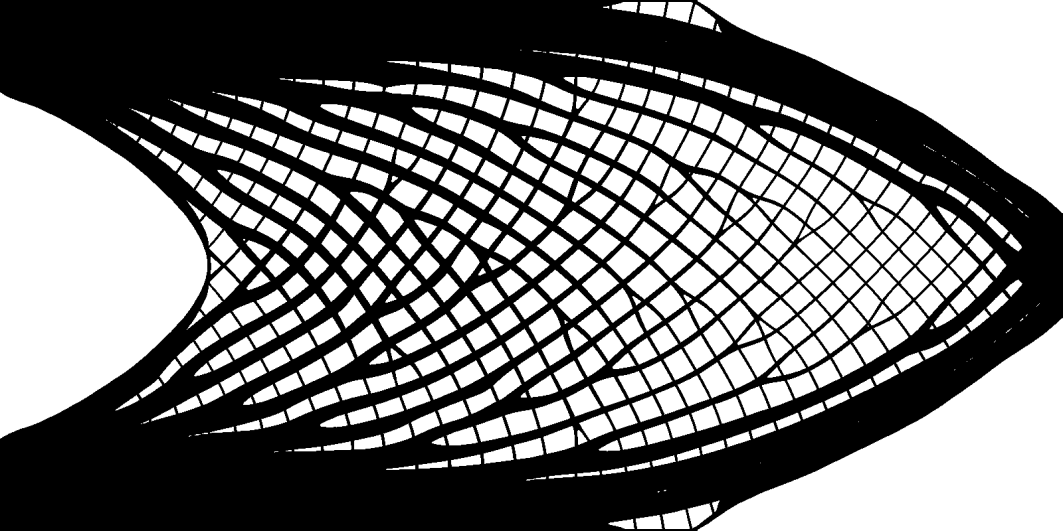}
    \captionof{figure}{Dehomogenisation structure of the Michell cantilever obtained with \deHomTopns.}
    \label{fig:cant_jeroen_dehom}
\end{Figure}

\subsection{Optimisation convergence}
A crucial part in establishing the value of the on-the-fly dehomogenisation option, is to consider how well the intermediate multi-scale designs are captured by the generated single-scale interpretations during the optimisation iterations. 

To this end, the baseline \deHomTop run with \verb|deHomFrq=1| is considered, including FE-analysis of the dehomogenised design in each iteration. The obtained iteration history is compared to the corresponding intermediate multi-scale iterate in \autoref{fig:conv_itr}. It is evident that the performance of the single-scale design deviates significantly from the multi-scale iterates, with higher compliance and lower volume fractions, until $\beta=4$ is reached, and the measure of non-discreteness (mnd) \citep{Sigmund2007} stabilises below 10\%. For the remainder of the iteration process, the single-scale projection captures the multi-scale iterate within a 10\% error.

\begin{Figure}
    \centering
%
%

%
\begin{tikzpicture}[declare function={
                betacont(\x)= (\x<=49) * (0.1)   +
                and(\x>49, \x<=97) * (1)     +
                and(\x>97,  \x<=143) * (2) +
                and(\x>143,  \x<=187) * (4) +
                and(\x>187,  \x<=229) * (8) +
                and(\x>229,  \x<=269) * (16) +
                (\x>269) * (32);
            }]

    \begin{axis}[%
            width=1.0\linewidth,
            height=0.8\linewidth,
            xmin=1,
            xmax=300,
            ymin=-0.05,
            ymax=2,
            xlabel={Itr},
            ylabel={$\epsilon$},
            axis background/.style={fill=white},
            legend style={at={(0.5,0.97)}, anchor=north,
                    nodes = {anchor=base, yshift=-0.5ex},
                    draw=white!15!black},
            legend columns=-1,
            xmajorgrids,
            ymajorgrids,
        ]


        \addplot[color=mycolor4,line width=1pt] table[row sep=crcr] {
                x	y\\
                1.000000 0.109952 \\
                2.000000 0.092895 \\
                3.000000 0.051833 \\
                4.000000 0.045344 \\
                5.000000 0.033803 \\
                6.000000 0.042408 \\
                7.000000 0.038263 \\
                8.000000 0.042192 \\
                9.000000 0.032772 \\
                10.000000 0.031511 \\
                11.000000 0.029720 \\
                12.000000 0.022866 \\
                13.000000 0.013836 \\
                14.000000 -0.002866 \\
                15.000000 -0.007993 \\
                16.000000 -0.013759 \\
                17.000000 -0.024828 \\
                18.000000 -0.034093 \\
                19.000000 -0.026799 \\
                20.000000 -0.049407 \\
                21.000000 -0.024079 \\
                22.000000 -0.040952 \\
                23.000000 -0.027593 \\
                24.000000 -0.036876 \\
                25.000000 -0.042422 \\
                26.000000 -0.041974 \\
                27.000000 -0.041616 \\
                28.000000 -0.037138 \\
                29.000000 -0.041441 \\
                30.000000 -0.035259 \\
                31.000000 -0.036995 \\
                32.000000 -0.029773 \\
                33.000000 -0.033731 \\
                34.000000 -0.029202 \\
                35.000000 -0.030350 \\
                36.000000 -0.015063 \\
                37.000000 -0.013927 \\
                38.000000 -0.008043 \\
                39.000000 -0.016628 \\
                40.000000 -0.010076 \\
                41.000000 -0.010666 \\
                42.000000 -0.013214 \\
                43.000000 -0.006857 \\
                44.000000 -0.007615 \\
                45.000000 -0.005101 \\
                46.000000 -0.004620 \\
                47.000000 -0.004555 \\
                48.000000 -0.002234 \\
                49.000000 -0.003059 \\
                50.000000 0.007313 \\
                51.000000 0.009943 \\
                52.000000 0.001615 \\
                53.000000 0.005937 \\
                54.000000 0.008926 \\
                55.000000 0.009311 \\
                56.000000 0.013575 \\
                57.000000 0.014899 \\
                58.000000 0.017283 \\
                59.000000 0.012545 \\
                60.000000 0.012659 \\
                61.000000 0.012626 \\
                62.000000 0.010672 \\
                63.000000 0.010080 \\
                64.000000 0.010372 \\
                65.000000 0.012264 \\
                66.000000 0.010661 \\
                67.000000 0.017350 \\
                68.000000 0.013636 \\
                69.000000 0.013347 \\
                70.000000 0.009780 \\
                71.000000 0.013281 \\
                72.000000 0.010022 \\
                73.000000 0.017607 \\
                74.000000 0.017725 \\
                75.000000 0.012928 \\
                76.000000 0.013675 \\
                77.000000 0.011600 \\
                78.000000 0.013375 \\
                79.000000 0.013525 \\
                80.000000 0.016916 \\
                81.000000 0.016031 \\
                82.000000 0.015561 \\
                83.000000 0.019050 \\
                84.000000 0.017904 \\
                85.000000 0.017959 \\
                86.000000 0.021098 \\
                87.000000 0.006952 \\
                88.000000 0.012012 \\
                89.000000 0.003176 \\
                90.000000 -0.002351 \\
                91.000000 0.003352 \\
                92.000000 0.001314 \\
                93.000000 0.003651 \\
                94.000000 0.008275 \\
                95.000000 0.005186 \\
                96.000000 -0.001070 \\
                97.000000 0.002212 \\
                98.000000 0.006754 \\
                99.000000 0.003357 \\
                100.000000 0.002768 \\
                101.000000 0.014575 \\
                102.000000 0.011934 \\
                103.000000 0.010593 \\
                104.000000 0.004442 \\
                105.000000 0.003825 \\
                106.000000 0.003848 \\
                107.000000 0.003819 \\
                108.000000 0.006311 \\
                109.000000 0.004642 \\
                110.000000 0.003695 \\
                111.000000 0.001908 \\
                112.000000 0.001394 \\
                113.000000 0.001025 \\
                114.000000 0.004433 \\
                115.000000 0.003189 \\
                116.000000 -0.000894 \\
                117.000000 -0.001789 \\
                118.000000 -0.000628 \\
                119.000000 0.000042 \\
                120.000000 -0.001002 \\
                121.000000 -0.001844 \\
                122.000000 -0.001919 \\
                123.000000 -0.000699 \\
                124.000000 0.000640 \\
                125.000000 0.004235 \\
                126.000000 0.006544 \\
                127.000000 0.009157 \\
                128.000000 0.018046 \\
                129.000000 0.012346 \\
                130.000000 0.018170 \\
                131.000000 0.012813 \\
                132.000000 0.010392 \\
                133.000000 0.019988 \\
                134.000000 0.018012 \\
                135.000000 0.018610 \\
                136.000000 0.016804 \\
                137.000000 0.013067 \\
                138.000000 0.013404 \\
                139.000000 0.014143 \\
                140.000000 0.015776 \\
                141.000000 0.011831 \\
                142.000000 0.012236 \\
                143.000000 0.010333 \\
                144.000000 0.038640 \\
                145.000000 0.037238 \\
                146.000000 0.039642 \\
                147.000000 0.038300 \\
                148.000000 0.040636 \\
                149.000000 0.034981 \\
                150.000000 0.035106 \\
                151.000000 0.036396 \\
                152.000000 0.038433 \\
                153.000000 0.034822 \\
                154.000000 0.037318 \\
                155.000000 0.039078 \\
                156.000000 0.040117 \\
                157.000000 0.036301 \\
                158.000000 0.036362 \\
                159.000000 0.036255 \\
                160.000000 0.036512 \\
                161.000000 0.038918 \\
                162.000000 0.040532 \\
                163.000000 0.038171 \\
                164.000000 0.033243 \\
                165.000000 0.032932 \\
                166.000000 0.036177 \\
                167.000000 0.038484 \\
                168.000000 0.037358 \\
                169.000000 0.030875 \\
                170.000000 0.032109 \\
                171.000000 0.030947 \\
                172.000000 0.036604 \\
                173.000000 0.037440 \\
                174.000000 0.036109 \\
                175.000000 0.040261 \\
                176.000000 0.042775 \\
                177.000000 0.039173 \\
                178.000000 0.047651 \\
                179.000000 0.045708 \\
                180.000000 0.047991 \\
                181.000000 0.044321 \\
                182.000000 0.047665 \\
                183.000000 0.045263 \\
                184.000000 0.046818 \\
                185.000000 0.047403 \\
                186.000000 0.040892 \\
                187.000000 0.045169 \\
                188.000000 0.057756 \\
                189.000000 0.062095 \\
                190.000000 0.060033 \\
                191.000000 0.057183 \\
                192.000000 0.056667 \\
                193.000000 0.059482 \\
                194.000000 0.061711 \\
                195.000000 0.061988 \\
                196.000000 0.062321 \\
                197.000000 0.058117 \\
                198.000000 0.061318 \\
                199.000000 0.060225 \\
                200.000000 0.061977 \\
                201.000000 0.060381 \\
                202.000000 0.061555 \\
                203.000000 0.060518 \\
                204.000000 0.061001 \\
                205.000000 0.060362 \\
                206.000000 0.061390 \\
                207.000000 0.060840 \\
                208.000000 0.061039 \\
                209.000000 0.061243 \\
                210.000000 0.061013 \\
                211.000000 0.060851 \\
                212.000000 0.058205 \\
                213.000000 0.058854 \\
                214.000000 0.059715 \\
                215.000000 0.060111 \\
                216.000000 0.060249 \\
                217.000000 0.060345 \\
                218.000000 0.061763 \\
                219.000000 0.061379 \\
                220.000000 0.061675 \\
                221.000000 0.064189 \\
                222.000000 0.063329 \\
                223.000000 0.064663 \\
                224.000000 0.065437 \\
                225.000000 0.065595 \\
                226.000000 0.066789 \\
                227.000000 0.066877 \\
                228.000000 0.068246 \\
                229.000000 0.067426 \\
                230.000000 0.071960 \\
                231.000000 0.071224 \\
                232.000000 0.070387 \\
                233.000000 0.070553 \\
                234.000000 0.071634 \\
                235.000000 0.070615 \\
                236.000000 0.071885 \\
                237.000000 0.069602 \\
                238.000000 0.071106 \\
                239.000000 0.070370 \\
                240.000000 0.069495 \\
                241.000000 0.068529 \\
                242.000000 0.071101 \\
                243.000000 0.067981 \\
                244.000000 0.068386 \\
                245.000000 0.067551 \\
                246.000000 0.067784 \\
                247.000000 0.068193 \\
                248.000000 0.067953 \\
                249.000000 0.067310 \\
                250.000000 0.067727 \\
                251.000000 0.067680 \\
                252.000000 0.067360 \\
                253.000000 0.067745 \\
                254.000000 0.068249 \\
                255.000000 0.068609 \\
                256.000000 0.068682 \\
                257.000000 0.068894 \\
                258.000000 0.069055 \\
                259.000000 0.068910 \\
                260.000000 0.069560 \\
                261.000000 0.069395 \\
                262.000000 0.068320 \\
                263.000000 0.068709 \\
                264.000000 0.067678 \\
                265.000000 0.069517 \\
                266.000000 0.071760 \\
                267.000000 0.073513 \\
                268.000000 0.074069 \\
                269.000000 0.073635 \\
                270.000000 0.067810 \\
                271.000000 0.068256 \\
                272.000000 0.067035 \\
                273.000000 0.065719 \\
                274.000000 0.066265 \\
                275.000000 0.066750 \\
                276.000000 0.066175 \\
                277.000000 0.068443 \\
                278.000000 0.067850 \\
                279.000000 0.067677 \\
                280.000000 0.067880 \\
                281.000000 0.068940 \\
                282.000000 0.068887 \\
                283.000000 0.069006 \\
                284.000000 0.068870 \\
                285.000000 0.068916 \\
                286.000000 0.071768 \\
                287.000000 0.071466 \\
                288.000000 0.071420 \\
                289.000000 0.071523 \\
                290.000000 0.071693 \\
                291.000000 0.071723 \\
                292.000000 0.072145 \\
                293.000000 0.071760 \\
                294.000000 0.073012 \\
                295.000000 0.072783 \\
                296.000000 0.072738 \\
                297.000000 0.072158 \\
                298.000000 0.071725 \\
                299.000000 0.074161 \\
                300.000000 0.073385 \\
            };
        \addlegendentry{f}

        \addplot[color=mycolor5,line width=1pt] table[row sep=crcr] {
                x	y\\
                1.000000 -0.532156 \\
                2.000000 0.072939 \\
                3.000000 0.064901 \\
                4.000000 0.168097 \\
                5.000000 0.395632 \\
                6.000000 0.450412 \\
                7.000000 0.549891 \\
                8.000000 0.579999 \\
                9.000000 0.642076 \\
                10.000000 0.934848 \\
                11.000000 0.892221 \\
                12.000000 1.126527 \\
                13.000000 1.203022 \\
                14.000000 1.223962 \\
                15.000000 1.452992 \\
                16.000000 1.460159 \\
                17.000000 1.560754 \\
                18.000000 1.342693 \\
                19.000000 1.407849 \\
                20.000000 1.607455 \\
                21.000000 1.499835 \\
                22.000000 1.514046 \\
                23.000000 1.198557 \\
                24.000000 1.616617 \\
                25.000000 1.499781 \\
                26.000000 1.558339 \\
                27.000000 1.525571 \\
                28.000000 1.521334 \\
                29.000000 1.652951 \\
                30.000000 1.645797 \\
                31.000000 1.609853 \\
                32.000000 1.526035 \\
                33.000000 1.606926 \\
                34.000000 1.711418 \\
                35.000000 1.640898 \\
                36.000000 1.587094 \\
                37.000000 1.644662 \\
                38.000000 1.499938 \\
                39.000000 1.561276 \\
                40.000000 1.402356 \\
                41.000000 1.477240 \\
                42.000000 1.385662 \\
                43.000000 1.452047 \\
                44.000000 1.443655 \\
                45.000000 1.439579 \\
                46.000000 1.441687 \\
                47.000000 1.428730 \\
                48.000000 1.429643 \\
                49.000000 1.483762 \\
                50.000000 1.133859 \\
                51.000000 0.582856 \\
                52.000000 1.170431 \\
                53.000000 0.991968 \\
                54.000000 1.252639 \\
                55.000000 1.233094 \\
                56.000000 1.187997 \\
                57.000000 1.267891 \\
                58.000000 1.203985 \\
                59.000000 1.340006 \\
                60.000000 1.282290 \\
                61.000000 1.150345 \\
                62.000000 1.203046 \\
                63.000000 1.232398 \\
                64.000000 1.233469 \\
                65.000000 1.228721 \\
                66.000000 1.297320 \\
                67.000000 1.243635 \\
                68.000000 1.229909 \\
                69.000000 1.254071 \\
                70.000000 1.216758 \\
                71.000000 1.248302 \\
                72.000000 1.247419 \\
                73.000000 1.202712 \\
                74.000000 1.227626 \\
                75.000000 1.201472 \\
                76.000000 1.196835 \\
                77.000000 1.202789 \\
                78.000000 1.218405 \\
                79.000000 1.200289 \\
                80.000000 1.181032 \\
                81.000000 1.203467 \\
                82.000000 1.211819 \\
                83.000000 1.186434 \\
                84.000000 1.172141 \\
                85.000000 1.212227 \\
                86.000000 1.157368 \\
                87.000000 1.172057 \\
                88.000000 1.079396 \\
                89.000000 0.996016 \\
                90.000000 1.003705 \\
                91.000000 0.980245 \\
                92.000000 0.976279 \\
                93.000000 0.970293 \\
                94.000000 0.934371 \\
                95.000000 0.962419 \\
                96.000000 0.987017 \\
                97.000000 0.936993 \\
                98.000000 0.475520 \\
                99.000000 0.827185 \\
                100.000000 0.696108 \\
                101.000000 0.810840 \\
                102.000000 0.834437 \\
                103.000000 0.787418 \\
                104.000000 0.862821 \\
                105.000000 0.826674 \\
                106.000000 0.844206 \\
                107.000000 0.829247 \\
                108.000000 0.828112 \\
                109.000000 0.821612 \\
                110.000000 0.808575 \\
                111.000000 0.838784 \\
                112.000000 0.864155 \\
                113.000000 0.866479 \\
                114.000000 0.826885 \\
                115.000000 0.812124 \\
                116.000000 0.841591 \\
                117.000000 0.841942 \\
                118.000000 0.840654 \\
                119.000000 0.850990 \\
                120.000000 0.837093 \\
                121.000000 0.833111 \\
                122.000000 0.839041 \\
                123.000000 0.844585 \\
                124.000000 0.840095 \\
                125.000000 0.827917 \\
                126.000000 0.821332 \\
                127.000000 0.807881 \\
                128.000000 0.794542 \\
                129.000000 0.811630 \\
                130.000000 0.811254 \\
                131.000000 0.837126 \\
                132.000000 0.829512 \\
                133.000000 0.837698 \\
                134.000000 0.819684 \\
                135.000000 0.836846 \\
                136.000000 0.794532 \\
                137.000000 0.806630 \\
                138.000000 0.828559 \\
                139.000000 0.831410 \\
                140.000000 0.834629 \\
                141.000000 0.862857 \\
                142.000000 0.856628 \\
                143.000000 0.845523 \\
                144.000000 0.269193 \\
                145.000000 0.267773 \\
                146.000000 0.303189 \\
                147.000000 0.311016 \\
                148.000000 0.363258 \\
                149.000000 0.340552 \\
                150.000000 0.312635 \\
                151.000000 0.322963 \\
                152.000000 0.321322 \\
                153.000000 0.308432 \\
                154.000000 0.304179 \\
                155.000000 0.302645 \\
                156.000000 0.286614 \\
                157.000000 0.298016 \\
                158.000000 0.292193 \\
                159.000000 0.285097 \\
                160.000000 0.277628 \\
                161.000000 0.283095 \\
                162.000000 0.270755 \\
                163.000000 0.275067 \\
                164.000000 0.271703 \\
                165.000000 0.263507 \\
                166.000000 0.254743 \\
                167.000000 0.255754 \\
                168.000000 0.252887 \\
                169.000000 0.239203 \\
                170.000000 0.230427 \\
                171.000000 0.223013 \\
                172.000000 0.232288 \\
                173.000000 0.229012 \\
                174.000000 0.224113 \\
                175.000000 0.228970 \\
                176.000000 0.246729 \\
                177.000000 0.245030 \\
                178.000000 0.237227 \\
                179.000000 0.230390 \\
                180.000000 0.233879 \\
                181.000000 0.214693 \\
                182.000000 0.212846 \\
                183.000000 0.205577 \\
                184.000000 0.203485 \\
                185.000000 0.199992 \\
                186.000000 0.205475 \\
                187.000000 0.200508 \\
                188.000000 0.038806 \\
                189.000000 0.088837 \\
                190.000000 0.077630 \\
                191.000000 0.081194 \\
                192.000000 0.090612 \\
                193.000000 0.088122 \\
                194.000000 0.085250 \\
                195.000000 0.084964 \\
                196.000000 0.084990 \\
                197.000000 0.088665 \\
                198.000000 0.090341 \\
                199.000000 0.089772 \\
                200.000000 0.087886 \\
                201.000000 0.091583 \\
                202.000000 0.089223 \\
                203.000000 0.090482 \\
                204.000000 0.089843 \\
                205.000000 0.089714 \\
                206.000000 0.089787 \\
                207.000000 0.089454 \\
                208.000000 0.089896 \\
                209.000000 0.090271 \\
                210.000000 0.089873 \\
                211.000000 0.087681 \\
                212.000000 0.089628 \\
                213.000000 0.089237 \\
                214.000000 0.087282 \\
                215.000000 0.087420 \\
                216.000000 0.087754 \\
                217.000000 0.087342 \\
                218.000000 0.091560 \\
                219.000000 0.088752 \\
                220.000000 0.087124 \\
                221.000000 0.083643 \\
                222.000000 0.088823 \\
                223.000000 0.090942 \\
                224.000000 0.089603 \\
                225.000000 0.095225 \\
                226.000000 0.096204 \\
                227.000000 0.096966 \\
                228.000000 0.093318 \\
                229.000000 0.094540 \\
                230.000000 0.077625 \\
                231.000000 0.088323 \\
                232.000000 0.085224 \\
                233.000000 0.089142 \\
                234.000000 0.090279 \\
                235.000000 0.090326 \\
                236.000000 0.088211 \\
                237.000000 0.090480 \\
                238.000000 0.088632 \\
                239.000000 0.089016 \\
                240.000000 0.090143 \\
                241.000000 0.090979 \\
                242.000000 0.089317 \\
                243.000000 0.091440 \\
                244.000000 0.091532 \\
                245.000000 0.091815 \\
                246.000000 0.091355 \\
                247.000000 0.091388 \\
                248.000000 0.087327 \\
                249.000000 0.087283 \\
                250.000000 0.087454 \\
                251.000000 0.086189 \\
                252.000000 0.086516 \\
                253.000000 0.086092 \\
                254.000000 0.085849 \\
                255.000000 0.085643 \\
                256.000000 0.085683 \\
                257.000000 0.085760 \\
                258.000000 0.087298 \\
                259.000000 0.089490 \\
                260.000000 0.089981 \\
                261.000000 0.092273 \\
                262.000000 0.090180 \\
                263.000000 0.089980 \\
                264.000000 0.089916 \\
                265.000000 0.088958 \\
                266.000000 0.084664 \\
                267.000000 0.093097 \\
                268.000000 0.095489 \\
                269.000000 0.094528 \\
                270.000000 0.087990 \\
                271.000000 0.082686 \\
                272.000000 0.090737 \\
                273.000000 0.085521 \\
                274.000000 0.087310 \\
                275.000000 0.085116 \\
                276.000000 0.085924 \\
                277.000000 0.088928 \\
                278.000000 0.089349 \\
                279.000000 0.089628 \\
                280.000000 0.089766 \\
                281.000000 0.088749 \\
                282.000000 0.088217 \\
                283.000000 0.087719 \\
                284.000000 0.087731 \\
                285.000000 0.087530 \\
                286.000000 0.094069 \\
                287.000000 0.093909 \\
                288.000000 0.094227 \\
                289.000000 0.094013 \\
                290.000000 0.093976 \\
                291.000000 0.093811 \\
                292.000000 0.093834 \\
                293.000000 0.093917 \\
                294.000000 0.093786 \\
                295.000000 0.094118 \\
                296.000000 0.093973 \\
                297.000000 0.094018 \\
                298.000000 0.093723 \\
                299.000000 0.094544 \\
                300.000000 0.094868 \\
            };
        \addlegendentry{s}

        \addplot[color=mycolor3,line width=1pt] table[row sep=crcr] {
                x	y\\
                1.000000 0.000000 \\
                2.000000 0.040082 \\
                3.000000 0.074735 \\
                4.000000 0.096151 \\
                5.000000 0.122772 \\
                6.000000 0.145173 \\
                7.000000 0.167452 \\
                8.000000 0.182543 \\
                9.000000 0.199283 \\
                10.000000 0.216581 \\
                11.000000 0.236631 \\
                12.000000 0.256104 \\
                13.000000 0.277517 \\
                14.000000 0.298304 \\
                15.000000 0.319519 \\
                16.000000 0.341946 \\
                17.000000 0.362599 \\
                18.000000 0.412623 \\
                19.000000 0.397913 \\
                20.000000 0.489352 \\
                21.000000 0.435963 \\
                22.000000 0.540620 \\
                23.000000 0.471853 \\
                24.000000 0.523292 \\
                25.000000 0.498496 \\
                26.000000 0.505769 \\
                27.000000 0.505231 \\
                28.000000 0.507706 \\
                29.000000 0.510469 \\
                30.000000 0.513190 \\
                31.000000 0.518469 \\
                32.000000 0.521019 \\
                33.000000 0.525701 \\
                34.000000 0.530651 \\
                35.000000 0.535332 \\
                36.000000 0.549453 \\
                37.000000 0.574690 \\
                38.000000 0.574843 \\
                39.000000 0.586045 \\
                40.000000 0.587078 \\
                41.000000 0.583527 \\
                42.000000 0.583904 \\
                43.000000 0.588501 \\
                44.000000 0.590922 \\
                45.000000 0.596019 \\
                46.000000 0.600147 \\
                47.000000 0.605988 \\
                48.000000 0.610191 \\
                49.000000 0.618531 \\
                50.000000 0.597725 \\
                51.000000 0.601557 \\
                52.000000 0.598571 \\
                53.000000 0.601011 \\
                54.000000 0.595755 \\
                55.000000 0.595753 \\
                56.000000 0.593073 \\
                57.000000 0.591675 \\
                58.000000 0.590818 \\
                59.000000 0.591034 \\
                60.000000 0.589156 \\
                61.000000 0.586350 \\
                62.000000 0.585292 \\
                63.000000 0.584708 \\
                64.000000 0.584262 \\
                65.000000 0.583621 \\
                66.000000 0.583093 \\
                67.000000 0.582484 \\
                68.000000 0.581819 \\
                69.000000 0.581041 \\
                70.000000 0.580193 \\
                71.000000 0.579313 \\
                72.000000 0.578142 \\
                73.000000 0.577185 \\
                74.000000 0.575617 \\
                75.000000 0.574956 \\
                76.000000 0.573391 \\
                77.000000 0.572675 \\
                78.000000 0.571173 \\
                79.000000 0.570201 \\
                80.000000 0.569492 \\
                81.000000 0.568002 \\
                82.000000 0.569424 \\
                83.000000 0.565575 \\
                84.000000 0.568743 \\
                85.000000 0.562467 \\
                86.000000 0.564890 \\
                87.000000 0.563399 \\
                88.000000 0.562089 \\
                89.000000 0.562103 \\
                90.000000 0.563997 \\
                91.000000 0.561088 \\
                92.000000 0.560627 \\
                93.000000 0.559738 \\
                94.000000 0.559089 \\
                95.000000 0.558378 \\
                96.000000 0.557784 \\
                97.000000 0.557143 \\
                98.000000 0.507518 \\
                99.000000 0.514002 \\
                100.000000 0.512495 \\
                101.000000 0.510353 \\
                102.000000 0.509205 \\
                103.000000 0.507585 \\
                104.000000 0.506723 \\
                105.000000 0.506496 \\
                106.000000 0.505374 \\
                107.000000 0.504923 \\
                108.000000 0.503721 \\
                109.000000 0.502627 \\
                110.000000 0.501233 \\
                111.000000 0.499526 \\
                112.000000 0.497616 \\
                113.000000 0.495299 \\
                114.000000 0.492517 \\
                115.000000 0.489304 \\
                116.000000 0.485142 \\
                117.000000 0.480345 \\
                118.000000 0.474553 \\
                119.000000 0.467574 \\
                120.000000 0.459364 \\
                121.000000 0.451572 \\
                122.000000 0.439109 \\
                123.000000 0.426383 \\
                124.000000 0.410222 \\
                125.000000 0.394000 \\
                126.000000 0.375272 \\
                127.000000 0.360907 \\
                128.000000 0.344622 \\
                129.000000 0.344642 \\
                130.000000 0.338807 \\
                131.000000 0.345567 \\
                132.000000 0.345133 \\
                133.000000 0.332071 \\
                134.000000 0.332812 \\
                135.000000 0.327082 \\
                136.000000 0.326118 \\
                137.000000 0.324572 \\
                138.000000 0.323284 \\
                139.000000 0.321967 \\
                140.000000 0.320694 \\
                141.000000 0.320257 \\
                142.000000 0.318999 \\
                143.000000 0.317679 \\
                144.000000 0.227728 \\
                145.000000 0.233628 \\
                146.000000 0.231369 \\
                147.000000 0.230431 \\
                148.000000 0.229560 \\
                149.000000 0.229170 \\
                150.000000 0.228828 \\
                151.000000 0.228568 \\
                152.000000 0.228236 \\
                153.000000 0.227917 \\
                154.000000 0.227504 \\
                155.000000 0.227001 \\
                156.000000 0.226400 \\
                157.000000 0.225657 \\
                158.000000 0.224828 \\
                159.000000 0.223796 \\
                160.000000 0.222594 \\
                161.000000 0.219987 \\
                162.000000 0.218299 \\
                163.000000 0.216556 \\
                164.000000 0.214616 \\
                165.000000 0.212249 \\
                166.000000 0.209405 \\
                167.000000 0.205922 \\
                168.000000 0.201553 \\
                169.000000 0.196415 \\
                170.000000 0.190655 \\
                171.000000 0.184154 \\
                172.000000 0.177647 \\
                173.000000 0.171658 \\
                174.000000 0.166777 \\
                175.000000 0.163247 \\
                176.000000 0.160985 \\
                177.000000 0.160125 \\
                178.000000 0.160187 \\
                179.000000 0.160535 \\
                180.000000 0.160805 \\
                181.000000 0.160770 \\
                182.000000 0.160804 \\
                183.000000 0.161145 \\
                184.000000 0.161203 \\
                185.000000 0.161426 \\
                186.000000 0.161648 \\
                187.000000 0.161674 \\
                188.000000 0.085712 \\
                189.000000 0.085864 \\
                190.000000 0.085785 \\
                191.000000 0.085735 \\
                192.000000 0.085720 \\
                193.000000 0.085712 \\
                194.000000 0.085714 \\
                195.000000 0.085718 \\
                196.000000 0.085727 \\
                197.000000 0.085736 \\
                198.000000 0.085748 \\
                199.000000 0.085761 \\
                200.000000 0.085778 \\
                201.000000 0.085771 \\
                202.000000 0.085792 \\
                203.000000 0.085815 \\
                204.000000 0.085842 \\
                205.000000 0.085873 \\
                206.000000 0.085910 \\
                207.000000 0.085953 \\
                208.000000 0.086005 \\
                209.000000 0.086062 \\
                210.000000 0.086136 \\
                211.000000 0.086211 \\
                212.000000 0.086301 \\
                213.000000 0.086389 \\
                214.000000 0.086479 \\
                215.000000 0.086563 \\
                216.000000 0.086625 \\
                217.000000 0.086659 \\
                218.000000 0.086679 \\
                219.000000 0.086684 \\
                220.000000 0.086646 \\
                221.000000 0.086594 \\
                222.000000 0.086557 \\
                223.000000 0.086563 \\
                224.000000 0.086539 \\
                225.000000 0.086514 \\
                226.000000 0.086469 \\
                227.000000 0.086438 \\
                228.000000 0.086386 \\
                229.000000 0.086346 \\
                230.000000 0.041203 \\
                231.000000 0.041197 \\
                232.000000 0.041200 \\
                233.000000 0.041198 \\
                234.000000 0.041199 \\
                235.000000 0.041198 \\
                236.000000 0.041199 \\
                237.000000 0.041199 \\
                238.000000 0.041199 \\
                239.000000 0.041198 \\
                240.000000 0.041198 \\
                241.000000 0.041186 \\
                242.000000 0.041185 \\
                243.000000 0.041184 \\
                244.000000 0.041182 \\
                245.000000 0.041179 \\
                246.000000 0.041175 \\
                247.000000 0.041170 \\
                248.000000 0.041163 \\
                249.000000 0.041154 \\
                250.000000 0.041143 \\
                251.000000 0.041129 \\
                252.000000 0.041115 \\
                253.000000 0.041097 \\
                254.000000 0.041078 \\
                255.000000 0.041054 \\
                256.000000 0.041030 \\
                257.000000 0.041000 \\
                258.000000 0.040965 \\
                259.000000 0.040929 \\
                260.000000 0.040892 \\
                261.000000 0.040856 \\
                262.000000 0.040813 \\
                263.000000 0.040774 \\
                264.000000 0.040737 \\
                265.000000 0.040701 \\
                266.000000 0.040666 \\
                267.000000 0.040633 \\
                268.000000 0.040600 \\
                269.000000 0.040582 \\
                270.000000 0.016607 \\
                271.000000 0.016606 \\
                272.000000 0.016602 \\
                273.000000 0.016596 \\
                274.000000 0.016589 \\
                275.000000 0.016579 \\
                276.000000 0.016568 \\
                277.000000 0.016554 \\
                278.000000 0.016536 \\
                279.000000 0.016514 \\
                280.000000 0.016487 \\
                281.000000 0.016440 \\
                282.000000 0.016405 \\
                283.000000 0.016364 \\
                284.000000 0.016320 \\
                285.000000 0.016274 \\
                286.000000 0.016226 \\
                287.000000 0.016177 \\
                288.000000 0.016133 \\
                289.000000 0.016089 \\
                290.000000 0.016050 \\
                291.000000 0.016010 \\
                292.000000 0.015974 \\
                293.000000 0.015939 \\
                294.000000 0.015911 \\
                295.000000 0.015884 \\
                296.000000 0.015859 \\
                297.000000 0.015832 \\
                298.000000 0.015809 \\
                299.000000 0.015787 \\
                300.000000 0.015768 \\
            };
        \addlegendentry{mnd}

        \addplot[name path=f,color=bycolor7,line width=1pt,opacity=0.1,forget plot] table[row sep=crcr] {
                x	y\\
                1. 0.00625 \\
                49. 0.00625 \\
                50. 0.0625 \\
                98 0.0625 \\
                98   0.125\\
                144   0.125\\
                144   0.25\\
                188   0.25\\
                188   0.5\\
                230   0.5\\
                230   1\\
                270   1\\
                270   2\\
                300   2\\
            };
        \path[name path=axis0] (axis cs:1,-0.05) -- (axis cs:300,-0.05);
        \addplot [
            thick,
            opacity=0.25,
            color=bycolor7,
            fill=bycolor7,
            fill opacity=0.25
        ]
        fill between[
                of=f and axis0,
            ];
        \addlegendentry{$\beta = 16\epsilon$}

        \draw [dashed,thin] (0,0.1) -- (300,0.1);

    \end{axis}

\end{tikzpicture}%
    \captionof{figure}{Evaluation of volume fraction error (\texttt{f}), volume-weighted compliance error (\texttt{s}) and measure of non-discreteness of the material indicator field (\texttt{mnd}) during optimisation. The shaded area indicates the $\beta$-continuation updates and the dotted line the threshold for a 10\% error ($\epsilon=0.1$).}
    \label{fig:conv_itr}
\end{Figure}
\begin{Figure}
\begin{minipage}{0.495\textwidth}
    \begin{center}
        \includegraphics[width=\linewidth]{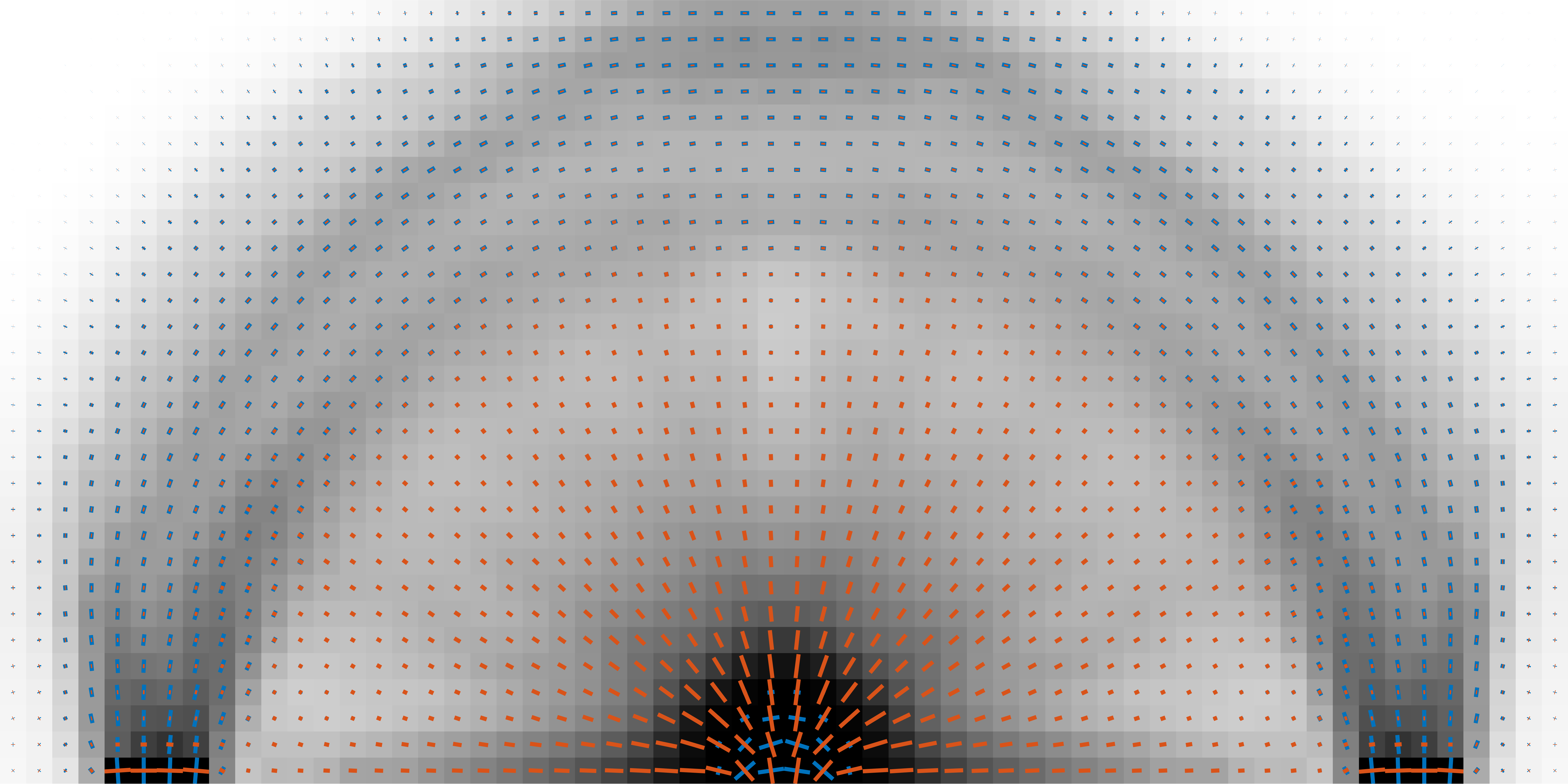}
    \end{center}
\end{minipage}
\hfill
\begin{minipage}{0.495\textwidth}
 \begin{center}
    \includegraphics[width=\linewidth]{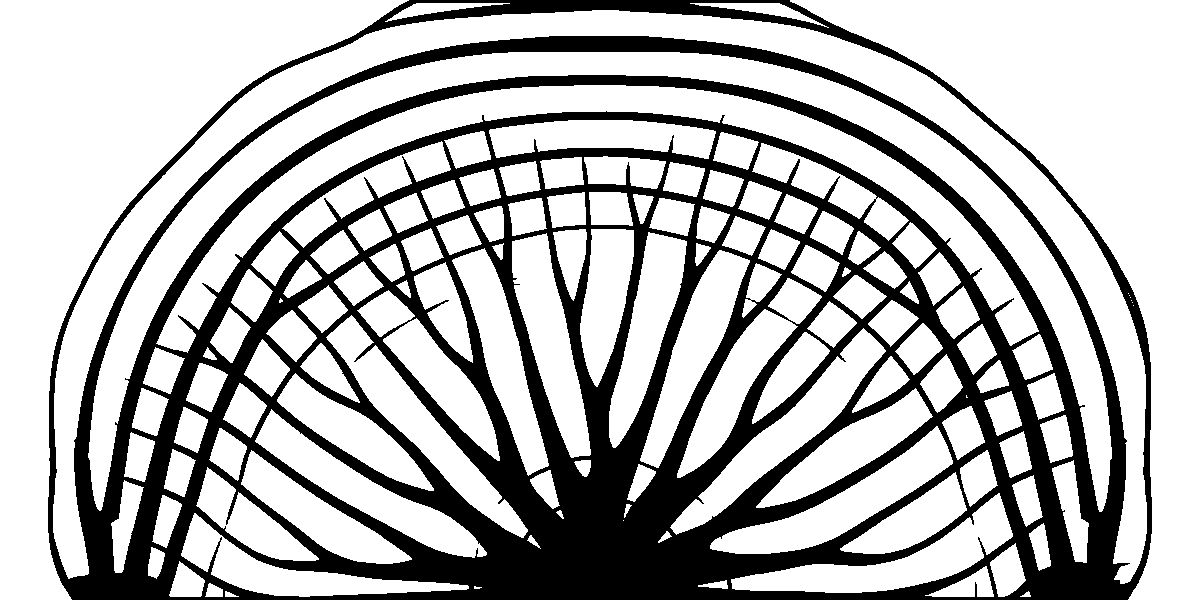}
 \end{center}
\end{minipage}
 \begin{center}
 (a) \texttt{Itr. 29}
 \end{center}
\vfill
\begin{minipage}{0.495\textwidth}
    \begin{center}
        \includegraphics[width=\linewidth]{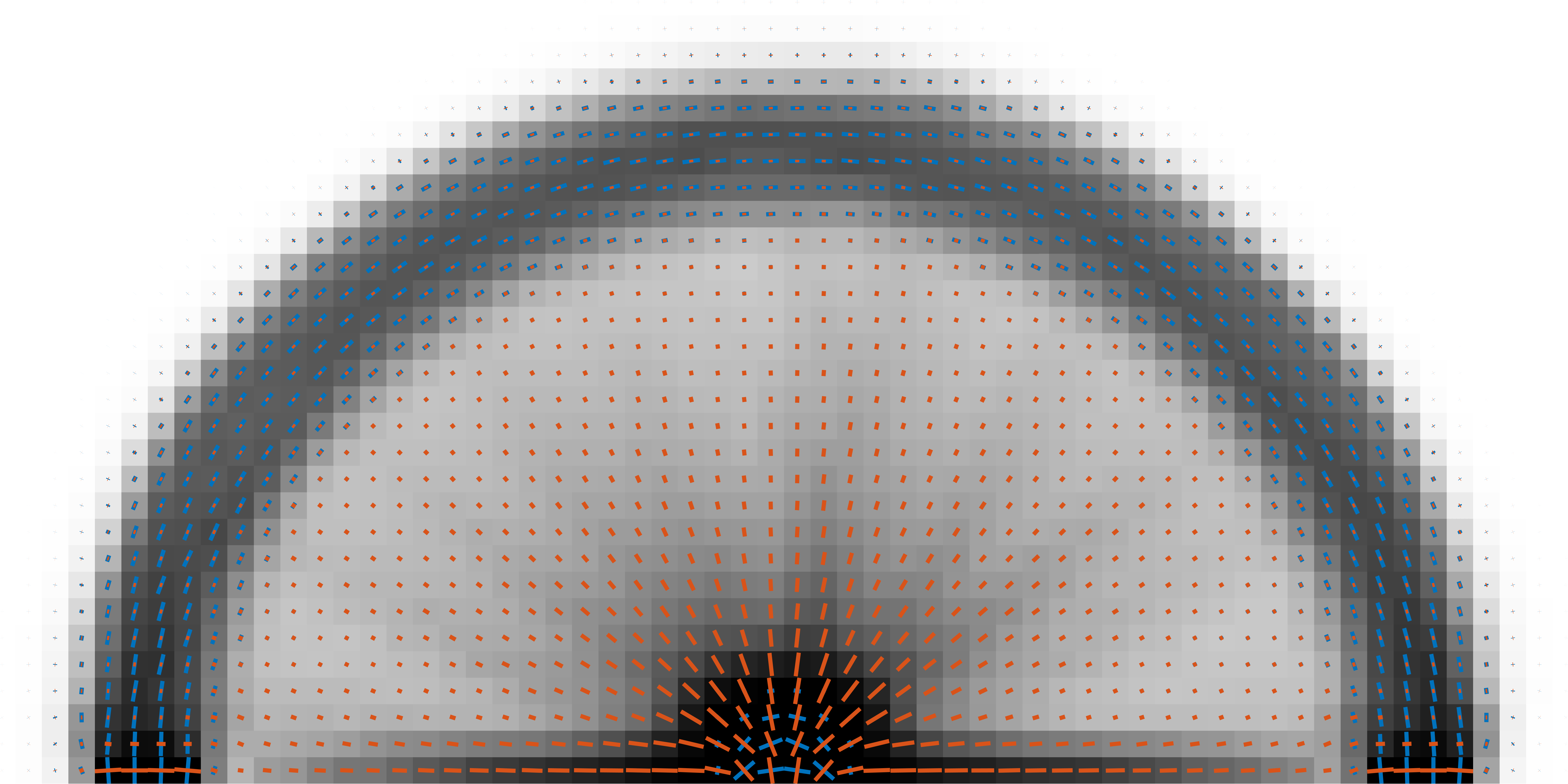}
    \end{center}
\end{minipage}
\hfill
\begin{minipage}{0.495\textwidth}
 \begin{center}
    \includegraphics[width=\linewidth]{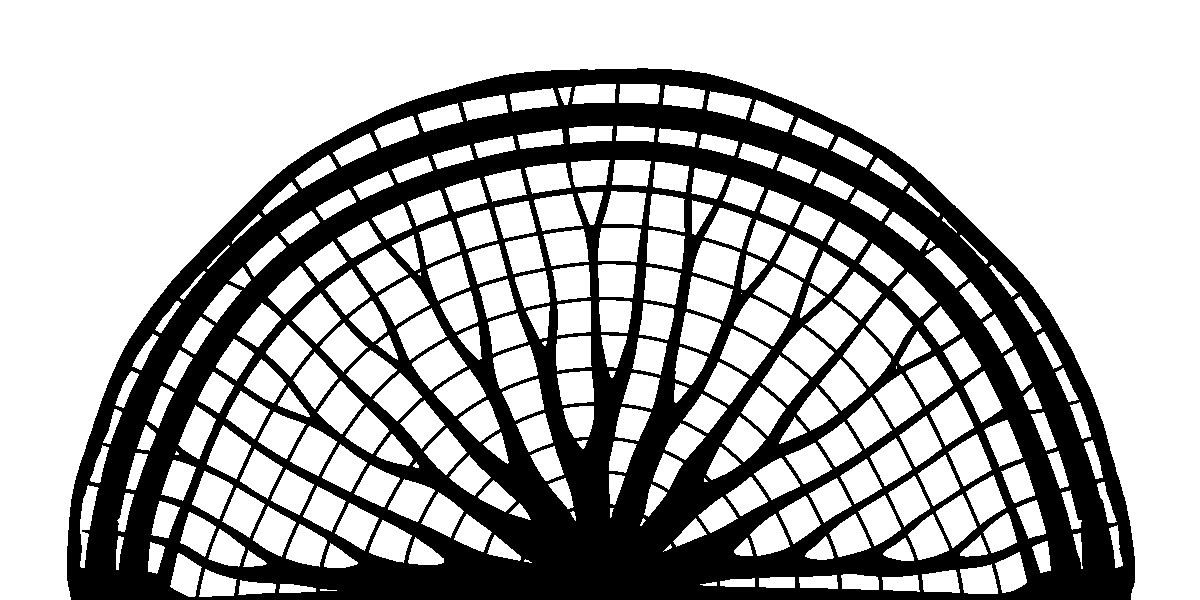}
 \end{center}
\end{minipage}
\begin{center}
 (b) \texttt{Itr. 200}
 \end{center}
\begin{center}
\captionof{figure}{Illustrating the 29th (a) and 200th (b) optimisation iteration in terms of the multi-scale designs and corresponding single-scale projections.}
    \label{fig:iter29}
\end{center}

\end{Figure}

The reason for the poor performance of the dehomogenised structures for large values of \verb|mnd|, is that the multi-scale iterate has not converged sufficiently to adhere to the minimum relative thickness constraint. To facilitate a single-scale interpretation of a multi-scale design \verb|wMin>0| must be defined. 

\autoref{fig:iter29} reveals the effect of the material indicator field convergence before and after stabilisation. At the 29th iteration the multi-scale design contains a large degree of relative thicknesses below the lower bound \verb|wMin|, imposed in the dehomogenisation, resulting in limited overlap between the single-scale projections of the lamination layers. When reaching 200 iterations, only the boundary region of the structure is affected by lower relative thicknesses, for which the lacking layer-overlap is corrected for by the threshold tolerances and location of the structural boundary.

\subsection{Rotation invariance and orientation seamlines}








Rotation invariance is a common issue for dehomogenisation due to the non-uniqueness of the Rank microtructures~\citep{Jensen2022}. For the Rank-2 microstructure, the rotation invariance is $\pi/2$, but occurrences of $\pi$-rotation discontinuities are most common. The orientation starting guess is one of the primary causes of these discontinuities in the multi-scale solution. In \deHomTopns, the starting guess is obtained from the principle stress directions, which have a $\pi$-rotation invariance, often located about symmetry lines in the design. For the bridge model, this invariance is manifested as a discontinuity in the orientations as illustrated with the vector fields in \autoref{fig:seamline_field_1}, where a ``seamline'' is visible at the centre of the bridge along the symmetry line.

To mitigate this artefact, $\pi$ is added to any negative angles in the starting guess (line 27).
\begin{lstlisting}[stepnumber=1,firstnumber=27,columns=fullflexible]
    a(not(a > 0)) = a(not(a > 0)) + pi;                              
\end{lstlisting}
This will not resolve the issue but will shift the invariance artefacts and make them less prevalent. In \autoref{fig:seamline_field_0}, the vector field with invariance shift is seen where the seamline occurrences are moved towards the domain boundary.
\begin{Figure}
    \centering
    \includegraphics[width=\linewidth]{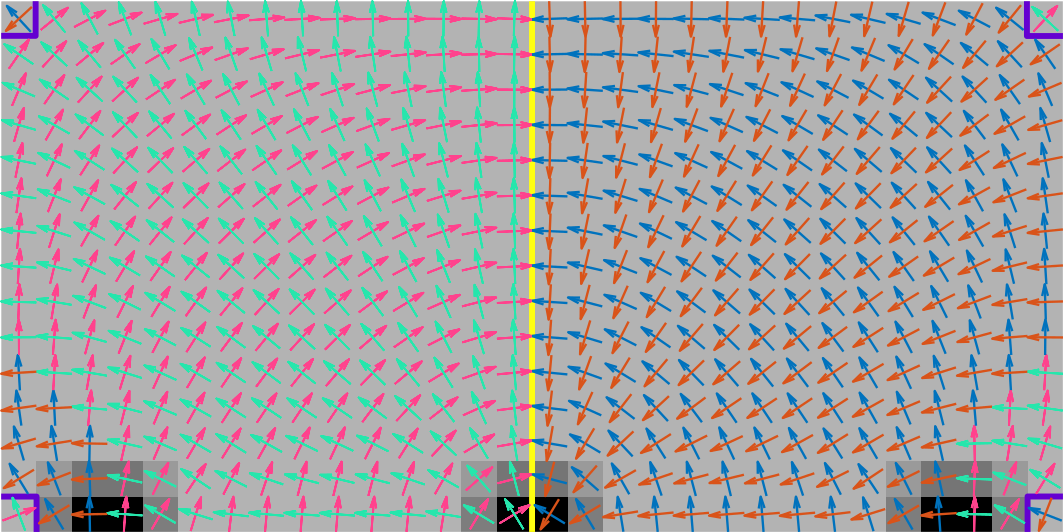}
    \captionof{figure}{Vector field of layer normals, yellow lines indicate orientation seamlines, blue lines indicate $\pi/2$-shifts in orientation. The inverted colours indicate negative angles.}
    \label{fig:seamline_field_1}
\end{Figure}

In the dehomogenisation routine, the \texttt{filterVectorField()} function further corrects local invariance artefacts as discussed in \autoref{sec:dehom_code_struct}. The heuristic correction maximises the dot-product between neighbouring phasor kernels by selecting from a set of four candidate orientation cases, consisting of unmodified, $\pi$-flip, $\pi/2$-flip, or $- \pi/2$-flip. This works well for localised discrepancies or small patches of orientation invariance, but enlarged patches may still cause issues in the dehomogenisation. %
\begin{Figure}
    \centering
    \includegraphics[width=\linewidth]{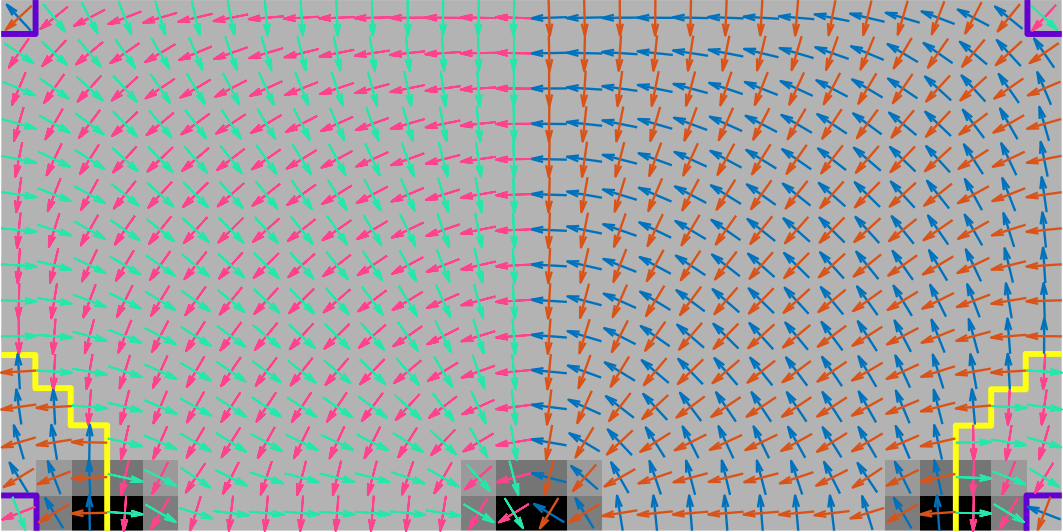}
    \captionof{figure}{Vector field of layer normals, yellow lines indicate orientation seamlines, blue lines indicate $\pi/2$-shifts in orientation. The inverted colours indicate negative angles.}
    \label{fig:seamline_field_0}
\end{Figure}
To demonstrate the effect of the seamline-discontinuity, the code is executed with parameters from \autoref{sec:howto}, removing the corrections to the starting guess (line 27). The results are presented in \autoref{tab:seamline} and the post-analysed dehomogenised structure is illustrated in \autoref{fig:seamline_eng}.

\begin{Table}
\centering
    \captionof{table}{Result from optimisation (TO) and dehomogenisation (Dehom) without angle invariance shift.}
    \begin{tabular}{@{}llll@{}}
    \toprule
                     & $J$     & $f$    & $\epsilon_\mathcal{S}$  \\ \midrule
    TO     &  10.24  & 0.300 & -              \\
    Dehom &  10.40  & 0.321 &  8.63\%         \\ \bottomrule
    \end{tabular}
    \label{tab:seamline}
\end{Table}
There is a minor improvement to both the optimised and dehomogenised results. However, the seamline causes artefacts in the post-analysed structure, and due to the heuristic nature of the phase alignment procedure, the extent of such artefacts may cause partly disconnected structural members. Therefore, to ensure a more stable performance, it is recommended to modify the starting guess or to perform post-optimisation combing as suggested in \citep{Stutz2020}.

\begin{Figure}
    \centering
    \includegraphics[width=\linewidth]{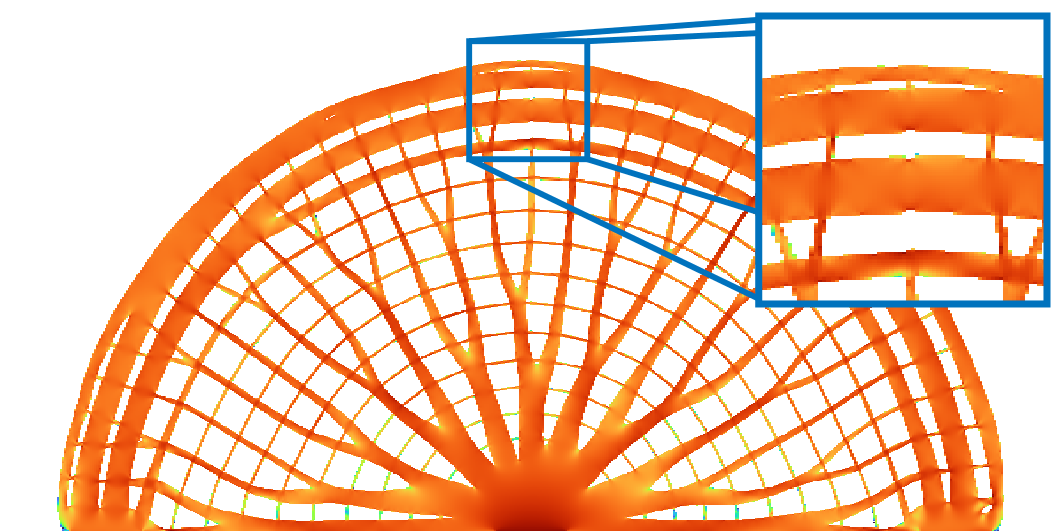}
    \captionof{figure}{Highlight of the analysed dehomogenised structure without angle invariance corrections of the optimisation starting guess.}
    \label{fig:seamline_eng}
\end{Figure}

\subsection{Additional models: MBB and DB}
\subsubsection{MBB beam model}
In addition to the bridge model and the Michell cantilever, models for the MBB beam and the double-clamped beam are included in this paper as extensions with the code. 




The $3{:}1$-aspect ratio MBB beam is slightly modified from the original model, with distributed simple supports and surface traction. To prevent undesirable mesh effects on the design, the distributed simple supports are moved away from the grid boundary, similarly to the two-load bridge from \cite{Jensen2022}. The MBB beam model is included in the code by replacing \texttt{prepFEA()} with \texttt{prepFEA\_mbb()}, and is for this demonstration executed with the following call to \deHomTopns;


\begin{lstlisting}[numbers=none,columns=fullflexible,xleftmargin=1em,framexleftmargin=0em]
deHomTop808(90,30,0.3,2,0.1,1.0,0.2,0,1);
\end{lstlisting}

A summary of the structural performance is presented in \autoref{tab:mbb}, where the dehomogenised design (\autoref{fig:mbb_dehom}) is well within 10\% of the multi-scale performance. 
\begin{Figure}
    \centering
    \includegraphics[width=\linewidth]{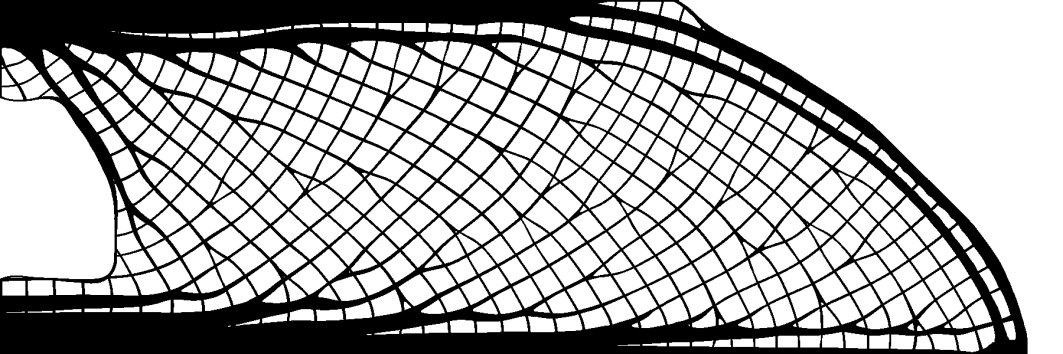}
    \captionof{figure}{Dehomogenised structure of the MBB beam model obtained by \deHomTopns.}
    \label{fig:mbb_dehom}
\end{Figure}
%
\begin{Table}
\centering
    \captionof{table}{Optimisation (TO) and dehomogenisation (Dehom) results for the MBB beam model obtained by \deHomTopns.}
    \begin{tabular}{@{}lllll@{}}
    \toprule
                     & $J$     & $f$    & $\epsilon_\mathcal{S}$ & Time [s] \\ \midrule
    TO    &  230.53  & 0.300 & -           & 13.16   \\
    Dehom &  236.45  & 0.318 &  8.70\%     & 0.95    \\ \bottomrule
    \end{tabular}
    \label{tab:mbb}
\end{Table}
%
The dehomogenised structure is seen in \autoref{fig:mbb_dehom}. The nature of the model requires many bifurcation branch points to maintain unit cell size control, which is one of the key strengths of phasor-based dehomogenisation. 
%



\subsubsection{Double-clamped beam model}
The $4{:}1$-aspect ratio double-clamped beam was also featured in the original paper on phasor-based dehomogenisation~\citep{Woldseth2023}, as a good benchmark case for orientation singularities, also exemplified in \cite{Stutz2020}. Due to the localised nature of the phasor-based approach, it offers a more stable solution in such cases, compared to alternative approaches. The double-clamped beam model is included in the code by replacing \texttt{prepFEA()} with \texttt{prepFEA\_db()}, and for exemplification executed in \deHomTop with the function call
\begin{lstlisting}[numbers=none,columns=fullflexible,xleftmargin=1em,framexleftmargin=0em]
deHomTop808(160,40,0.3,sqrt(2),0.1,1.0,0.2,0,1);
\end{lstlisting}
The structural performance obtained is presented in \autoref{tab:db} and the single-scale structure is illustrated in \autoref{fig:db_dehom}. Even though the \deHomTop code is a simplification of the proposed method of \cite{Woldseth2023}, the code is robust in handling orientation singularities and achieves solutions approaching the same quality level.

\begin{Table}
\centering
    \captionof{table}{Optimisation (TO) and dehomogenisation (Dehom)  results for the double-clamped beam model obtained by \deHomTopns.}
    \begin{tabular}{@{}lllll@{}}
    \toprule
                     & $J$     & $f$    & $\epsilon_\mathcal{S}$ & Time [s] \\ \midrule
    TO    &  22.84  & 0.300 & -           & 25.30   \\
    Dehom &  23.33  & 0.315 &  7.12\%     & 2.54    \\ \bottomrule
    \end{tabular}
    \label{tab:db}
\end{Table}

\begin{Figure}
    \centering
    \includegraphics[width=\linewidth]{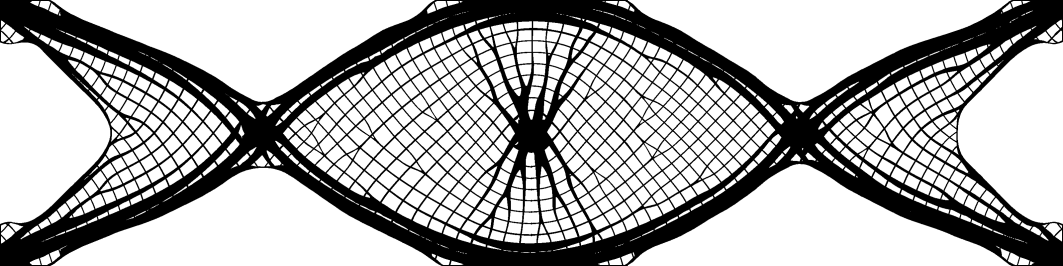}
    \captionof{figure}{Dehomogenised structure of the double-clamped beam model obtained by \deHomTopns.}
    \label{fig:db_dehom}
\end{Figure}


\section{Extensions}\label{sec:extension}
One of the goals of \deHomTop is to provide the user with a baseline code for further adaptions to application needs. This section will cover some examples of how to integrate different code extensions, showing how \deHomTop allows for additional problem requirements or dehomogenisation complexity.

\subsection{Passive design variables}
Extending the passive design beyond a solid domain, to allow for continuous design values, presents many use cases for a multi-scale design. One such use case is to predefine infill density or orientation, where the passive domain $\Omega_P$ is extended for any passive design variable, $w_1$, $w_2$, $a$, or $s$.

This control is achieved by replacing the array \texttt{pasE} with a \texttt{struct} containing member arrays \texttt{w1}, \texttt{w2}, \texttt{a}, and \texttt{s}, each of size $\mathbb{R}^{N_p \times 2}$, where $N_p$ are the individual number of passive variables. The first column of the arrays indicates the element index, while the second column indicates the passive value. The indicator variable is implicitly determined based on the values of \texttt{w1} and \texttt{w2}. Note that this can cause conflicts due to the requirement that the two layers share material indicator field. This means that if the user specifies that $w_1$ is passive solid and $w_2$ is passive void in the same element, the optimisation problem is ill-defined for the \deHomTop-implementation. 
\begin{Figure}
    \centering
    \includegraphics[width=\linewidth]{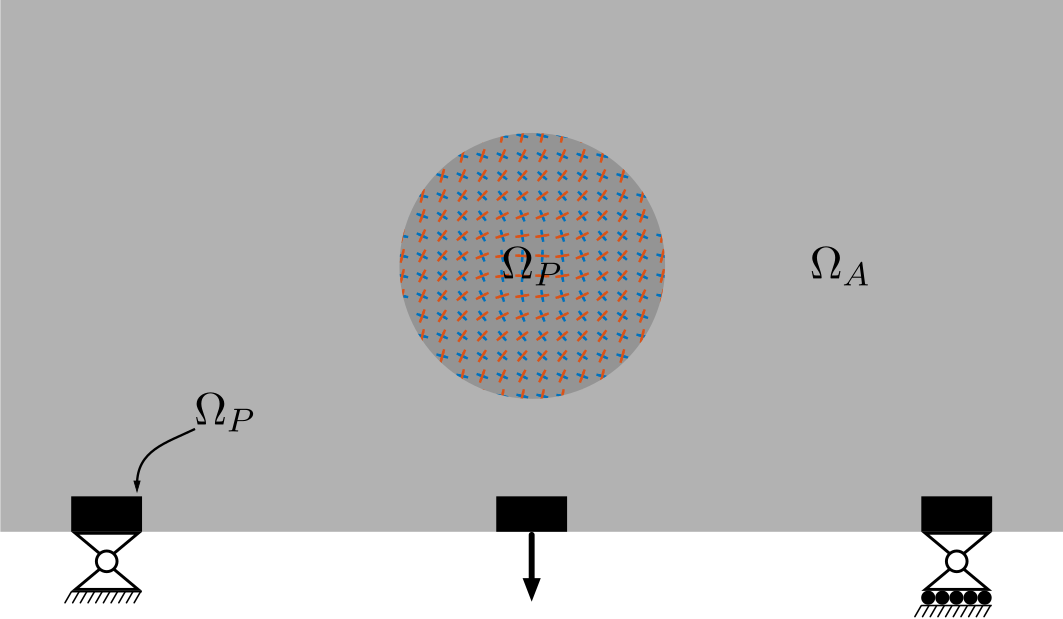}
    \captionof{figure}{Bridge model with a partial passive domain in the centre, where $w_1$, $s$ and $a$ are assigned pre-determined values.}
    \label{fig:model_pas}
\end{Figure}
\begin{Figure}
    \centering
    \includegraphics[width=\linewidth]{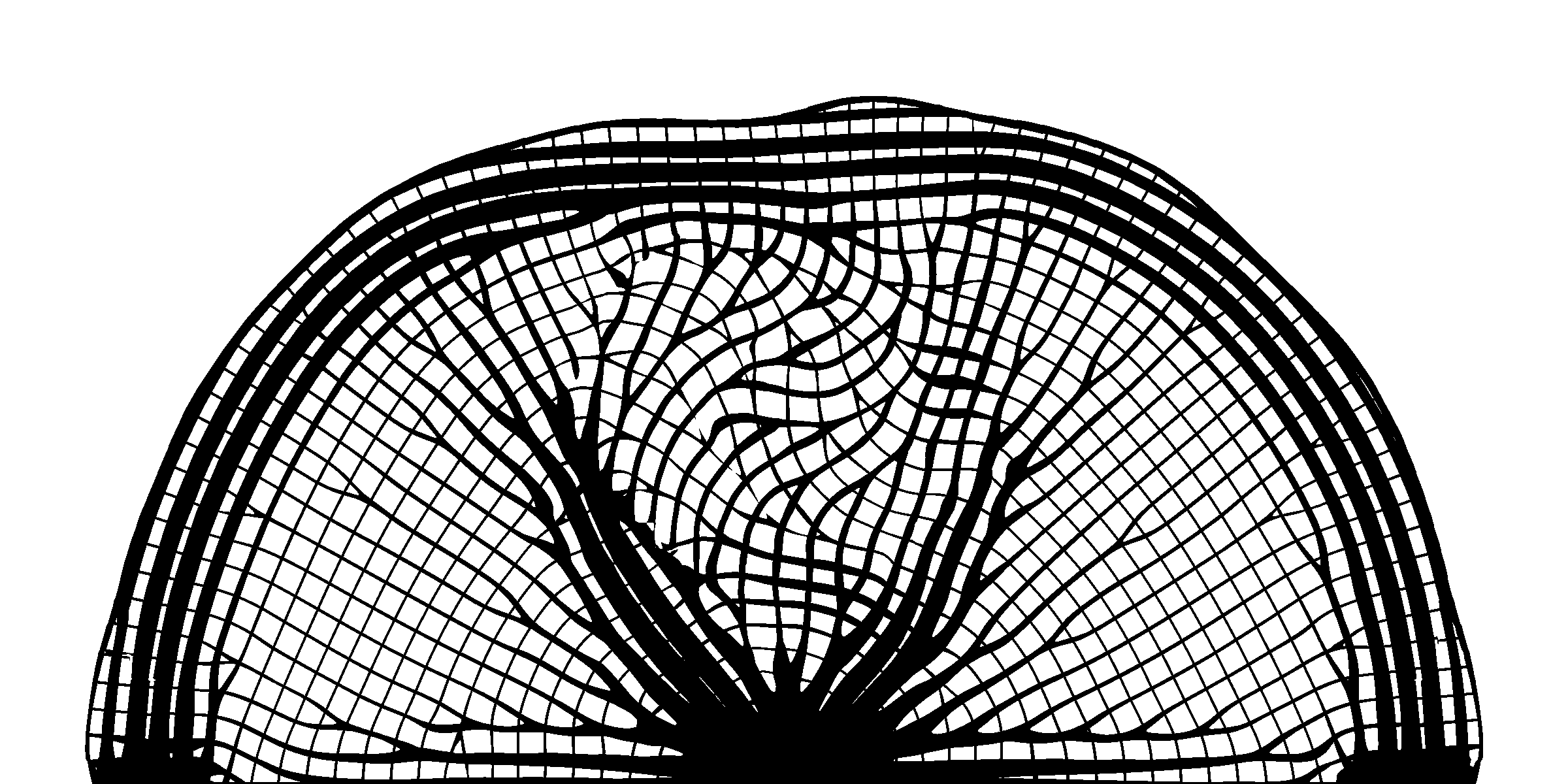}
    \captionof{figure}{Dehomogenised optimised bridge with a partial passive center domain according to the example case in \autoref{fig:model_pas}.}
    \label{fig:pas_extension}
\end{Figure}
The code has to be modified in six places in order to have multiple passive variables in an element. The modifications are here described in chronological order.
First to set active elements and design variables, replace line 11 and 12 with,
\begin{lstlisting}[numbers=none,columns=fullflexible,xleftmargin=1em,framexleftmargin=0em]
actE.w1 = setdiff(1:ne, pasE.w1(:,1))'; 
actE.w2 = setdiff(1:ne, pasE.w2(:,1))';
actE.s = setdiff(1:ne,[pasE.s(:,1);pasE.w1(:,1);pasE.w2(:,1)])';
dv =union(setdiff(1:ne, pasE.a(:,1))+2*ne,[actE.w1;actE.w2+ne;actE.s+3*ne]);
\end{lstlisting}
This replaces the \texttt{actE} array with a \texttt{struct} that holds information on active design elements for the later filtering operation. To assign the passive values, replace line 34 with,
\begin{lstlisting}[numbers=none,columns=fullflexible,xleftmargin=1em,framexleftmargin=0em]
s([pasE.w1(:,1);pasE.w2(:,1);pasE.s(:,1)]) = ...   
    [[pasE.w1(:,2); pasE.w2(:,2)]>0; pasE.s(:,2)]; 
[w(pasE.w1(:,1),1),w(pasE.w2(:,1),2)] = deal(pasE.w1(:,2),pasE.w2(:,2));  
a(pasE.a(:,1)) = pasE.a(:,2);                             
pasE.w1(:,2) = []; pasE.w2(:,2) = []; pasE.a(:,2) = []; pasE.s(:,2) = []; 
\end{lstlisting}
The final line removes the passive values, as they are unnecessary beyond this point. For the dehomogenisation, only solid passive elements are updated, insert the following between lines 45 and 46,
\begin{lstlisting}[numbers=none,columns=fullflexible,xleftmargin=1em,framexleftmargin=0em]
pasRhoPhys = union(pasRhoPhys.w1(pasRhoPhys.w1(:,2) == 1,1),...          
    pasRhoPhys.w2(pasRhoPhys.w2(:,2) == 1,1));
\end{lstlisting}

For the filling step, only active design variables should be updated, replace line 78 with,
\begin{lstlisting}[numbers=none,columns=fullflexible,xleftmargin=1em,framexleftmargin=0em]
    sTilde(actE.s) = sBar(actE.s);                                
    wTilde(actE.w1,1) = wBar(actE.w1,1);                                  
    wTilde(actE.w2,2) = wBar(actE.w2,2);               
\end{lstlisting}
Furthermore, for the sensitivities, only active design variables should be updated; replace line 107 with,
\begin{lstlisting}[numbers=none,columns=fullflexible,xleftmargin=1em,framexleftmargin=0em]
    [dmudw(pasE.w1,1,:),dfdw(pasE.w1,1),dSds(pasE.s)]=deal(0);
    [dmudw(pasE.w2,2,:),dfdw(pasE.w2,2),dJda(pasE.a)]=deal(0);
\end{lstlisting}

Finally, the \texttt{struct} with information on passive variables must be introduced for the FE-model. If a variable has no passive domain, then the member array must be initialised with \texttt{zeros(0,2)}. For the example shown in \autoref{fig:model_pas} with a semi-passive circle, replace line 218 with,
\begin{lstlisting}[numbers=none,columns=fullflexible,xleftmargin=1em,framexleftmargin=0em]
pasBC = reshape([forceElms, pasBC1, pasBC2] - (0:(py-1))',[],1);     
sdfCircle =sqrt(((1:nelX)-nelX/2-1/2).^2+ ((1:nelY)-nelY/2-1/2)'.^2)-nelY/4;
pasCircle = find(sdfCircle<=0);                                            
passiveElms.w1 =[[pasBC, 1.0*(pasBC>0)]; [pasCircle, 0.25*(pasCircle>0)]]; 
passiveElms.w2 =[pasBC, 1.0*(pasBC>0)];  
passiveElms.a = [pasCircle, pi/2+atand(sdfCircle(pasCircle)./(4*nelX))];   
passiveElms.s = zeros(0,2);                                                
\end{lstlisting}

Executing the code with 
\begin{lstlisting}[numbers=none,columns=fullflexible,xleftmargin=1em,framexleftmargin=0em]
    deHomTop808(60,30,0.3,2,0.1,1.0,0.1,0,1);
\end{lstlisting}
will result in the dehomogenised result seen in \autoref{fig:pas_extension} and \autoref{tab:pas_extension}. Due to the rapid orientation changes within the passive circle compared to the dehomogenised length-scale, excess material is introduced in the branch closure procedure. This causes a larger volume fraction deviation between the homogenised and dehomogenised solutions, but the improvement in compliance ensures that the volume-weighted compliance error is within 10\%.

\begin{Table}
\centering
    \captionof{table}{Optimisation (TO) and dehomogenisation (Dehom)  results of bridge with a partial passive center domain.}
    \begin{tabular}{@{}lllll@{}}
    \toprule
                     & $J$     & $f$    & $\epsilon_\mathcal{S}$ & Time [s] \\ \midrule
    TO     &  11.87  & 0.300 & -           & 12.09   \\
    Dehom &  12.39  & 0.31 &  7.92\%     & 1.62    \\ \bottomrule
    \end{tabular}
    \label{tab:pas_extension}
\end{Table}





\subsection{Multiple loading cases for dehomgoenisation}
Homogenisation-based topology optimisation can also be applied to optimise multi-load problems, as described in \citet{Jensen2022}. \deHomTop is not designed to consider multiple loads for optimisation and analysis, but the phasor-based dehomogenisation procedure is directly applicable to Rank-$N$ multi-scale structures. Given a multi-scale solution obtained through appropriate optimisation, \deHomTop can be utilised to obtain a dehomogenised design at any desired finite minimum length-scale \verb|dmin|, requiring only modifications of the \verb|passiveElms| definition (line 207) within the \verb|prepFEA()| sub-function to fit the desired problem formulation. 
To exemplify this independency of the phasor-based dehomogenisation procedure on the number of lamination layers in the provided solution, the multi-scale solution to one of the 2-load bridge examples from \citet{Jensen2022}, \verb|twoLoadBridge_80_48_Rank3_data.mat|, is included with the code extensions. To account for the change in passive regions, compared to the default single-load bridge model, the function \verb|getPas_2loadbridge()| is included as an extension, and is incorporated as a replacement of the fine-scale passive elements definition \texttt{pasRhoPhys} (line 39).
\begin{lstlisting}[stepnumber=1,firstnumber=39,columns=fullflexible]
pasRhoPhys = getPas_2loadbridge(nelX,nelY,deHomGrid.fscale);
\end{lstlisting}
The setup and function call to execute the dehomogenisation procedure, given the correct \verb|pasRhoPhys| definition, is summarised as follows
\begin{lstlisting}[numbers=none,columns=fullflexible,,xleftmargin=1em,framexleftmargin=0em, deletekeywords={eval}]
load('twoLoadBridge_80_48_Rank3_data.mat') % load dataset
[rmin, deHomFrq, eval] = deal(2, 0, false); % suggested default
[nelX,nelY] = deal(nDim(1),nDim(2)); % domain dimensions of multi-scale solution
TO.w = w; % (nelY*nelX)x(#layers) array of layerwise thicknesses
TO.N = zeros(size(N,1),2,size(N,2)/2); % (nelY*nelX)x(2)x(#layers) array of orientation vectors
TO.N(:,1,:)=N(:,1:2:end); TO.N(:,2,:)=N(:,2:2:end); % insert x- and y- components
[TO.f, volFrac] = deal(0.3); % volume fraction of multiscale solution
TO.J = c; % multi-scale compliance if needed for comparison
[wMin, wMax] = deal(eta(1),eta(2)); % minimal and maximal relative thicknesses
dmin = 0.1; % desired physical minimum length scale
rhoPhys = deHomTop808(nelX,nelY,volFrac,rmin,wMin,wMax,dmin, deHomFrq,eval,TO);
\end{lstlisting}
Note that the \verb|rmin| and \verb|deHomFrq| inputs must be appropriately defined for the program to not fail during the initialisation process, and that \verb|eval=false| is necessary unless \deHomTop,\verb|prepFEA()| and \verb|doFEA()| have been modified to the considered multi-load example. 

An alternative approach of isolating the dehomogenisation procedure from \deHomTop is available, where a Rank-N multiscale structure can be dehomogenised by the following function-call sequence
\begin{lstlisting}[numbers=none,columns=fullflexible,,xleftmargin=1em,framexleftmargin=0em, deletekeywords={eval}]
deHomGrid = prepPhasor(nelX,nelY,dmin,wMin);
align = prepPhasorKernels(deHomGrid); 
alignItr = 20; 
rhoPhys = phasorDehomogenise(deHomGrid,wMin,TO.w,TO.N,alignItr,align); 
pasRhoPhys = getPas_2loadbridge(size(rhoPhys,2),size(rhoPhys,1),1);
rhoPhys(pasRhoPhys) = 1;
\end{lstlisting}
In this latter case, the passive regions are imposed to \verb|rhoPhys| after the function call to \verb|phasorDehomogenise()|. 


The resulting dehomogenised design for \verb|dmin=0.1| is illustrated in \autoref{fig:2loadbridge_dehom}. \autoref{tab:twoloadbridge} and its solution quality, obtained outside of \deHomTopns, compared to the phasor-based result obtained from \citet{Woldseth2023} in \autoref{tab:twoloadbridge}.
\begin{Table}
\centering
    \captionof{table}{Rank-N dehomogenisation utilising \deHomTop exemplified by the 2-load bridge example from \citet{Jensen2022} analysed by the FEA-setup from and compared to the dehomogenised solution from \citet{Woldseth2023}.}
    \begin{tabular}{@{}lcc@{}}
    \toprule
    & \deHomTopns &   \cite{Woldseth2023}\\\midrule
    $J$     & 5.785 & 5.334\\
    $f$    & 0.297 & 0.314\\
    $\epsilon_\mathcal{S}$ & 6.94\% & 4.35\%\\
    Time [s] & 4.233 & 5.530\\\bottomrule
    \end{tabular}
    \label{tab:twoloadbridge}
\end{Table}
The solution quality obtained utilising \deHomTop is reduced compared to the original phasor-based dehomogenisation procedure from \citet{Woldseth2023}, but the process is executed more efficiently when run on the same machinery. This loss in solution quality is a part of the trade-off between an on-the-fly compact version for educational purposes and the more heuristically detailed and adapted version presented in \citet{Woldseth2023}. For most problem cases, these changes instigate a moderate loss in structural performance at the benefit of simplicity and significant computational speed-up, but for extended applications it might be useful to consider some version of the extensions included in the original procedure.
\begin{Figure}
    \centering
    \includegraphics[width=\linewidth]{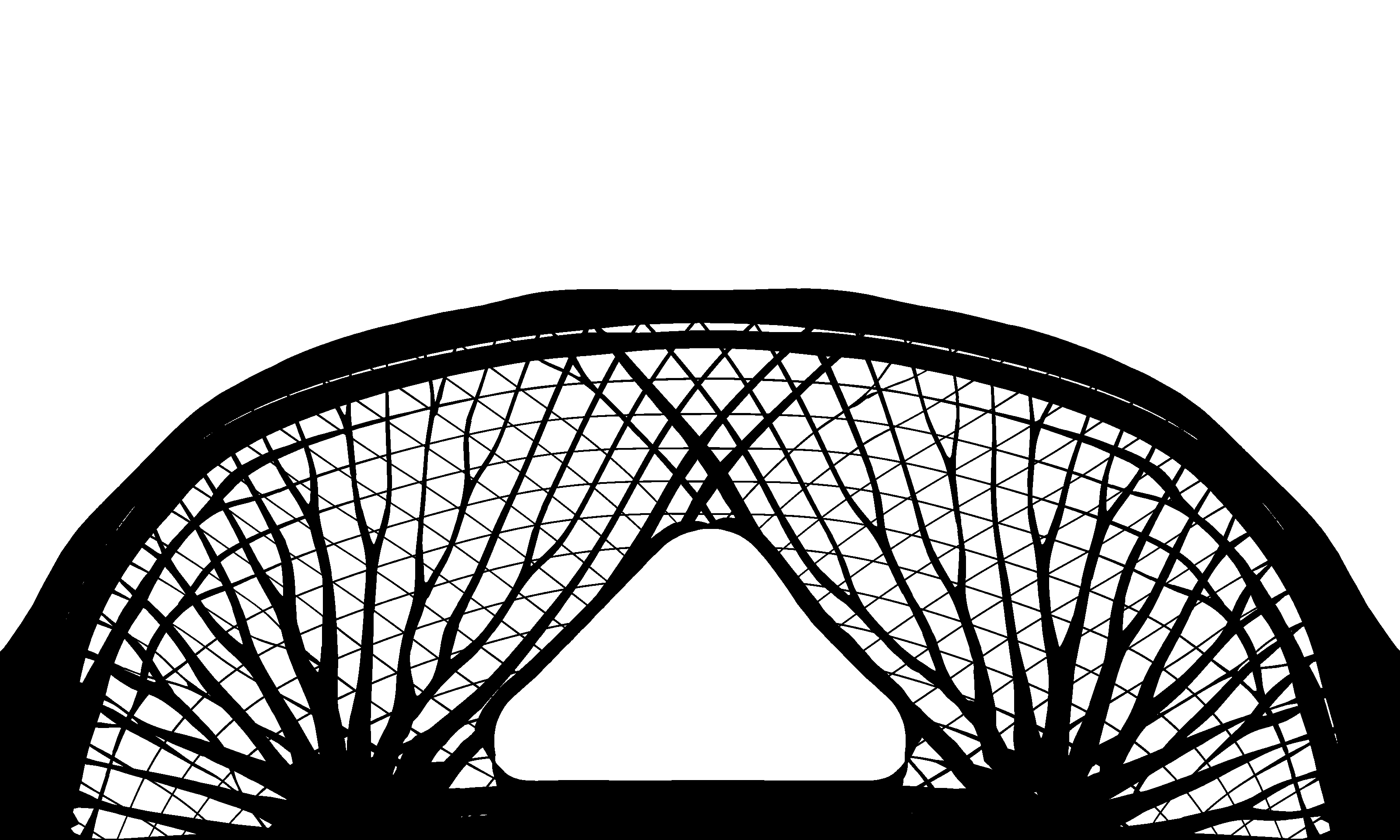}
    \captionof{figure}{The multi-scale 2-load bridge example from \citet{Jensen2022}, optimised using starting guess $SG_A$, dehomogenised by \deHomTopns.}
    \label{fig:2loadbridge_dehom}
\end{Figure}

\section{Discussion and concluding remarks}\label{sec:end}

This paper presents a Matlab code for multi-scale topology optimisation that maximises the stiffness of single-load case problems and allows for subsequent on-the-fly phasor-based dehomogenisation, including analysis options for verifying the obtained design. The code presents a highly efficient framework, capable of producing dehomogenised designs that, for realistic length-scales, performs within 10\% of the stiffness-optimised design, as proven by the length-scale convergence results. These solution qualities are on par with those obtained by established state-of-the-art dehomogenisation method, and are orders of magnitude faster to obtain.



Different model examples are provided with the code, which shows the ease of extending the code to other applications. This is further emphasised with the extension of partially passive domains, allowing for complex and advanced examples. The dehomogenisation code also allows for the realisation of any lamellar-based multi-scale structure.

The on-the-fly dehomogenisation feature of the code demonstrates real-time convergence of the mapped design. From the results, it is clear that the structural integrity of the dehomogenised structure is heavily influenced by the convergence of the multi-scale indicator field. Nevertheless, this suggests a potential future extension for interactive applications \citep{AageNobel2013}, opening up new possibilities for real-time design adjustments and optimisations.

The process of multi-scale topology optimisation and dehomogenisation is quite complex and heuristic in nature. However, by presenting this procedure as an educational Matlab code, the goal is to encourage researchers and engineers to consider multi-scale topology optimisation as an alternative to conventional large-scale approaches.


  \section{Acknowledgments}
  The authors acknowledge the financial support from the InnoTop VILLUM investigator project through the Villum Foundation and nTopology inc. Furthermore, the authors would like to express their gratitude to Dr. Federico Ferrari for valuable discussions during the preparation of this work.
  \bibliographystyle{apalike}
  \bibliography{database}
\end{multicols}
\FloatBarrier
\appendix
\renewcommand{\thesection}{\Alph{section}}
\section{\deHomTop}\label{app:code}
The code, including the Michell cantilever beam, MBB beam and double-clamped beam model, can be downloaded from \url{https://github.com/peterdorffler/deHomTop808.git}. 




\end{document}